\documentclass[symmetry,article,submit,pdftex,moreauthors]{Definitions/mdpi} 
\usepackage{empheq}
\usepackage{mathrsfs} 
\usepackage{amsmath,bm}
\firstpage{1} 
\pubvolume{1}
\issuenum{1}
\articlenumber{0}
\pubyear{2025}
\copyrightyear{2025}
\datereceived{ } 
\daterevised{ } 
\dateaccepted{ } 
\datepublished{ } 
\hreflink{https://doi.org/} 

\Title{Nonlocal effective field theory and its applications}

\TitleCitation{Nonlocal effective field theory and its applications}

\Author{P. Wang $^{1,2\ *}$, Zhengyang Gao $^{1,2}$ Fangcheng He $^{3}$, Chueng-Ryong Ji $^{4}$, W. Melnitchouk $^{5}$ and Y. Salamu $^{6}$}

\address{%
$^{1}$ \quad Institute of High Energy Physics, Chinese Academy of Sciences, Beijing 100049, China\\
$^{2}$ \quad College of Physics Sciences, University of Chinese Academy of Sciences, Beijing 100049, China \\
$^{3}$ \quad Department of Physics and Astronomy, Stony Brook University, Stony Brook, New York 11790, USA\\
$^{4}$ \quad Department of Physics, North Carolina State University, Raleigh, North Carolina 27695, USA \\
$^{5}$ \quad Jefferson Lab, Newport News, Virginia 23606, USA \\
$^{6}$ \quad School of Physics and Electrical Engineering, Kashi University, Kashgar, 844000, Xinjiang, China}

\corres{Correspondence: pwang4@ihep.ac.cn}

\abstract{We review recent applications of nonlocal effective field theory, focusing in particular on nonlocal chiral effective theory and nonlocal quantum electrodynamics (QED), as well as an extension of nonlocal effective theory to curved spacetime. For the chiral effective theory, we discuss the calculation of generalized parton distributions (GPDs) of the nucleon at nonzero skewness, along with the corresponding gravitational (or mechanical) form factors, within the convolution framework. In the QED application, we extend the nonlocal formulation to construct the most general nonlocal QED interaction, in which both the propagator and fundamental QED vertex are modified due to the nonlocal Lagrangian, while preserving the Ward-Green-Takahashi identities. For consistency with the modified propagator, a solid quantization is proposed, and the nonlocal QED is applied to explain the lepton $g-2$ anomalies without the introduction of new particles or interactions. Finally, with an extension of the chiral effective action to curved spacetime, we investigate the nonlocal energy-momentum tensor and gravitational form factors of the nucleon with a nonlocal pion-nucleon interaction.}


\begin{document}

\section{Introduction}

Since the discovery of the finite size of the proton in the 1950s, extensive theoretical and experimental work has greatly expanded our knowledge of the inner structure of the nucleon. Examples that have challenged our understanding include the proton spin puzzle~\cite{Ashman, Anthony, Abe2, Anthony2, Anthony5, Ackerstaff1, Airapetian0, Airapetian1, Adeva, Adeva2, Alexakhin, Alekseev, Prok2, Fersch, Parno, Posik, Solvignon, Armstrong2, Ethier}, the flavor asymmetry in the proton sea~\cite{Arneodo, Ackerstaff2, Baldit, Towell, Dove01, Dove02, Thomas01, Signal, Kumano, Speth, Chang, Alberg, Thomas02, Chen1, Salamu01, Salamu02, Accardi1, Duan}, and more recently the proton electric radius puzzle~\cite{Xiong, Bezginov, Sick, Sick2, Hill, Mohr, Mohr2, Mohr3, Pohl, Antognini}. With the increased energies and luminosities available at new facilities and upgrading of experimental equipment, considerable progress has been made in recent years in extracting quantitative information about nucleon structure, characterized in the form of nucleon form factors, parton distribution functions (PDFs), and generalized parton distributions (GPDs).

GPDs in particular contain rich information about the 3-dimensional structure of hadrons, in terms of their fundamental quark and gluon constituents. These describe the distributions of partons carrying a specific fraction $x$ of the hadron’s light-front momentum, squared four-momentum transfer $t$, and skewness $\xi$. Information about GPDs of the nucleon, or more specifically about their Compton form factors, can be obtained through processes such as deeply-virtual Compton scattering (DVCS), deeply-virtual meson production (DVMP), and time-like Compton scattering (TCS). More recently, a new set of processes, referred to as single diffractive hard exclusive photoproduction (SDHEP), has been proposed as a way of accessing the $x$ dependence of the GPDs directly from experiment~\cite{Qiu:2022pla, Qiu:2023mrm}.

The determination of GPDs has become one of the important scientific goals at accelerator facilities around the world. Many measurements of observables sensitive to nucleon GPDs or their moments have been performed over a wide range of kinematics through unpolarized and polarized experiments at DESY (HERMES, H1, and ZEUS experiments) and Jefferson Lab (CLAS)~\cite{Camacho, Defurne, Airapetian, Airapetian6, Airapetian7, Airapetian8, Adloff, Aaron, Chekanov, Hadjidakis, Morrow, Bedlinskiy}. Future experiments are planned for Jefferson Lab's Halls~A, B and C, at J-PARC in Japan, COMPASS at CERN, and the planned Electron-Ion Collider (EIC) and Large Hadron-electron Collider (LHeC)~\cite{Gao2, Kubarovsky, Armstrong3, d'Hose, Silva, Kouznetsov, Sawada, Kroll, Accardi2, Fernandez}.

The GPDs are closely related to the electromagnetic elastic and transition form factors. Integrating over the momentum fraction $x$, one obtains the Mellin moments of the GPDs, or generalized form factors (GFFs). For the lowest order form factors there has been considerable theoretical and experimental study over the past decades. Higher order generalized form factors, such as the gravitational or mechanical form factors, have attracted growing interest recently~\cite{Freese:2025tqd}. The gravitational form factors are related to the matrix elements of the energy-momentum tensor, reflecting the bulk properties of hadrons, such as mass, spin, and pressure, and can be extracted from DVCS and exclusive meson production~\cite{Burkert, Collins}. Rather than integrating over the GPDs, the gravitational form factors can also be obtained directly from lattice QCD or chiral effective theory calculations~\cite{Hackett, Shanahan1, Alharazin, Alharazin2}. The GPDs provide information about the distribution of partons inside the nucleon, while the GFFs reflect information about the overall properties of the nucleon. These two types of functions provide us with important insights into the structure of the nucleon from different perspectives.

On the theory side, lattice QCD is the most rigorous approach that is based directly on the fundamental QCD theory. Since PDFs are defined in Minkowski space, it has not been possible to simulate PDFs directly on a Euclidean lattice. In recent years, however, quasi-PDFs have been proposed as a means of indirectly accessing the light-front distributions from lattice simulations within the large momentum effective theory (LaMET) approach~\cite{Ji}. Alternatively, pseudo-PDFs have also been suggested, involving ratios of equal-time matrix elements of the Wilson line between quarks to the rest-frame density matrix elements, and parametrized in terms of the Ioffe time~\cite{Radyushkin, Radyushkin2, Ioffe, Braun, Orginos2017, Joo2020}. A number of lattice groups have been engaged in computing unpolarized PDFs \cite{Lin4, Alexandrou11, Chen2}, polarized PDFs \cite{Chen1, Alexandrou12, Green3} and transversity PDFs \cite{Chen3, Alexandrou12, Alexandrou13} using either the quasi-PDF or pseudo-PDF approach, although the efforts are generally still in their early stages. In addition, the higher order GFFs, such as the gravitational form factors, have also been simulated on the lattice \cite{Hackett, Shanahan1}.

Another systematical approach used in hadronic physics applications has been chiral effective field theory (EFT), a specific example of which is chiral perturbation theory ($\chi$PT). This allows the description of low energy properties and processes within a perturbative approach, emphasizing the chiral symmetry aspects of QCD. Historically, most formulations of $\chi$PT have utilized dimensional or infrared regularization~\cite{Becher, Ellis2, Kubis, Borasoy} and the extended on-mass-shell renormalization scheme~\cite{Bauer, Fuchs, Schindler}. Although $\chi$PT has been a fairly successful approach, it can only describe physical quantities at low momentum transfers. For example, for the nucleon form factors it is only valid at relatively small $Q^2$ values, $Q^2 \lesssim 0.1$~GeV$^2$~\cite{Fuchs2}. The range can be extended up to $Q^2 \lesssim 0.4$~GeV$^2$ by explicitly including vector meson degrees of freedom into the theory~\cite{Kubis, Kubis2}. An alternative regularization method, finite range regularization (FRR), has been argued to achieve better convergence than dimensional regularization in the calculation of many hadronic observables~\cite{Young, Young2, Leinweber}. EFT with FRR has been applied in the investigation of the vector meson mass, nucleon magnetic moments, electric and magnetic radii, the $Q^2$ dependence of electroweak form factors, and moments of PDFs and GPDs~\cite{Wang, Wang2, Hall, Li, Shanahan2}. Earlier treatments with nonrelativistic regulators in the FRR EFT have also more recently been generalized to relativistic ones, by making the regulators four-dimensional functions or using Pauli-Villars regularization~\cite{Wang:2016ndh, XGWang2}.

In recent years, a nonlocal chiral effective theory has been proposed, which reflects the non-point-like properties of hadrons. In the nonlocal theory, the local meson-baryon interactions are replaced by nonlocal interactions, where the baryon field is defined at a spacetime point $x$ and the meson or photon field is displaced by a distance $a$ to spacetime point $x + a$~\cite{Review}. A correlation function, $F(a)$, parametrizes the nonlocal interaction. To guarantee local gauge invariance, a gauge link is introduced, generating additional Feynman diagrams that are essential for preserving charge conservation. The nonlocal chiral effective theory has been successfully applied to the calculation of nucleon form factors, strange form factors, unpolarized and polarized PDFs, and GPDs with zero and nonzero skewness~\cite{pwang0,h1, He:2017viu, He:2018eyz, Salamu:2018cny, Salamu:2019dok, Yang:2020rpi, He:2022leb, h6, Gaozhengyang}.

Beyond chiral effective theory, in Refs.~\cite{Li1,Li2} a nonlocal formulation of QED was proposed and applied to the study of the lepton $g-2$ anomaly. The anomalous magnetic moments of the electron and muon, $a_e$ and $a_\mu$, are some of the most precisely determined quantities in particle physics~\cite{Aoyama3}. Recent measurement of the muon anomalous magnetic moment in the E989 experiment at Fermilab showed a $3.3\sigma$ discrepancy from the standard model (SM) prediction \cite{Abi}. Combined with the previous E821 result from BNL \cite{Bennett}, the measurements revealed a $4.2\sigma$ deviation from the SM. For the electron, the most accurate measurement of $a_e$ was carried out by the Harvard group, and the discrepancy from the SM with the fine structure constant $\alpha$ measured at Berkeley was $2.4 \sigma$ \cite{Aoyama,Parker,Hanneke}. However, a new determination of $\alpha$ from the Laboratoire Kastler Brossel (LKB) improves the accuracy by a factor of 2.5 compared to the previous Berkeley measurement~\cite{Morel}. With this new $\alpha$, the SM prediction for the electron magnetic moment is $1.6\sigma$ below experiment. Possible solutions to the discrepancy have been proposed, typically by introducing new particles, symmetries, and interactions beyond the SM \cite{Ansta, Balkin, Bai, Borah, ZLi, WangF, Aboubrahim, Dey, Arcadi, Perez, Ban, Athron, Endo, Badziak, JjCao, Calibbi, Chen4, Botella, Chun, LiSP, Han, seesaw, seesaw2, Cadeddu, Aebischer, Crivellin, Davoudiasl, Liu, Dutta, Dorsner}. In contrast, Refs.~\cite{Li1, Li2} focused on potential explanations of the lepton $g-2$ anomalies in terms of nonlocal QED, without the introduction of new particles.

In the most general form of nonlocal QED, the propagators are modified due to the free part of the nonlocal Lagrangian~\cite{Li2}. Since the propagators are related to the canonical quantization, the modified propagator corresponds to the new quantization (solid quantization) conditions \cite{Quantization1,Quantization2}. Here, the fields of the non-point particles are expanded with wave packets rather than with plane wave functions. The nonlocal Lagrangian, as well as the solid quantization for the fundamental interaction, provide a gauge invariant method to deal with ultraviolet divergences, and in fact there are no divergences appearing in the loop integral due to the correlators.

As a final application of nonlocal field theory, we consider an extension of the theory to curved spacetime, and the investigation of the corresponding nonlocal energy-momentum tensor. The curved spacetime formulation of the energy-momentum tensor can then be applied to compute the gravitational form factors of the nucleon using a nonlocal interaction.

In this review article, we will review these nonlocal theories and their applications. We begin in Sec.~\ref{sec:nonlocal} with a general discussion of nonlocal EFT, where we introduce both local and nonlocal EFT Lagrangians, and apply them to the calculation of GPDs with zero and nonzero skewness. In Sec.~\ref{sec:qed}, we present nonlocal QED and its application to the possible explanation of the lepton $g-2$ anomaly. We derive the most general nonlocal QED Lagrangian, and summarize the Feynman rules for the propagators and vertices. The proof of the modified Ward-Green-Takahashi identity will be discussed, along with the corresponding quantization conditions. In Sec.~\ref{sec:gravity}, we discuss the nonlocal action in curved spacetime and nonlocal energy-momentum tensor, and investigate the gravitational form factors of the nucleon with a nonlocal interaction. Finally, in Sec.~\ref{sec:summary} we summary the main points discussed in this review. \\ \\

\begin{adjustwidth}{-\extralength}{0cm}

\section{Nonlocal chiral effective theory}\label{sec:nonlocal}

\subsection{Local chiral effective Lagrangian} 
\label{ssec.chirallag}

In this section we introduce the basic chiral Lagrangian which defines effective pion-nucleon interactions at low energies. The nonlocal generalization of the chiral Lagrangian naturally generates the ultraviolet regulator for loop integrals, which respects Lorentz and gauge invariance. The nonlocal formulation relevant for PDFs was presented in Refs.~\cite{Salamu:2018cny, Salamu02}; here we discuss the application of the formalism to the case of nonforward matrix elements needed to compute GPDs. To lowest order, the local Lagrangian for the chiral SU(3)$_L\times$SU(3)$_R$ effective theory that describes the interaction of pseudoscalar mesons ($\phi$) with octet ($B$) and decuplet ($T_\mu$) baryons is given by~\cite{Jenkins:1991ts, Ledwig:2014rfa}, \\
\begin{eqnarray}
\label{eq:ch8}
{\cal L}
&=& {\rm Tr} \big[ \bar B (i\not\!\!{D} - M_B) B \big]
 -  \frac{D}{2} \, {\rm Tr} \big[ \bar B \gamma^\mu \gamma_5 \{u_\mu, B\} \big]
 -  \frac{F}{2}\, {\rm Tr} \big[ \bar B \gamma^\mu \gamma_5 [ u_\mu ,B ] \big]
\nonumber\\
&+& \overline{T}_\mu^{ijk}
   (i\gamma^{\mu \nu \alpha} D_\alpha - M_T \gamma^{\mu\nu}) T_\nu^{ijk}
 -\ \frac{{\cal H}}{2}\,
(\overline{T}_\mu)^{ijk} \gamma^{\alpha} \gamma_5 (u_\alpha)^{kl}\,
	      (T^\mu)^{ijl}
\nonumber\\
&-& \frac{{\cal C}}{2}
   \big[ \epsilon^{ijk}\, \overline{T}_\mu^{ilm}
	  \Theta^{\mu\nu} (u_\nu)^{lj} B^{mk} + {\rm H.c.}
   \big]
 +\ \frac{f^2}{4}
    {\rm Tr} \big[ D_\mu U (D^\mu U)^\dag \big],
\end{eqnarray}
where $M_B$ and $M_T$ are the octet and decuplet masses, $D$, $F$, $\cal C$ and $\cal H$ are the baryon-meson coupling constants, and $f=93$~MeV is the pseudoscalar decay constant. The octet–decuplet transition operator $\Theta^{\mu\nu}$ is given by
\begin{equation}
\Theta^{\mu\nu}
= g^{\mu\nu} - \Big(Z+\frac12\Big) \gamma^\mu \gamma^\nu,
\label{eq:Theta}
\end{equation}
where $Z$ is the decuplet off-shell parameter, usually chosen to be 1/2~\cite{Nath:1971wp}, and the tensors are given by
  $\gamma^{\mu\nu}
   = \frac{1}{2} [\gamma^\mu,\gamma^\nu] = -i \sigma^{\mu\nu}$,
  $\gamma^{\mu\nu\alpha}
   = \frac{1}{2} \{\gamma ^{\mu\nu}, \gamma^\alpha\}$,
with $\epsilon^{ijk}$ the antisymmetric tensor in flavor space. The SU(3) baryon octet fields $B^{ij}$ are represented by the matrix
\begin{eqnarray}
\label{e.B}
B =
\left(
\begin{array}{ccc}
  \frac{1}{\sqrt 2} \Sigma^0 + \frac{1}{\sqrt 6} \Lambda
& \Sigma^+
& p					\\
  \Sigma^-
&-\frac{1}{\sqrt 2} \Sigma^0 + \frac{1}{\sqrt 6} \Lambda
& n					\\
  \Xi^-
& \Xi^0
&-\frac{2}{\sqrt6} \Lambda
\end{array}
\right),
\end{eqnarray}
and the decuplet fields $T^{ijk}_\mu$ are represented by symmetric tensors with components
\begin{eqnarray}
\label{e.T}
&&T^{111} = \Delta^{++},\ \
T^{112} = \frac{1}{\sqrt 3} \Delta^+,\ \
T^{122} = \frac{1}{\sqrt 3} \Delta^0,\ \
T^{222} = \Delta^-,				\notag\\
&&T^{113} = \frac{1}{\sqrt 3} \Sigma^{*+},\ \
T^{123} = \frac{1}{\sqrt 6} \Sigma^{*0},\ \
T^{223} = \frac{1}{\sqrt 3} \Sigma^{*-},	\\
&&T^{133} = \frac{1}{\sqrt 3} \Xi^{*0},\ \
T^{233} = \frac{1}{\sqrt 3} \Xi^{*-},		\notag\\
&&T^{333} = \Omega^-.           \notag
\end{eqnarray}
The operator $U$ is defined in terms of the matrix of pseudoscalar meson fields~$\phi$,
\begin{equation}
U \equiv u^2 = \exp\bigg(i \frac{\sqrt2\phi}{f}\bigg),
\end{equation}
where
\begin{eqnarray}
\label{e.phi}
\phi =
\left(
{\begin{array}{*{20}{c}}
  \frac{1}{\sqrt 2} \pi^0 + \frac{1}{\sqrt 6} \eta
& \pi^+
& K^+						\\
  \pi^-
& -\frac{1}{\sqrt 2} \pi^0 + \frac{1}{\sqrt 6} \eta
& K^0						\\
  K^-
& \overline K^0
& -\frac{2}{\sqrt 6} \eta
\end{array}}
\right)
\end{eqnarray}
represents the $\pi$, $K$ and $\eta$ mesons. The covariant derivatives of the octet and decuplet baryon fields in Eq.~(\ref{eq:ch8}) are given by~\cite{Hemmert:1998pi, Hemmert:1999mr}
\begin{subequations}
\begin{eqnarray}
D_\mu B
&=& \partial_\mu B
 + [\Gamma_\mu, B]
 - i \langle \lambda^0 \rangle \upsilon_\mu^0\, B,	\\
D_\mu T_\nu^{ijk}
&=& \partial_\mu T_\nu^{ijk}
 + (\Gamma_\mu, T_\nu )^{ijk}
 - i \langle \lambda^0 \rangle \upsilon_\mu^0\, T_\nu^{ijk},
\label{eq:11}
\end{eqnarray}
\end{subequations}
respectively, where $\upsilon_\mu^0$ denotes an external singlet vector field, $\lambda^0$ is the unit matrix, and $\langle\, \cdots \rangle$ represents a trace in flavor space. For the covariant derivative of the decuplet field, the shorthand notation
\begin{equation}
(\Gamma_\mu, T_\nu)^{ijk}
= (\Gamma_\mu)_l^i\, T_\nu^{ljk}
+ (\Gamma_\mu)_l^j\, T_\nu^{ilk}
+ (\Gamma_\mu)_l^k\, T_\nu^{ijl}
\end{equation}
is used, while for the meson fields the covariant derivarive is given by
\begin{eqnarray}
D_\mu U
&=& \partial_\mu U
 + (iU \lambda^a - i \lambda^a U)\, \upsilon_\mu^a.
\end{eqnarray}
The mesons couple to the baryon fields through the vector and axial vector combinations
\begin{eqnarray}
\Gamma_\mu
&=& \frac12
    \left( u \partial_\mu u^\dagger + u^\dagger \partial_\mu u \right)
 -  \frac{i}{2}
    \left( u \lambda^a u^\dagger + u^\dagger \lambda^a u \right) \upsilon_\mu^a,
\\
  u_\mu
&=& i
    \left( u^\dagger \partial_\mu u - u \partial_\mu u^\dagger \right)
 +  \left( u^\dagger \lambda^a u - u \lambda^a u^\dagger \right) \upsilon_\mu^a,
\label{eq:22}
\end{eqnarray}
where $\upsilon_\mu^a$ corresponds to an external octet vector field,
and $\lambda^a$ ($a=1, \ldots, 8$) are the Gell-Mann matrices.

The unpolarized twist-two GPD $H$ receives contributions from each quark flavor from the lowest-order Lagrangian in Eq.~(\ref{eq:ch8}). To compute the effects of meson loops on the magnetic GPD $E$, on the other hand, requires an additional contribution to the Lagrangian for the magnetic interaction, which enters at higher order. The magnetic Lagrangian at ${\cal O}(q^2)$ for the octet, decuplet and octet-decuplet transition interaction is given by~\cite{He:2017viu, He:2018eyz, Yang:2020rpi, Jones:1972ky, Geng:2009ys}
\begin{eqnarray}
\label{lomag}
{\cal L}_{\rm mag}
&=& \frac{1}{4 M_B} 
\Big( c_1{\rm Tr}\left[\bar{B} \sigma^{\mu\nu}
  \left\{F^+_{\mu\nu},B\right\}\right]+c_2{\rm Tr}\left[\bar{B}
  \sigma^{\mu\nu} \left[F^+_{\mu\nu},B \right]\right]+c_3{\rm Tr}\left[\bar{B}
  \sigma^{\mu\nu}B \right]{\rm Tr}\left[F^+_{\mu\nu}\right]
\Big) \nonumber \\
&+&
\frac{i}{4 M_B} c_4 F_{\mu\nu}\Big(\epsilon_{ijk}(\lambda_q)^i_l\bar{B}^j_m\gamma^\mu\gamma_5(T^\nu)^{klm} 
+\epsilon^{ijk}(\lambda_q)^l_i(\overline{T}^\mu)_{klm}\gamma^\nu\gamma_5 B^m_j\Big)
  \nonumber \\
&+&
\frac{F_2^T}{2 M_T}
(\overline{T}_\mu)^{abc}\sigma^{\rho\sigma}\partial_{\sigma} \upsilon_\rho^q(\lambda_q)^a_e (T^\mu)^{ebc}.
\end{eqnarray}
Here we adopt the notation $c_1$, $c_2$ and $c_3$ for the octet baryon interaction from Ref.~\cite{Yang:2020rpi} and $c_4$ for the octet-decuplet transition, which corresponds to the constant $\mu_T$ in Refs.~\cite{He:2017viu, He:2018eyz}. Also, following Refs.~\cite{He:2017viu, He:2018eyz} we denote by $F_2^T$ the coupling for the decuplet interaction. The electromagnetic interaction with the individual quark flavors in Eq.~(\ref{lomag}) is introduced by the field strength tensor
\begin{equation}
F^+_{\mu\nu} = \frac12
\left(u^\dag F^q_{\mu\nu}\lambda_q u + u F^q_{\mu\nu}\lambda_q u^\dag\right),
\end{equation}
where $F^q_{\mu\nu} = \partial_\mu \upsilon^q_\nu-\partial_\nu \upsilon^q_\mu$ for an external field $\upsilon^q_\mu$ interacting with the quark flavor $q=u,d,s$ having unit charge, and the matrix $\lambda_q$ is the diagonal quark flavor matrix defined as
\mbox{$\lambda_q = {\rm diag}\{\delta_{qu},\delta_{qd},\delta_{qs}\}$}.
At this order, the magnetic Lagrangian ${\cal L}_{\rm mag}$ in Eq.~(\ref{lomag}) generates a quark flavor decomposition for the proton anomalous magnetic moment given by the proton's Pauli form factor $F_2^p(t)$ at $t=0$,
\begin{subequations}
\begin{eqnarray}
\label{treemag}
F^{p(u)}_2(0) &=& c_1+c_2+c_3, \\
F^{p(d)}_2(0) &=& c_3,         \\
F^{p(s)}_2(0) &=& c_1-c_2+c_3.
\end{eqnarray}
\end{subequations}
Since there is no strange quark contribution to the proton at tree level, we take $c_3 = c_2 - c_1$. Furthermore, from SU(3) symmetry one obtains relationships between the octet and decuplet constants~\cite{He:2017viu, He:2018eyz} given by
\begin{subequations}
\label{eq:F2}
\begin{eqnarray}
c_4   &=& 4 c_1,  \\
F_2^T &=& c_1+3c_2.
\end{eqnarray}
\end{subequations}
Within the flavor SU(3) framework, the magnetic moments of the octet and decuplet baryons and the transition moments between the octet and decuplet baryons can be expressed in terms of quark magnetic moments, $\mu_q$. For the proton and neutron, for example, one would have
    $\mu_p = \frac43 \mu_u - \frac13 \mu_d$
and $\mu_n = \frac43 \mu_d - \frac13 \mu_u$,
respectively, while for the $\Delta^{++}$ baryon $\mu_{\Delta^{++}} = 3 \mu_u$.

Including the higher order magnetic Lagrangian ${\cal L}_{\rm mag}$ in Eq.~(\ref{lomag}), for consistency in the power counting we also need to consider the next-to-leading order Lagrangian for the baryon-meson interaction. Generalizing Eq.~(\ref{eq:ch8}), and using the notation from Ref.~\cite{Kubis:2000aa}, we include the additional baryon contribution involving two derivatives~\cite{Kubis:2000aa},
\begin{eqnarray}
\label{adm}
\!\!{\cal L}_{B\phi}' &=& \frac{i}{2}\sigma^{\mu\nu} 
\Big( 
  b_9\, {\rm Tr} \left[\bar{B}u_\mu\right]{\rm Tr}\left[u_\nu B\right]
+ b_{10}\, {\rm Tr} \left[\bar{B}\{[u_\mu,u_\nu],B\}\right]
+ b_{11}\, {\rm Tr} \left[\bar{B}[[u_\mu,u_\nu],B]\right]\!
\Big),~~~
\end{eqnarray}
where the values of the coefficients have been determined to be
    $b_9=1.36$~GeV$^{-1}$,
    $b_{10}=1.24$~GeV$^{-1}$,
and
    $b_{11}=0.46$~GeV$^{-1}$~\cite{Kubis:2000aa}.
Expanding the Lagrangians $\cal L$ in Eq.~(\ref{eq:ch8}) and ${\cal L}_{B\phi}'$ in Eq.~(\ref{adm}), the lowest order baryon-meson interaction involving the proton can then be written as
\begin{eqnarray}
{\cal L}_{\rm int}
&=& \frac{(D+F)}{2f}
  \left( \bar p\, \gamma^\mu \gamma_5 p\, \partial_\mu \pi^0
       + \sqrt2\, \bar p\, \gamma^\mu \gamma_5 n\, \partial_\mu \pi^+
  \right)
- \frac{(D-3F)}{\sqrt{12} f}
  \bar p\, \gamma^\mu \gamma_5 p\, \partial_\mu \eta       \notag \\
&+& \frac{(D-F)}{2 f}
  \left( \sqrt2\, \bar p\, \gamma^\mu \gamma_5 \Sigma^+\, \partial_\mu K^0
       + \bar p\, \gamma^\mu \gamma_5 \Sigma^0\, \partial_\mu K^+
  \right)                                                   
- \frac{(D+3F)}{\sqrt{12} f}
  \bar p\, \gamma^\mu \gamma_5 \Lambda\, \partial_\mu K^0   \notag \\
&+& \frac{\cal C}{\sqrt{12} f}
  \left(
  - 2\, \bar p\, \Theta^{\nu\mu} \Delta_\mu^+\, \partial_\nu \pi^0
  - \sqrt2\, \bar p\, \Theta^{\nu\mu} \Delta_\mu^0\, \partial_\nu \pi^+
  + \sqrt6\, \bar p\, \Theta^{\nu\mu} \Delta_\mu^{++}\, \partial_\nu \pi^-
  \right.						\notag \\
&&
  \left.
  - \bar p\, \Theta^{\nu\mu} \Sigma_\mu^{*0}\, \partial_\nu K^+
  + \sqrt2\, \bar p\, \Theta^{\nu\mu} \Sigma_\mu^{*+}\, \partial_\nu K^0
  + {\rm H.c.}
  \right)						\notag \\
&+& \frac{i}{4f^2} \bar p\, \gamma^\mu p
  \Big[
    (\pi^+ \partial_\mu \pi^-  -  \pi^- \partial_\mu \pi^+)
    + 2 (K^+ \partial_\mu K^-  -  K^- \partial_\mu K^+)
    + (K^0 \partial_\mu \bar K^0  -  \bar K^0 \partial_\mu K^0)
  \Big]   \notag \\
&+& \frac{i}{f^2} \bar p\, \sigma^{\mu\nu} p
  \Big( 2(b_{10}+b_{11})
    \partial_\mu\pi^+ \partial_\nu \pi^-
    + (4b_{11}+b_9) \partial_\mu K^+ \partial_\nu K^-  
    + 2(b_{10}-{b_{11}}) \partial_\mu K^0 \partial_\nu \bar K^0 
  \Big).
\label{eq:j1}
\end{eqnarray}
For the interactions with an external field $\upsilon_\mu^a$, from the Lagrangian $\cal L$ in Eq.~(\ref{eq:ch8}) one obtains the vector current 
\begin{eqnarray}
J^\mu_a
&=& \frac12 {\rm Tr}
   \big[
   \bar B \gamma^\mu
   \left[ u \lambda^a u^\dagger + u^\dagger \lambda^a u, B
   \right]	
      + \frac{D}{2}{\rm Tr}
   \big[
   \bar B \gamma^\mu \gamma_5
   \left\{ u \lambda^a u^\dagger - u^\dagger \lambda^a u, B
   \right\}
   \big]						\notag\\
&+& \frac{F}{2}{\rm Tr}
   \big[
   \bar B \gamma^\mu \gamma_5
   \left[ u \lambda^a u^\dagger - u^\dagger \lambda^a u, B
   \right]
   \big]						\notag\\
&+& \frac12\,
   \overline{T}_\nu \gamma^{\nu\alpha\mu}
   \left( u \lambda^a u^\dagger + u^\dagger \lambda^a u, T_\alpha
   \right)
   + \frac{\cal C}{2}
   \left[
   \overline{T}_\nu \Theta^{\nu\mu}
   (u \lambda^a u^\dagger - u^\dagger \lambda^a u) B
   + {\rm H.c.}
   \right]						\notag\\
&+& \frac{f^2}{4}{\rm Tr}
   \big[
   \partial^\mu U
   (U^\dagger i \lambda^a
    - i \lambda^a U^\dagger)
+  (U i \lambda^a
    - i \lambda^a U)
   \partial^\mu U^\dagger
   \big].
\label{eq:ch1}
\end{eqnarray}
For the SU(3) flavor singlet case, the current coupling to the external field $\upsilon_\mu^0$ can be written as
\begin{eqnarray}
J^\mu_0
&=& \langle \lambda^0 \rangle\,
    {\rm Tr}[\bar B \gamma^\mu B]
 +  \langle \lambda^0 \rangle\,
    \overline{T}_\nu \gamma^{\nu\alpha\mu}\, T_\alpha.
\label{eq:ch2}
\end{eqnarray}
The magnetic current coupling to the external field $\upsilon_\mu^q$ can be obtained from the magnetic Lagrangian in Eq.~(\ref{lomag}) as,
\begin{eqnarray}
J_{q, \rm mag}^\mu
&=&
\frac{\partial_\nu}{4 M_B}
\Big(
  c_1 {\rm Tr}\bar{B} \sigma^{\mu\nu}
    \left\{u^\dag \lambda_q u + u \lambda_q u^\dag,B\right\}
+ c_2 {\rm Tr}\bar{B} \sigma^{\mu\nu}
    \left[u^\dag \lambda_q u + u \lambda_q u^\dag,B \right]
\nonumber \\
& & \qquad
+\, c_3 {\rm Tr}\bar{B} \sigma^{\mu\nu} B\, 
        {\rm Tr}(u^\dag \lambda_q u + u \lambda_q u^\dag)
\Big)
-\
\frac{F_2^T}{2 M_T}
\partial_\sigma
\Big( (\overline{T}_\nu)^{abc}\sigma^{\mu\sigma}(\lambda_q)^a_e (T^\nu)^{ebc}
\Big)
\nonumber \\
&-&
\frac{ic_4}{4 M_B} 
(g^{\mu\nu} \partial^\sigma - g^{\mu\sigma} \partial^\nu)
\Big( \epsilon_{ijk} 
      (\lambda_q)^i_l\, \bar{B}^j_m \gamma_\sigma \gamma_5 T_\nu^{klm}
    + \epsilon^{ijk}
      (\lambda_q)^l_i\, \overline{T}_{\sigma,klm} \gamma_\nu \gamma_5 B^m_j
\Big),
\end{eqnarray}
and satisfies current conservation,
    $\partial_\mu J_{q, \rm mag}^\mu = 0$.
The quark flavor currents can be written in terms of the SU(3) singlet ($a=0$) and octet ($a=3, 8$) and quark magnetic currents as,
\begin{subequations}
\label{eq:ch3}
\begin{eqnarray}
J^\mu_u
&=& \frac13 J^\mu_0
 +  \frac12 J^\mu_3
 +  \frac{1}{2\sqrt3} J^\mu_8 
 + J_{u,\rm mag}^\mu,			\\
J^\mu_d
&=& \frac13 J^\mu_0
 -  \frac12 J^\mu_3
 +  \frac{1}{2\sqrt3} J^\mu_8 
 + J_{d,\rm mag}^\mu,			\\
J^\mu_s
&=& \frac13 J^\mu_0
 -  \frac{1}{\sqrt3} J^\mu_8 
 + J_{s,\rm mag}^\mu.
\end{eqnarray}
\end{subequations}
From Eqs.~(\ref{eq:ch1}), (\ref{eq:ch2}) and (\ref{eq:ch3}) the quark flavor currents can be written more explicitly in the form
\begin{subequations}
\label{eq:jq}
\begin{eqnarray}
J_u^\mu
&=& 2 \bar p \gamma^\mu p + \bar n \gamma^\mu n
+ \bar\Lambda \gamma^\mu \Lambda
+ 2 \overline{\Sigma}^+ \gamma^\mu \Sigma^+
+ {\overline\Sigma}^0 \gamma^\mu \Sigma^0
- \frac{1}{2f^2}\, \bar p \gamma^\mu p\, 
  \big( \pi^+ \pi^- + 2 K^+ K^- \big)
\notag\\
&+&
  3 \overline{\Delta}_\alpha^{++} \gamma^{\alpha\beta\mu} \Delta_\beta^{++}
+ 2 \overline{\Delta}_\alpha^+    \gamma^{\alpha\beta\mu} \Delta_\beta^+
+   \overline{\Delta}_\alpha^0    \gamma^{\alpha\beta\mu} \Delta_\beta^0
+ 2 \overline{\Sigma}_\alpha^{*+} \gamma^{\alpha\beta\mu} \Sigma_\beta^{*+}
+   \overline{\Sigma}_\alpha^{*0} \gamma^{\alpha\beta\mu} \Sigma_\beta^{*0}
\notag\\
&+&
  i \left( \pi^- \partial^\mu\pi^+ - \pi^+ \partial^\mu\pi^- \right)
+ i \left( K^-   \partial^\mu K^+  - K^+   \partial^\mu K^-   \right)
\notag\\
&-&
  \frac{i(D+F)}{\sqrt 2f} \bar p \gamma^\mu \gamma_5 n\, \pi^+
+ \frac{i(D+3F)}{\sqrt{12}f} \bar p \gamma^\mu \gamma_5 \Lambda\, K^+
- \frac{i(D-F)}{2f} \bar p \gamma^\mu \gamma_5 \Sigma^0\, K^+
\notag\\
&+&
  \frac{i\, \cal C}{\sqrt{12}f}
  \left(
    \sqrt{6}\, \bar p\, \Theta^{\mu\nu} \Delta_\nu^{++}\, \pi^-
  + \sqrt{2}\, \bar p\, \Theta^{\mu\nu} \Delta_\nu^0\, \pi^+
  + \,\bar p\, \Theta^{\mu\nu} \Sigma_\nu^{*0}\, K^+
  + {\rm H.c.}
  \right)
\notag\\
&+& \frac{1}{4M_B} \partial_\nu (\bar{p}\sigma^{\mu\nu}p)
    \bigg[
      4 c_2 \Big( 1 - \frac{1}{2 f^2} K^+K^- \Big) - \frac{(c_1+c_2)}{f^2} \pi^+\pi^- 
    \bigg]
 +  \frac{c_2-c_1}{2M_B} \partial_\nu(\bar{n}\sigma^{\mu\nu}n)
\notag\\
&+& \frac{3c_2-2c_1}{6M_B} \partial_\nu (\bar{\Lambda}\sigma^{\mu\nu}\Lambda) 
 +  \frac{c_1}{2\sqrt3 M_B} \partial_\nu (\overline{\Lambda}\sigma^{\mu\nu}\Sigma^0) 
 +  \frac{c_2}{M_B} \partial_\nu (\overline{\Sigma}^+\sigma^{\mu\nu}\Sigma^+)
 +  \frac{c_2}{2M_B} \partial_\nu (\overline{\Sigma}^0\sigma^{\mu\nu}\Sigma^0) 
\notag\\
&+&
\frac{i c_4}{4\sqrt3 M_B} \partial^\nu
\bigg[
  \bar{p}(\gamma_\nu \gamma_5 \Delta^{+\mu} - \gamma^\mu\gamma_5 \Delta^+_\nu)
+ \bar{n}(\gamma_\nu \gamma_5 \Delta^{0\mu} - \gamma^\mu\gamma_5 \Delta^0_\nu)
- \overline{\Sigma}^+ (\gamma_\nu \gamma_5 \Sigma^{*+\mu} - \gamma^\mu \gamma_5 \Sigma^{*+}_\nu)
\notag\\
& & \hspace{2cm}
- \frac{\sqrt3}{2} 
  \bar{\Lambda} (\gamma_\nu \gamma_5 \Sigma^{*0\mu} - \gamma^\mu \gamma_5 \Sigma^{*0}_\nu)
+ \frac12 
  \overline{\Sigma}^0 (\gamma_\nu \gamma_5 \Sigma^{*0\mu} - \gamma^\mu \gamma_5 \Sigma^{*0}_\nu)   
\bigg]
\notag\\
&-& \frac{F_2^T}{6 M_T}\partial_\nu
\Big[
  3\overline{\Delta}^{++}_\alpha \sigma^{\mu\nu} \Delta^{++\alpha}
+ 2\overline{\Delta}^{+}_\alpha  \sigma^{\mu\nu} \Delta^{+\alpha}   
+  \overline{\Delta}^{0}_\alpha  \sigma^{\mu\nu} \Delta^{0\alpha}
+ 2\overline{\Sigma}^{*+}_\alpha \sigma^{\mu\nu} \Sigma^{*+\alpha}
+  \overline{\Sigma}^{*0}_\alpha \sigma^{\mu\nu} \Sigma^{*0\alpha}
\Big], 
\label{eq:ju}
\end{eqnarray}
\begin{eqnarray}
J^\mu_d
&=& \bar p \gamma^\mu p
+ 2 \bar n \gamma^\mu n
+ 2 \overline{\Sigma}^- \gamma^\mu \Sigma^-
+ \overline{\Sigma}^0 \gamma^\mu \Sigma^0
+ \bar\Lambda \gamma^\mu \Lambda
+ \frac{1}{2f^2}\, \bar p \gamma^\mu p\, 
  \big( \pi^+ \pi^- - \overline{K}^0 K^0 \big)
\notag\\
&+&
    \overline{\Delta}_\alpha^+ \gamma^{\alpha\beta\mu} \Delta_\beta^+
+ 2 \overline{\Delta}_\alpha^0 \gamma^{\alpha\beta\mu} \Delta_\beta^0
+ 3 \overline{\Delta}_\alpha^- \gamma^{\alpha\beta\mu} \Delta_\beta^-
+   \overline{\Sigma}_\alpha^{*0}  \gamma^{\alpha\beta\mu} \Sigma_\beta^{*0}
+ 2 \overline{\Sigma}_\alpha^{*0-} \gamma^{\alpha\beta\mu} \Sigma_\beta^{*-}
\notag\\
&-&
  i (\pi^- \partial^\mu \pi^+  -  \pi^+ \partial^\mu \pi^-)
+ i (\overline{K}^0 \partial^\mu K^0  -  K^0 \partial^\mu \overline{K}^0)
\notag\\
&+&
  \frac{i(D+F)}{\sqrt2 f}
  \bar p \gamma^\mu \gamma_5 n\, \pi^+
- \frac{i(D-F)}{\sqrt2 f}
  \bar p \gamma^\mu \gamma_5 \Sigma^+\, K^0
\notag\\
&-&
\frac{i\, \cal C}{\sqrt6 f}
  \left(
    \sqrt3\, \bar p\, \Theta^{\mu\nu} \Delta_\nu^{++}\, \pi^-
  + \bar p\, \Theta^{\mu\nu} \Delta_\nu^0\, \pi^+
  + \bar p\, \Theta^{\mu\nu} \Sigma_\nu^{*+}\, K^0
  + {\rm H.c.}
  \right)
\notag\\
&+& \frac{1}{4M_B} \partial_\nu (\bar{p}\sigma^{\mu\nu} p)
    \bigg[ (c_2-c_1) \Big( 2 - \frac{1}{f^2} \overline{K}^0 K^0 \Big)
        + \frac{(c_1+c_2)}{f^2} \pi^+\pi^-
    \bigg]
 +  \frac{c_2}{M_B} \partial_\nu (\bar{n}\sigma^{\mu\nu}n)
\notag\\
&+& \frac{3c_2-2c_1}{6M_B} \partial_\nu (\bar{\Lambda}\sigma^{\mu\nu}\Lambda) 
 +  \frac{c_2}{M_B} \partial_\nu (\overline{\Sigma}^-\sigma^{\mu\nu}\Sigma^-)    
 +  \frac{c_2}{2M_B} \partial_\nu (\overline{\Sigma}^0\sigma^{\mu\nu}\Sigma^0)
 -  \frac{c_1}{2\sqrt3 M_B} \partial_\nu (\bar{\Lambda}\sigma^{\mu\nu}\Sigma^0) 
\notag\\
&-& \frac{i c_4}{4\sqrt3 M_B} \partial^\nu
\bigg[
  \bar{p}(\gamma_\nu \gamma_5 \Delta^{+\mu} - \gamma^\mu \gamma_5 \Delta^+_\nu)
+ \bar{n}(\gamma_\nu \gamma_5 \Delta^{0\mu} - \gamma^\mu \gamma_5 \Delta^0_\nu)
- \overline{\Sigma}^-(\gamma_\nu \gamma_5 \Sigma^{*-\mu} - \gamma^\mu \gamma_5 \Sigma^{*-}_\nu)
\notag\\  
& & \hspace{2cm}
- \frac{\sqrt3}{2} \bar{\Lambda}(\gamma_\nu \gamma_5 \Sigma^{*0\mu} - \gamma^\mu \gamma_5 \Sigma^{*0}_\nu)
- \frac12 \overline{\Sigma}^0(\gamma_\nu \gamma_5 \Sigma^{*0\mu} - \gamma^\mu \gamma_5 \Sigma^{*0}_\nu)
\bigg]
\notag\\
&-& \frac{F_2^T}{6 M_T}\partial_\nu
\Big[
  3\overline{\Delta}^-_\alpha \sigma^{\mu\nu} \Delta^{-\alpha}
+ 2\overline{\Delta}^0_\alpha \sigma^{\mu\nu} \Delta^{0\alpha}
+  \overline{\Delta}^+_\alpha \sigma^{\mu\nu} \Delta^{+\alpha}
+ 2\overline{\Sigma}^{*-}_\alpha \sigma^{\mu\nu} \Sigma^{*-\alpha}
+  \overline{\Sigma}^{*0}_\alpha \sigma^{\mu\nu} \Sigma^{*0\alpha}
\Big], 
\label{eq:jd}
\end{eqnarray}
\begin{eqnarray}
\label{eq:js}
J^\mu_s
&=&\overline{\Sigma}^+ \gamma^\mu \Sigma^+
 + \overline{\Sigma}^0 \gamma^\mu \Sigma^0
 + \bar\Lambda \gamma^\mu \Lambda
 + \frac{1}{2f^2}\, \bar p \gamma^\mu p\, 
   \big( 2 K^+ K^- + \overline{K}^0 K^0 \big)
\notag\\
&+&
   \overline{\Sigma}_\alpha^{*+} \gamma^{\alpha\beta\mu} \Sigma_\beta^{*+}
 + \overline{\Sigma}_\alpha^{*0} \gamma^{\alpha\beta\mu} \Sigma_\beta^{*0}
 - i (K^- \partial^\mu K^+  -  K^+ \partial^\mu K^-)
 - i (\overline{K}^0 \partial^\mu K^0 - K^0 \partial^\mu \overline{K}^0)
\notag\\
&+&
   \frac{i(D-F)}{\sqrt2 f}
   \bar p \gamma^\mu \gamma_5 \Sigma^+\, K^0
 + \frac{i(D-F)}{2f}
   \bar p \gamma^\mu \gamma_5 \Sigma^0\, K^+
 - \frac{i(D+3F)}{\sqrt{12} f}
   \bar p \gamma^\mu \gamma_5 \Lambda\, K^+		\notag\\
&-&
   \frac{i\, \cal C}{\sqrt{12} f}
   \left(
     \bar p\, \Theta^{\mu\nu} \Sigma_\nu^{*0}\, K^+
   - \sqrt2\, \bar p\, \Theta^{\mu\nu} \Sigma_\nu^{*+}\, K^0
   + {\rm H.c.}
   \right)                                              \notag\\
&+& \frac{1}{4 M_B f^2} \partial_\nu (\bar{p}\sigma^{\mu\nu} p) 
\bigg[ 2 c_2 K^+K^- + (c_2-c_1) \overline{K}^0 K^0 \bigg]
 +  \frac{(c_1+3c_2)}{6M_B} \partial_\nu (\bar{\Lambda}\sigma^{\mu\nu}\Lambda)
\notag\\
&+& \frac{(c_2-c_1)}{2M_B} \partial_\nu (\overline{\Sigma}^+ \sigma^{\mu\nu}  \Sigma^+)
 +  \frac{(c_2-c_1)}{2M_B} \partial_\nu (\overline{\Sigma}^- \sigma^{\mu\nu} \Sigma^-) 
 +  \frac{(c_2-c_1)}{2M_B} \partial_\nu (\overline{\Sigma}^0 \sigma^{\mu\nu} \Sigma^0)
\notag\\
&-& \frac{\, ic_4}{4\sqrt3 M_B} \partial^\nu
\bigg[
  \overline{\Sigma}^0 (\gamma_\nu \gamma_5 \Sigma^{*0\mu} - \gamma^\mu \gamma_5 \Sigma^{*0}_\nu)
+ \overline{\Sigma}^- (\gamma_\nu \gamma_5 \Sigma^{*-\mu} - \gamma^\mu \gamma_5 \Sigma^{*-}_\nu)
\notag\\
& & \hspace*{2cm}
- \overline{\Sigma}^+ (\gamma_\nu \gamma_5 \Sigma^{*+\mu} - \gamma^\mu \gamma_5 \Sigma^{*+}_\nu) 
\bigg],
\notag\\
&-& \frac{F_2^T}{6 M_T} \partial_\nu 
\Big[
  \overline{\Sigma}^{*-}_\alpha \sigma^{\mu\nu} \Sigma^{*-\alpha}   
+ \overline{\Sigma}^{*0}_\alpha \sigma^{\mu\nu} \Sigma^{*0\alpha}
+ \overline{\Sigma}^{*+}_\alpha \sigma^{\mu\nu} \Sigma^{*+\alpha}
\Big],
\end{eqnarray}
\end{subequations}
for the $u$, $d$ and $s$ quark flavors, respectively. As in Ref.~\cite{Salamu:2018cny}, terms involving the doubly strange $\Xi^{0,-}$ and $\Xi^{*0,-}$ hyperons and the triply-strange $\Omega^-$ baryon do not couple directly to the proton, and are therefore not included here.

\subsection{Nonlocal chiral Lagrangian}

In this section we outline the generalization of the effective local chiral Lagrangian to the case of nonlocal interactions. Taking the traces in Eqs.~(\ref{eq:ch8}), (\ref{lomag}) and (\ref{adm}) in Sec.~\ref{ssec.chirallag}, we can write the local Lagrangian density as
\begin{eqnarray}
{\cal L}^{\rm (local)}(x)
&=&\bar B(x)(i \gamma^\mu \mathscr{D}_{\mu} - M_B) B(x)
 + \frac{C_{B\phi}}{f}
   \big[ \bar{p}(x) \gamma^\mu \gamma_5 B(x)\,
	  \mathscr{D}_{\mu} \phi(x) + {\rm H.c.}
   \big]						\notag\\
&+&\overline{T}_\mu(x)
   (i \gamma^{\mu\nu\alpha} \mathscr{D}_{\alpha} - M_T \gamma^{\mu\nu})\,
   T_\nu(x)
 + \frac{C_{T\phi}}{f}
   \left[\, \bar{p}(x) \Theta^{\mu\nu} T_\nu(x)\,
	 \mathscr{D}_{\mu} \phi(x) + {\rm H.c.}
   \right]						\notag\\
&+&\mathscr{D}_{\mu} \phi(x) (\mathscr{D}_{\mu} \phi)^\dag(x)\,
 + \frac{i C_{\phi\phi^\dag}}{2 f^2}
   \bar p(x) \gamma^\mu p(x)
   \left[ \phi(x) (\mathscr{D}_\mu \phi)^\dag(x)
	- \mathscr{D}_{\mu} \phi(x) \phi^\dag(x)
   \right]
\notag\\
&+&\frac{i C_{\phi\phi^\dag}'}{2 f^2}
   \bar p(x) \sigma^{\mu\nu} p(x)\mathscr{D}_\mu \phi(x)
    (\mathscr{D}_{\nu} \phi)^\dag(x)
\notag\\
&+&\frac{C_{\phi\phi^\dag}^{\rm mag}}{4 M_B f^2}
   \bar p(x) \sigma^{\mu\nu} p(x) F_{\mu\nu}(x) \phi(x)\phi^\dag(x)
 + \frac{C_{B}^{\rm mag}}{4 M_B}
   \bar B(x) \sigma^{\mu\nu} B(x) F_{\mu\nu}(x) 
\notag\\
&+&\frac{i C_{BT}^{\rm mag}}{4 M_B}
   \bar B(x) \gamma^\mu\gamma_5 T^\nu (x) F_{\mu\nu}(x)		
 - \frac{C_{T}^{\rm mag}}{4 M_T}
   \overline T_\alpha(x) \sigma^{\mu\nu} T^\alpha(x) F_{\mu\nu}(x)	
 + \cdots ,
\label{eq:Llocal}
\end{eqnarray}
where the dependence on the spacetime coordinate $x$ is shown explicitly. For the interaction part only those terms that contribute to the proton GPDs are shown. The covariant derivatives in Eq.~(\ref{eq:Llocal}) are given by
\begin{subequations}
\begin{eqnarray}
\mathscr{D}_{\mu} B(x)
&=& \left[ \partial_\mu - i e^q_B\, \mathscr{A_\mu}(x)
    \right] B(x),
\\
\mathscr{D}_{\mu} T^\nu(x)
&=& \left[ \partial_\mu - i e^q_T\, \mathscr{A_\mu}(x)
    \right] T^\nu(x),
\\
\mathscr{D}_{\mu} \phi(x)
&=& \left[ \partial_\mu - i e^q_\phi\, \mathscr{A_\mu}(x)
    \right] \phi(x),
\end{eqnarray}
\end{subequations}
where $\mathscr{A_\mu}$ is the electromagnetic gauge field, and $e^q_B$, $e^q_T$ and $e^q_\phi$ denote the quark flavor charges for the octet baryon $B$, decuplet baryon $T$, and meson $\phi$, respectively. In the case of the proton, one has the flavor charges
$e^u_p = 2 e^d_p = 2$, $e^s_p = 0$,
while for the $\Sigma^+$ hyperon
$e^u_{\Sigma^+} = 2 e^s_{\Sigma^+} = 2$, $e^d_{\Sigma^+} = 0$,
and similarly for the other baryons.
For the mesons, the flavor charges are
$e^u_{\pi^+} = -e^d_{\pi^+} = 1$, $e^q_{\pi^0} = 0$ (for all $q$) for pions,
and 
$e^u_{K^+} = -e^s_{K^+} = 1$, $e^d_{K^+} = 0$ for kaons,
with the values for other mesons obtained by charge conjugation. The coefficients $C_{B\phi}$, $C_{T\phi}$, $C_{\phi\phi^\dag}$, $C_{\phi\phi^\dag}'$, $C_B^{\rm mag}$, $C_{BT}^{\rm mag}$, $C_T^{\rm mag}$ and $C_{\phi\phi^\dag}^{\rm mag}$ in Eq.~(\ref{eq:Llocal}) are given in Table~\ref{tab:C} for the various processes discussed here.

Following Salamu {\it et al.}~\cite{Salamu:2018cny}, we sketch here the derivation of the nonlocal Lagrangian from the local Lagrangian in Eq.~(\ref{eq:Llocal}) (further details can be found in Refs.~\cite{terning, Holdom:1992fn, Faessler:2003yf, Wang:1996zu, He:2017viu, He:2018eyz}). The nonlocal analog of the local Lagrangian (\ref{eq:Llocal}) can be written as
\begin{eqnarray}
{\cal L}^{\rm (nonloc)}(x)
&=& \bar B(x) (i\gamma^\mu \mathscr{D}_\mu - M_B) B(x)
+ \overline{T}_\mu(x)
  (i\gamma^{\mu\nu\alpha} \mathscr{D}_\alpha - M_T \gamma^{\mu\nu})
  T_\nu(x)
\notag\\
& & \hspace*{-2.2cm}
+\ \bar{p}(x)
  \left[
    \frac{C_{B\phi}}{f} \gamma^\mu \gamma_5 B(x)\,
  + \frac{C_{T\phi}}{f} \Theta^{\mu\nu} T_\nu(x)
  \right]
  \mathscr{D}_\mu\!
  \int d^4a\, {\cal G}_\phi^q(x,x+a) F(a) \phi(x+a)
  + {\rm H.c.}
\notag\\
& & \hspace*{-2.2cm}
+\ \frac{i C_{\phi\phi^\dag}}{2 f^2}\,
  \bar{p}(x) \gamma^\mu p(x)\!                           
  \int d^4a\, {\cal G}_\phi^q(x,x+a) F(a) \phi(x+a)\,
  \mathscr{D}_\mu\!\!
  \int d^4b\, {\cal G}_\phi^q(x+b,x) F(b) \phi^\dag(x+b)
\notag\\
& & \hspace*{-2.2cm}
+\ \frac{i C_{\phi\phi^\dag}'}{2 f^2}\,
  \bar{p}(x) \sigma^{\mu\nu} p(x)\,				
  \mathscr{D}_\mu\! \int d^4a\, {\cal G}_\phi^q(x,x+a) F(a) \phi(x+a)\,
  \mathscr{D}_\nu\! \int d^4b\, {\cal G}_\phi^q(x+b,x) F(b) \phi^\dag(x+b)
\notag\\
& & \hspace*{-2.2cm}
+\ \frac{C_B^{\rm mag}}{4 M_B}\,
   \bar B(x) \sigma^{\mu\nu} B(x) F_{\mu\nu}(x)
+  \frac{i C_{BT}^{\rm mag}}{4 M_B}
   \bar B(x) \gamma^\mu\gamma_5 T^\nu(x) F_{\mu\nu}(x)
-  \frac{C_T^{\rm mag}}{4 M_T}
   \overline T_\alpha(x) \sigma^{\mu\nu} T^\alpha(x) F_{\mu\nu}(x)	
\notag\\
& & \hspace*{-2.2cm}
+\ \frac{C_{\phi\phi^\dag}^{\rm mag}}{4 M_B f^2}\,
  \bar{p}(x) \sigma^{\mu\nu} p(x) \int d^4a \int d^4b\, F_{\mu\nu}(x)\,
  {\cal G}_\phi^q(x+b,x+a) F(a) F(b)\, \phi(x+a) \phi^\dag(x+b)
\notag\\
& & \hspace*{-2.2cm}
+\ \mathscr{D}_\mu \phi(x) (\mathscr{D}_\mu \phi)^\dag(x)
+\ \cdots,
\label{eq:j4}
\end{eqnarray}
where the gauge link ${\cal G}_\phi^q$ is introduced to maintain local gauge invariance,
\begin{equation}
{\cal G}_\phi^q(x,y)=\text{exp}\left[-ie^q_\phi\int_x^y dz^\mu \int d^4l\, F(l)\, A_\mu(z+l)\right]
\label{eq:link}
\end{equation}
and $F(a)$ is the meson--baryon vertex form factor in coordinate space. The Fourier transformation of $F(a)$ gives the form factor $\widetilde{F}(k)$ in momentum space, where $k$ is the momentum of the corresponding meson. In previous numerical calculations a dipole form has typically been chosen,
\begin{equation}
\widetilde{F}(k)=\left(\frac{\Lambda^2-m_\phi^2}{\Lambda^2-k^2}\right)^2,
\end{equation}
where $\Lambda$ is the cutoff parameter and $m_\phi$ is the meson mass. Note that both the local Lagrangian in Eq.~(\ref{eq:Llocal}) and the nonlocal Lagrangian in Eq.~(\ref{eq:j4}) are invariant under the gauge transformations
\begin{subequations}
\begin{eqnarray}
B(x) &\to& 
B'(x) = B(x) \exp\big[i e^q_B\, \theta(x)\big],
\\
T_\mu(x) &\to&
T_\mu'(x) = T_\mu(x) \exp\big[i e^q_T\, \theta(x)\big],
\\
\phi(x) &\to&
\phi'(x) = \phi(x) \exp\big[i e^q_\phi\, \theta(x)\big],
\\
\mathscr{A}^\mu(x) &\to&
\mathscr{A}'^\mu(x) = \mathscr{A}^\mu(x) + \partial^\mu \theta(x),
\end{eqnarray}
\end{subequations}
where $\theta(x)$ is an auxiliary function.

\begin{table}[] 
\begin{adjustwidth}{-\extralength}{0cm}
\begin{center}
\caption{Coupling constants $C_{B\phi}$ and $C_{T\phi}$ for the $p B \phi$ and $p T \phi$ interactions, respectively, and $C_{\phi\phi^\dag}$ and $C'_{\phi\phi^\dag}$ for the $p p \phi \phi^\dag$ coupling, and the tree level magnetic moments $C_B^{\rm mag}$, $C_T^{\rm mag}$, $C_{BT}^{\rm mag}$ and $C_{\phi\phi^\dag}^{\rm mag}$, respectively, for all the allowed flavor channels.}
{\normalsize
\begin{tabular}{c|ccccccc} \hline
&&&&&&& \\
\hspace*{0.1cm}$\bm B$ \hspace*{0.1cm}
& \hspace*{0.1cm}$\bm p$ \hspace*{0.1cm}
& \hspace*{0.1cm}$\bm n$ \hspace*{0.1cm}
& \hspace*{0.1cm}$\bm \Sigma^+$\hspace*{0.1cm}
& \hspace*{0.1cm}$\bm \Sigma^0$\hspace*{0.1cm}
& \hspace*{0.1cm}$\bm \Sigma^-$\hspace*{0.1cm}
& \hspace*{0.1cm}$\bm \Lambda$\hspace*{0.1cm}	
& \hspace*{0.1cm}$\bm{\Lambda\Sigma^0}$\hspace*{0.1cm}
\\
\hspace*{0.1cm}$C_{B}^{\text {mag}}$\hspace*{0.1cm}
& \hspace*{0.1cm}$\frac13c_1 +c_2$\hspace*{0.1cm}
& \hspace*{0.1cm}$-\frac23 c_1$\hspace*{0.1cm}
& \hspace*{0.1cm}$\frac13c_1 +c_2$\hspace*{0.1cm}
& \hspace*{0.1cm}$\frac13 c_1$\hspace*{0.1cm}
& \hspace*{0.1cm}$\frac13 c_1-c_2$\hspace*{0.1cm}
& \hspace*{0.1cm}$-\frac13 c_1$\hspace*{0.1cm}	
& \hspace*{0.1cm}$\frac{1}{\sqrt{3}} c_1$\hspace*{0.1cm}	 
\\ 
&&&&&&& \\
\hline
&&&&&&& \\
\hspace*{0.1cm}$\bm T$\hspace*{0.1cm}
& \hspace*{0.1cm}$\bm \Delta^{++}$\hspace*{0.1cm}
& \hspace*{0.1cm}$\bm \Delta^+$\hspace*{0.1cm}
& \hspace*{0.1cm}$\bm \Delta^0$\hspace*{0.1cm}
& \hspace*{0.1cm}$\bm \Delta^-$ \hspace*{0.1cm}
& \hspace*{0.1cm}$\bm \Sigma^{*+}$\hspace*{0.1cm}
& \hspace*{0.1cm}$\bm \Sigma^{*0}$\hspace*{0.1cm}
& \hspace*{0.1cm}$\bm \Sigma^{*-}$\hspace*{0.1cm}
\\
\hspace*{0.1cm}$C_T^{\text {mag}}$\hspace*{0.1cm}
& \hspace*{0.1cm}$\frac23 F_2^T$\hspace*{0.1cm}
& \hspace*{0.1cm}$\frac13 F_2^T$\hspace*{0.1cm}
& \hspace*{0.1cm}$0$\hspace*{0.1cm}
& \hspace*{0.1cm}$-\frac13 F_2^T$\hspace*{0.1cm}
& \hspace*{0.1cm}$\frac13 F_2^T$\hspace*{0.1cm}
& \hspace*{0.1cm}$0$\hspace*{0.1cm}
& \hspace*{0.1cm}$-\frac13 F_2^T$\hspace*{0.1cm}	
\\ 
&&&&&&& \\
\hline
&&&&&&& \\
\hspace*{0.1cm}$\bm{BT}$\hspace*{0.1cm}
& \hspace*{0.1cm}$\bm{p\Delta^+}$\hspace*{0.1cm}
& \hspace*{0.1cm}$\bm{\Delta^0}$\hspace*{0.1cm}
& \hspace*{0.1cm}$\bm{\Sigma^+\Sigma^{*+}}$ \hspace*{0.1cm}
& \hspace*{0.1cm}$\bm{\Sigma^0\Sigma^{*0}}$\hspace*{0.1cm}
& \hspace*{0.1cm}$\bm{\Lambda\Sigma^{*0}}$\hspace*{0.1cm}
& \hspace*{0.1cm}$\bm{\Sigma^-\Sigma^{*-}}$\hspace*{0.1cm}
\\
\hspace*{0.1cm}$C_{BT}^{\text {mag}}$\hspace*{0.1cm}
& \hspace*{0.1cm}$-\frac{1}{\sqrt3}c_4$\hspace*{0.1cm}
& \hspace*{0.1cm}$-\frac{1}{\sqrt3}c_4$\hspace*{0.1cm}
& \hspace*{0.1cm}$\frac{1}{\sqrt3}c_4$\hspace*{0.1cm}
& \hspace*{0.1cm}$\frac{1}{2\sqrt3}c_4$\hspace*{0.1cm}
& \hspace*{0.1cm}$\frac12 c_4$\hspace*{0.1cm}
& \hspace*{0.1cm}$0$\hspace*{0.1cm}	
\\ 
&&&&&&& \\
\hline
&&&&&&& \\
\hspace*{0.1cm}$\bm{B\phi}$\hspace*{0.1cm}
& \hspace*{0.1cm}$\bm{p \pi^0}$\hspace*{0.1cm}
& \hspace*{0.1cm}$\bm{n \pi^+}$\hspace*{0.1cm}
& \hspace*{0.1cm}$\bm{\Sigma^+ K^0}$\hspace*{0.1cm}
& \hspace*{0.1cm}$\bm{\Sigma^0 K^+}$\hspace*{0.1cm}
& \hspace*{0.1cm}$\bm{\Lambda K^+}$\hspace*{0.1cm}	
\\
\hspace*{0.1cm}$C_{B\phi}$\hspace*{0.1cm}
& \hspace*{0.1cm}$\frac12 (D+F)$\hspace*{0.1cm}
& \hspace*{0.1cm}$\frac{1}{\sqrt2} (D+F)$\hspace*{0.1cm}
& \hspace*{0.1cm}$\frac{1}{\sqrt2} (D-F)$\hspace*{0.1cm}
& \hspace*{0.1cm}$\frac12 (D-F)$\hspace*{0.1cm}
& \hspace*{0.1cm}$-\frac{1}{\sqrt{12}} (D+3F)$\hspace*{0.1cm} 
\\ 
&&&&&&& \\
\hline
&&&&&&& \\
$\bm{T\phi}$
& $\bm{\Delta^0 \pi^+}$
& $\bm{\Delta^+ \pi^0}$
& $\bm{\Delta^{++} \pi^-}$
& $\bm{\Sigma^{*+} K^0}$
& $\bm{\Sigma^{*0} K^+}$				
\\
$C_{T\phi}$
& $-\frac{1}{\sqrt 6} {\cal C}$
& $-\frac{1}{\sqrt 3} {\cal C}$
& $\frac{1}{\sqrt 2} {\cal C}$
& $\frac{1}{\sqrt 6} {\cal C}$
& $-\frac{1}{\sqrt {12}} {\cal C}$
\\ 
&&&&&&& \\
\hline
&&&&&&& \\
$\bm{\phi\phi^\dag}$
& $\bm{\pi^+\pi^-}$
& $\bm{K^0 \overline{K}^0}$
& $\bm{K^+ K^-}$
&
&						\\
$C_{\phi\phi^\dag}$
& $\frac{1}{2}$
& $\frac{1}{2}$
& 1
&
&						\\
$C_{\phi\phi^\dag}'$
& $4(b_{10}+b_{11})$
& $4(b_{11}-b_{10})$
& $8b_{11}+2b_{9}$
&
&						\\
$C_{\phi\phi^\dag}^{\rm mag}$
& $-\frac{1}{2}(c_1+c_2)$
& $0$
& $-c_2$
&
&						\\
&&&&&&& \\
\hline
\end{tabular}
}
\label{tab:C}
\end{center}
\end{adjustwidth}
\end{table}

Using a change of variables	$z^\mu \to x^\mu + a^\mu\, t + b^\mu\, (1-t)$, the gauge link ${\cal G}_\phi^q$ in Eq.~(\ref{eq:link}) can be expanded in powers of the charge $e_\phi^q$,
\begin{eqnarray}
{\cal G}^q_\phi(x+b,x+a)
&=& \exp \Big[ -i e^q_\phi\, (a-b)^\mu
	      \int_0^1 dt\, \mathscr{A}_\mu\big(x+at+b(1-t)\big)
	 \Big]
\notag\\
&=& 1\ +\ \delta {\cal G}^q_\phi\
       +\ \cdots,
\label{eq:linkexpand}
\end{eqnarray}
where
\begin{eqnarray}
\delta {\cal G}^q_\phi
&=& -\ i e^q_\phi\, (a-b)^\mu
       \int_0^1 dt\, \mathscr{A}_\mu\big(x+at+b(1-t)\big).
\label{eq:deltaG}
\end{eqnarray}
The nonlocal Lagrangian ${\cal L}^{\rm (nonloc)}$ in Eq.~(\ref{eq:j4}) can be further decomposed into free and interacting parts, with the interacting parts consisting of purely hadronic (${\cal L}^{\rm (nonloc)}_{\rm had}$), electromagnetic (${\cal L}^{\rm (nonloc)}_{\rm em}$), and gauge link (${\cal L}^{\rm (nonloc)}_{\rm link}$) contributions. The~hadronic and electromagnetic interaction parts of ${\cal L}^{\rm (nonloc)}$ are obtained from the first term in the gauge link in Eq.~(\ref{eq:linkexpand}), and are given by
\begin{eqnarray}
{\cal L}^{\rm (nonloc)}_{\rm had}(x)
&=& \bar{p}(x)
    \left[ \frac{C_{B\phi}}{f}\, \gamma^\mu \gamma_5 B(x)
	 + \frac{C_{T\phi}}{f}\, \Theta^{\mu\nu} T_\nu(x)
    \right]
    \!\int d^4a\, F(a)\, \partial_\mu \phi(x+a)
	+ {\rm H.c.}
\notag\\
&+& 
    \frac{iC_{\phi\phi^\dag}}{2f^2}
    \bar{p}(x) \gamma^\mu p(x)
    \int d^4a \int d^4b\ F(a)\, F(b)
    \left[ \phi(x+a) \partial_\mu \phi^\dag(x+b)
	 - \partial_\mu \phi(x+a) \phi^\dag(x+b)
    \right]
\notag \\
&+& 
    \frac{iC_{\phi\phi^\dag}'}{2f^2}
    \bar{p}(x) \sigma^{\mu\nu} p(x)
    \int d^4a \int d^4b\ F(a)\, F(b)
    \left[ \partial_\mu\phi(x+a) \partial_\nu \phi^\dag(x+b)
	 - \partial_\mu \phi(x+a) \partial_\nu \phi^\dag(x+b)
    \right],
\label{eq:Lnonloc_had}
\end{eqnarray}
and 
\begin{eqnarray}
{\cal L}^{\rm (nonloc)}_{\rm em}(x)
&=& e^q_B\, \bar{B}(x) \gamma^\mu B(x)\, \mathscr{A}_\mu(x)\
 +\ e^q_T\, \overline{T}_\mu(x) \gamma^{\mu\nu\alpha} T_\nu(x)\,
	    \mathscr{A}_\alpha(x)
\notag\\
& & \hspace*{-2.3cm}
 + i e^q_\phi \left[ \partial^\mu \phi(x) \phi^\dag(x)
		    - \phi(x) \partial^\mu \phi^\dag(x)
	       \right] \mathscr{A}_\mu(x)
\notag\\
& & \hspace*{-2.3cm}
 - i e^q_\phi\, \bar{p}(x)
    \left[
      \frac{C_{B\phi}}{f}\, \gamma^\mu \gamma_5 B(x)
    + \frac{C_{T\phi}}{f}\, \Theta^{\mu\nu} T_\nu(x)
    \right]
    \int d^4a\, F(a)\, 
    \phi(x+a) \mathscr{A}_\mu(x) + {\rm H.c.}
\notag\\
& & \hspace*{-2.3cm}
 -  \frac{e^q_\phi C_{\phi\phi^\dag}}{2f^2}\,
    \bar{p}(x) \gamma^\mu p(x)
    \int d^4a \int d^4b\ 
    F(a)\, F(b)\, \phi(x+a) \phi^\dag(x+b)\mathscr{A}_\mu(x)
\notag\\
& & \hspace*{-2.3cm}
 -  \frac{C_{\phi\phi^\dag}'}{f^2}\,
    \bar{p}(x) \sigma^{\mu\nu} p(x) 
    \int d^4a \int d^4b\ 
    F(a)\, F(b)\, \phi(x+a) \partial_\nu\phi^\dag(x+b) \mathscr{A}_\mu(x)
\notag\\
& & \hspace*{-2.3cm}
 +  \frac{C_B^{\rm mag}}{4 M_B}
    \bar B(x) \sigma^{\mu\nu} B(x) F_{\mu\nu}(x)
 +  \frac{i C_{BT}^{\rm mag}}{4 M_B}
    \bar B(x) \gamma^\mu\gamma_5 T^\nu (x) F_{\mu\nu}(x)	
 -  \frac{C_T^{\rm mag}}{4 M_T}
    \overline T_\alpha(x) \sigma^{\mu\nu} T^\alpha(x) F_{\mu\nu}(x)	
\notag\\
& & \hspace*{-2.3cm}
 +  \frac{C_{\phi\phi^\dag}^{\rm mag}}{4 M_B f^2}\,
    \bar{p}(x) \sigma^{\mu\nu} p(x) \int d^4a \int d^4b\,
    F(a) F(b)\, \phi(x+a) \phi^\dag(x+b) F_{\mu\nu}(x),
\label{eq:Lnonloc_em}
\end{eqnarray}
respectively.
The second term in Eq.~(\ref{eq:linkexpand}) explicitly depends on the gauge link and gives rise to an additional contribution to the Lagrangian density that can be expanded as
\begin{eqnarray}
\hspace*{-1cm}{\cal L}^{\rm (nonloc)}_{\rm link}(x)
&=& -i e^q_\phi\, \bar{p}(x)
    \left[ \frac{C_{B\phi}}{f}\, \gamma^\rho \gamma_5 B(x)
	     + \frac{C_{T\phi}}{f}\, \Theta^{\rho\nu} T_\nu(x)
    \right]
\notag\\
& & \hspace*{0.3cm} \times
    \int_0^1 dt \int d^4a\,   
    F(a) a^\mu\, \partial_\rho (\phi(x+a) \mathscr{A}_\mu\big(x+at)\big)
    + {\rm H.c.}
\notag\\
&+& \frac{e^q_\phi C_{\phi\phi^\dag}}{2f^2}\,
    \bar{p}(x) \gamma^\rho p(x)
    \int_0^1 dt \int d^4a\! \int d^4b\, 
	F(a)\, F(b)\, (a-b)^\mu\,
\notag\\
& & \hspace*{0.3cm} \times
    \left[ \phi(x+a) \partial_\rho \phi^\dag(x+b)
	     - \partial_\rho \phi(x+a) \phi^\dag(x+b)
    \right] \mathscr{A}_\mu\big(x+at+b(1-t)\big).
\label{eq:Lnonloc_link}
\end{eqnarray}
Finally, the quark current for the nonlocal theory can be written as a sum of two terms, arising from the usual electromagnetic current $J_{q, \rm em}^\mu$ obtained from Eq.~(\ref{eq:Lnonloc_em}) with minimal substitution, and from the additional term associated with the gauge link $\delta J_q^\mu$,
\begin{subequations}
\begin{eqnarray}
J_{q, \rm em}^\mu(x)
&\equiv& \frac{\delta \int d^4y\,
		{\cal L}_{\rm em}^{\rm (nonloc)}(y)}
	      {\delta \mathscr{A_\mu}(x)}, 
\\
\delta J_q^\mu(x)
&\equiv& \frac{\delta \int d^4y\, {\cal L}^{\rm (nonloc)}_{\rm link}(y)}
              {\delta \mathscr{A_\mu}(x)},
\end{eqnarray}
\end{subequations}
where, explicitly,
\begin{eqnarray}
\label{eq:Jem}
J_{q, \rm em}^\mu(x)
&=& e^q_B\,\bar{B}(x) \gamma^\mu B(x)
 +  e^q_T\, \overline{T}_\alpha(x) \gamma^{\alpha\nu\mu} T_\nu(x)
 +  ie^q_\phi \big[ \partial^\mu \phi(x) \phi^\dag(x)
                  - \phi(x) \partial^\mu \phi^\dag(x)
              \big]
\notag\\
&-& i e^q_\phi\,
\bigg(    \int d^4a F(a)\, \bar{p}(x)
    \left[ \frac{C_{B\phi}}{f}\, \gamma^\mu \gamma_5 B(x)
	     + \frac{C_{T\phi}}{f}\, \Theta^{\mu\nu} T_\nu(x)
    \right] \phi(x+a)\ +\ {\rm H.c.}
\bigg)
\notag\\
&-& \frac{e^q_\phi C_{\phi\phi^\dag}}{2f^2}
    \int d^4a \int d^4b F(a) F(b)\,
    \bar{p}(x) \gamma^\mu p(x)\, \phi(x+a) \phi^\dag(x+b)
\notag\\
&+& \frac{C_B^{\rm mag}}{2 M_B}\, \int d^4a F(a)\,
    \partial_\nu\big(\bar{p}(x) \sigma^{\mu\nu} p(x) \big)
 -  \frac{C_T^{\rm mag}}{2 M_T}\, \int d^4a F(a)\,
    \partial_\nu\big(\overline {T}_\alpha(x) \sigma^{\mu\nu} T^\alpha(x) \big )
\notag\\
&+& \frac{i C_{BT}^{\rm mag}}{4 M_B}\, 
    ( \partial_\nu\big(\bar{p}(x) \gamma^\nu \gamma_5 T^\mu(x) \big)
    - \partial_\nu\big(\bar{p}(x) \gamma^\mu \gamma_5 T^\nu(x) \big)
\notag\\
&+& \frac{C_{\phi\phi^\dag}^{\rm mag}}{2 M_B f^2}\,
    \int d^4a \int d^4b F(a) F(b)\,
   \partial_\nu \big(\bar{p}(x) \sigma^{\mu\nu} p(x) \phi(x+a) \phi^\dag(x+b) \big),
\\
& &
\notag\\
\delta J_q^\mu(x)
&=& i e^q_\phi
    \int_0^1 dt \int d^4a F(a)\, a^\mu\,
\notag\\
& & \times
    \partial_\rho 
    \left( \bar{p}(x-at)
    \left[ \frac{C_{B\phi}}{f}\, \gamma^\rho \gamma_5 B(x-at)\,
       +\, \frac{C_{T\phi}}{f}\, \Theta^{\rho\nu} T_\nu(x-at)
    \right]
    \right) \phi(x+a\bar{t}\big) + {\rm H.c.}
\notag\\
&-& \frac{e^q_\phi C_{\phi\phi^\dag}}{2f^2}
    \int_0^1 dt \int d^4a \int d^4b\, F(a) F(b)\, (a-b)^\mu\,
\notag\\
& & \times
\bigg[ 
    \partial_\rho
    \Big(
        \bar{p}\big(x-at-b\bar{t}\big) \gamma^\rho p\big(x-at-b\bar{t}\big)
    \phi\,\big(x+(a-b)\bar{t}\big)
    \Big)
    \phi^\dag\big(x-(a-b)t)
\notag\\
& & \hspace*{0.2cm}
 -  \partial_\rho
    \Big( 
        \bar{p}\big(x-at-b\bar{t}\big) \gamma^\rho p\big(x-at-b\bar{t}\big)
    \phi^\dag\big(x-(a-b)t\big)
	\Big)
    \phi\big(x+(a-b)\bar{t}\big)
\bigg],
\label{eq:Jlink}
\end{eqnarray}
with $\bar t \equiv 1-t$.
Compared with the local theory, Eqs.~(\ref{eq:j1}) and (\ref{eq:jq}), the nonlocal formulation in Eqs.~(\ref{eq:Lnonloc_had})--(\ref{eq:Jlink}) includes the regulator function $F(a)$. In the limit where $F(a) \to \delta^{(4)}(a)$, which corresponds to taking the momentum space form factor to unity, the local limit can be obtained from the nonlocal result. \\

\subsection{GPDs with zero skewness}

The spin-averged GPDs for a quark flavor $q$ in a proton with initial momentum $p$ and final momentum $p'$ are defined by the Fourier transform of the matrix elements of the quark bilocal field operators $\psi_q$ as~\cite{Ji:1996nm}
\begin{eqnarray}
\label{eq:GPD}
\int_{-\infty}^{\infty}\frac{d\lambda}{2\pi} e^{-ix\lambda}
\langle p'| 
    \bar\psi_q (\tfrac12\lambda n) \not\!{n}\, \psi_q(-\tfrac12\lambda n)
|p \rangle 
&=& \bar u(p') 
  \Big[ \not\!{n} H^q(x,\xi,t) 
      + \frac{i\sigma^{\mu\nu}n_\mu \Delta_\nu}{2M}\, E^q(x,\xi,t)
  \Big] u(p),
\end{eqnarray}
where $n_\mu$ is the light-cone vector which projects the ``plus" component of momenta, and $\lambda$ is a dimensionless parameter. From Lorentz invariance, the Dirac ($H^q$) and Pauli ($E^q$) GPDs can be written as functions of the light-cone momentum fraction $x$ of the proton carried by the initial quark with momentum $k_q$ and the skewness parameter $\xi$, which are defined as
\begin{equation}
x\,   \equiv\, \frac{k_q^+}{P^+}, \qquad
\xi\, \equiv\, -\frac{\Delta^+}{2P^+},
\end{equation}
where
\begin{equation}
P\, =\, \frac12 \big(p + p'\big), \qquad
\Delta\, =\, p' - p,
\end{equation}
are the average and difference of the initial and final proton momenta, respectively. The light-front components $k^+$ and $k^-$ of any four-vector $k^\mu$ are defined as 
$k^+ = \frac{1}{\sqrt{2}}(k^0+k^3)$ and 
$k^- = \frac{1}{\sqrt{2}}(k^0-k^3)$.
The GPDs are also functions of the hadronic four-momentum transfer squared, $t \equiv \Delta^2$. The dependence of the GPDs on the fourth variable, typically taken to be the four-momentum transfer squared from the incident lepton, $Q^2$, is suppressed. For the case of zero skewness, $\xi=0$, the hadron momenta are parametrized as~\cite{Brodsky:1997de}
\begin{subequations}
\begin{eqnarray}
p^\mu  &=& \Big(P^+,P^-,-\frac12 {\bm \Delta_\perp}\Big), \\ 
p'^\mu &=& \Big(P^+,P^-,+\frac12 {\bm \Delta_\perp}\Big),
\end{eqnarray}
\end{subequations}
where the momentum transfer $\Delta^\mu$ is purely in the transverse direction.

\begin{figure}[] 
\begin{adjustwidth}{-\extralength}{0cm}
\begin{center}
\includegraphics[scale=0.85]{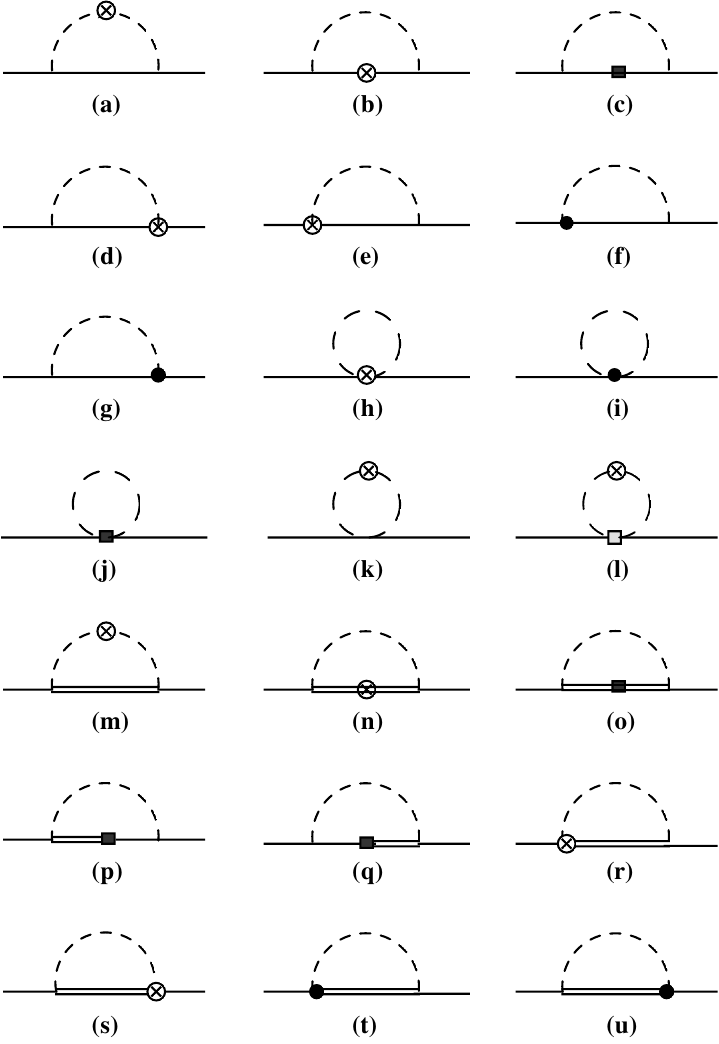}
\caption{One-loop diagrams for the proton to pseudoscalar meson (dashed lines) and octet baryon (solid lines) or decuplet baryon (double solid lines) splitting functions up to the fourth chiral order:
{\bf (a)--(c)} octet baryon rainbow diagrams,
{\bf (d)--(g)} octet baryon Kroll-Ruderman diagrams,
{\bf (h)--(j)} tadpole diagrams,
\mbox{{\bf (k)--(l)} bubble} diagrams,
{\bf (m)--(o)} decuplet baryon rainbow diagrams,
{\bf (p)--(q)} octet-decuplet transition rainbow diagrams,
{\bf (r)--(u)} decuplet baryon Kroll-Ruderman diagrams.
The crossed circles ($\otimes$) represent the interaction with external vector field from the minimal substitution, filled circles ({\Large$\bullet$}) denote additional gauge link interaction with the external field, black squares ($\blacksquare$) represent the magnetic interaction in Eq.~(\ref{lomag}), and gray squares (${\textcolor{gray}{\blacksquare}}$) denote the interaction in Eq.~(\ref{adm}).} 
\label{diagrams}
\end{center}
\end{adjustwidth}
\end{figure}

The diagrams relevant for the calculation of the one-meson loop contributions to GPDs up to the fourth chiral order are illustrated in Fig.~\ref{diagrams}. Assuming that meson loops are the only source of antiquarks in the proton, the convolution form for the antiquark electric and magnetic GPDs in the proton involves contributions only from the diagrams in Fig.~\ref{diagrams}(a), \ref{diagrams}(k), \ref{diagrams}(l) and \ref{diagrams}(m). In particular, for the $H^{\bar{q}}$ and $E^{\bar{q}}$ GPDs at zero skewness, one has
\begin{subequations}
\label{eq:HEqbar}
\begin{eqnarray}
H^{\bar{q}}(x,t)
&=& \sum_{\phi B T} 
    \Big[ \big( f_{\phi B}^{(\rm rbw)} 
              + f_{\phi T}^{(\rm rbw)} 
              + f^{(\rm bub)}_\phi 
          \big) \otimes H^{\bar{q}}_\phi
    \Big](x,t),
\\
E^{\bar{q}}(x,t)
&=& \sum_{\phi B T} 
    \Big[ \big( g_{\phi B}^{(\rm rbw)}
              + g_{\phi T}^{(\rm rbw)}
              + g'^{(\rm bub)}_\phi 
           \big) \otimes H^{\bar{q}}_\phi
    \Big](x,t),
\end{eqnarray}
\end{subequations}
where $H^{\bar{q}}_\phi$ is the electric GPD for quark flavor $\bar{q}$ in the meson $\phi$. The functions $f_{\phi B}^{(\rm rbw)}$ and $g_{\phi B}^{(\rm rbw)}$ are the splitting functions for Fig.~\ref{diagrams}(a), respectively, $f_{\phi T}^{(\rm rbw)}$ and $g_{\phi T}^{(\rm rbw)}$ are the decuplet recoil splitting functions for Fig.~\ref{diagrams}(m), respectively, and $f^{(\rm bub)}_\phi$ and $g'^{(\rm bub)}_\phi$ are the functions for Figs.~\ref{diagrams}(k) and (l), respectively. The splitting functions are given explicitly by He~{\it et~al.}~\cite{He:2022leb}.

The convolution form for the quark GPDs receive contributions from all diagrams in Fig.~\ref{diagrams} and hence have a more complicated structure,
\begin{subequations}
\label{eq:HEq}
\begin{eqnarray}
\label{eq:Hq}
H^q(x,t)
&=& Z_2\, H^q_0(x,t)\
 +\ \sum_{\phi B T}
\Big[ 
   \big( f_{\phi B}^{(\rm rbw)} + f_{\phi T}^{(\rm rbw)} + f^{(\rm bub)}_\phi \big) \otimes H^q_\phi 
\nonumber\\
& & \hspace*{3.3cm}
+\ \bar{f}_{B\phi}^{(\rm rbw)} \otimes H^q_B\
+\ \bar{f}^{(\rm KR)}_{B\phi} \otimes H^{q({\rm KR})}_B\
+\ \delta \bar{f}^{(\rm KR)}_{B} \otimes H^{q({\rm KR})}_B\
\nonumber\\
& & \hspace*{3.3cm}
+\ \bar{f}_{T\phi}^{(\rm rbw)} \otimes H^q_T\
+\ \bar{f}^{(\rm KR)}_{T\phi} \otimes H^{q({\rm KR})}_T\
+\ \delta\bar{f}^{(\rm KR)}_{T\phi} \otimes H^{q({\rm KR})}_T\
\nonumber\\
& & \hspace*{3.3cm}
+\ \bar{f}_{B\phi}^{(\rm rbw\,mag)} \otimes E^q_B\
+\ \bar{f}_{T\phi}^{(\rm rbw\,mag)} \otimes E^q_T\
+\ \bar{f}_{BT}^{(\rm rbw\,mag)} \otimes E^q_{BT}~~~~~~
\nonumber\\
& & \hspace*{3.3cm}
+\ \bar{f}^{(\rm tad)}_\phi \otimes H^{q({\rm tad})}_{\phi\phi^\dag}\
+\ \delta\bar{f}^{(\rm tad)}_\phi \otimes H^{q({\rm tad})}_{\phi\phi^\dag}
\Big](x,t),
\\
\nonumber \\
\label{eq:Eq}
E^q(x,t)
&=& Z_2\, E^q_0(x,t)\
 +\ \sum_{\phi B T}
\Big[ 
   \big( g_{\phi B}^{(\rm rbw)} + g_{\phi T}^{(\rm rbw)} + g'^{(\rm bub)}_\phi \big) \otimes H^q_\phi
\nonumber\\
& & \hspace*{3.3cm}
+\ \bar{g}_{B\phi}^{(\rm rbw)} \otimes H^q_B\
+\ \bar{g}^{(\rm KR)}_{B\phi} \otimes H^{q({\rm KR})}_B\
+\ \delta \bar{g}^{(\rm KR)}_{B} \otimes H^{q({\rm KR})}_B\
\nonumber\\
& & \hspace*{3.3cm}
+\ \bar{g}_{T\phi}^{(\rm rbw)} \otimes H^q_T\
+\ \bar{g}^{(\rm KR)}_{T\phi} \otimes H^{q({\rm KR})}_T\
+\ \delta\bar{g}^{(\rm KR)}_{T\phi} \otimes H^{q({\rm KR})}_T\
\nonumber\\
& & \hspace*{3.3cm}
+\ \bar{g}_{B\phi}^{(\rm rbw\,mag)} \otimes E^q_B\
+\ \bar{g}_{T\phi}^{(\rm rbw\,mag)} \otimes E^q_T\
+\ \bar{g}_{BT}^{(\rm rbw\,mag)} \otimes E^q_{BT}~~~~~~
\nonumber\\
& & \hspace*{3.3cm}
+\ \bar{g}^{(\rm tad\,mag)}_\phi \otimes E^{q({\rm tad})}_{\phi\phi^\dag}
\Big](x,t),
\end{eqnarray}
\end{subequations}
where $H^q_0$ and $E^q_0$ are the quark GPDs of the bare proton, and $Z_2$ is the wave function renormalization constant associated with the dressing of the bare proton by the meson loops. As shorthand, in Eqs.~(\ref{eq:HEq}) we use the notation $\bar{f}_j(y) \equiv f_j(1-y)$ and $\bar{g}_j(y) \equiv g_j(1-y)$ to denote the electric and magnetic splitting functions involving couplings to baryons. Note that both the electric and magnetic operators contribute to $H^q(x,t)$ and $E^q(x,t)$ at zero and finite momentum transfer. At zero momentum transfer, however, there is no contribution from the magnetic term to the matrix element, even though the GPD $E^q(x,0)$ itself is nonzero.

The expressions for the quark and antiquark GPDs in Eqs.~(\ref{eq:HEqbar}) and (\ref{eq:HEq}) form the basis for the calculations of meson loop contributions to GPD flavor asymmetries. For the case of $u$ and $d$ quarks, the intermediate states include the nucleon and $\Delta$ baryons and $\pi$ mesons. For the strange quark, on the other hand, the intermediate states that contribute are the $\Lambda$, $\Sigma$ and $\Sigma^*$ hyperons and $K$ mesons. To compute the quark and antiquark GPDs numerically requires information about the GPDs for the various hadronic configurations that contribute in Eqs.~(\ref{eq:HEqbar})--(\ref{eq:HEq}). As discussed in Refs.~\cite{He:2022leb, Salamu:2019dok}, the GPDs used in the calculation can be expressed in terms of the GPDs in the pion and proton, which can be parametrized as products of valence PDFs and $t$-dependent factors~\cite{He:2022leb, Diehl:2004cx}. With the calculated splitting functions and the valence quark distributions as input, one can evaluate the GPDs of the sea quarks from the convolution expressions (\ref{eq:HEqbar})--(\ref{eq:HEq}).

\begin{figure}[t]  
\begin{adjustwidth}{-\extralength}{0cm}
\begin{center}
    \begin{minipage}{0.45\linewidth}
        \centering
        \centerline{
        \includegraphics[width=1\textwidth]{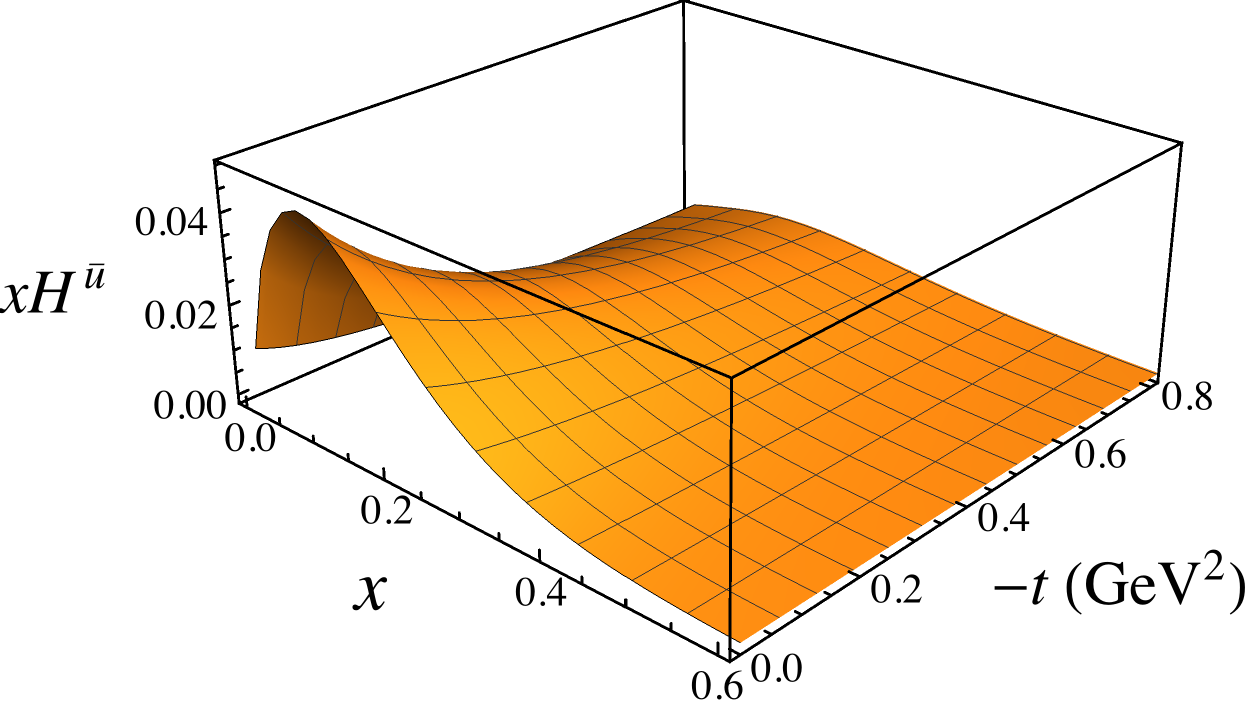}
        }
        \centerline{\small{\bf{(a)}}}
    \end{minipage}
    \begin{minipage}{0.45\linewidth}
        \centering
        \centerline{
        \includegraphics[width=1\textwidth]{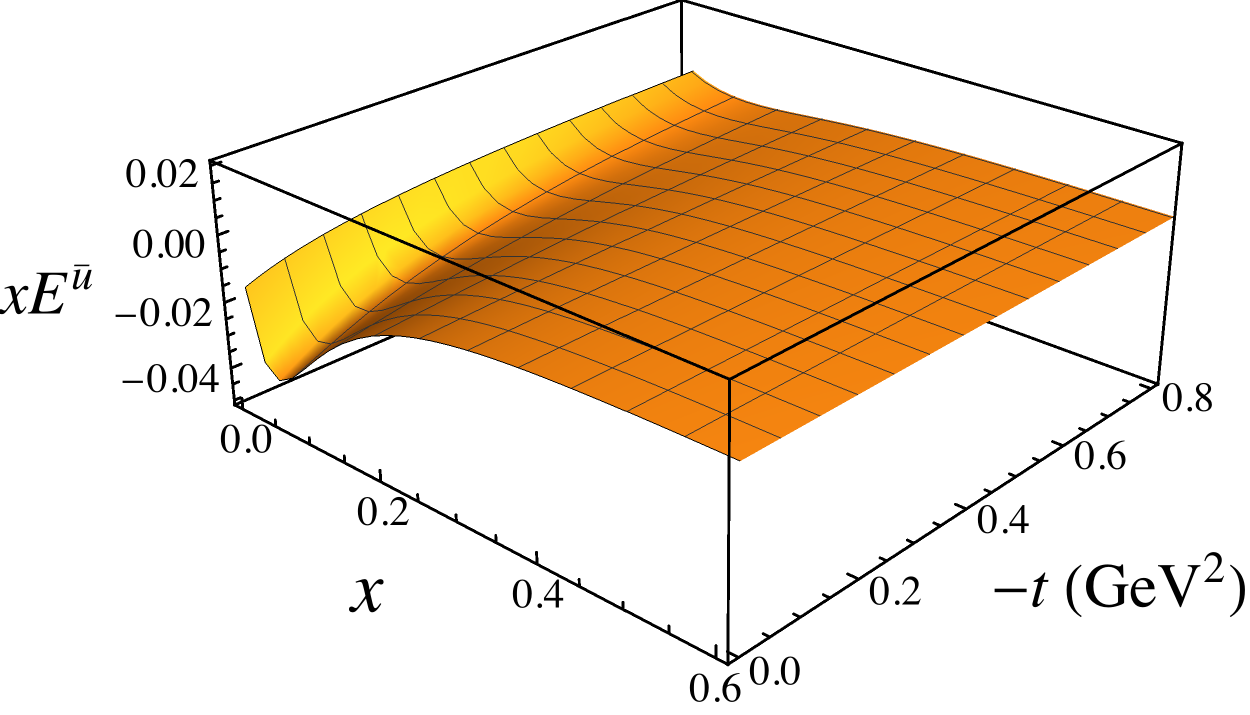}
        }
        \centerline{\small{\bf{(b)}}}
    \end{minipage}    
\\[0.4cm]
        \begin{minipage}{0.45\linewidth}
        \centering
        \centerline{
        \includegraphics[width=1\textwidth]{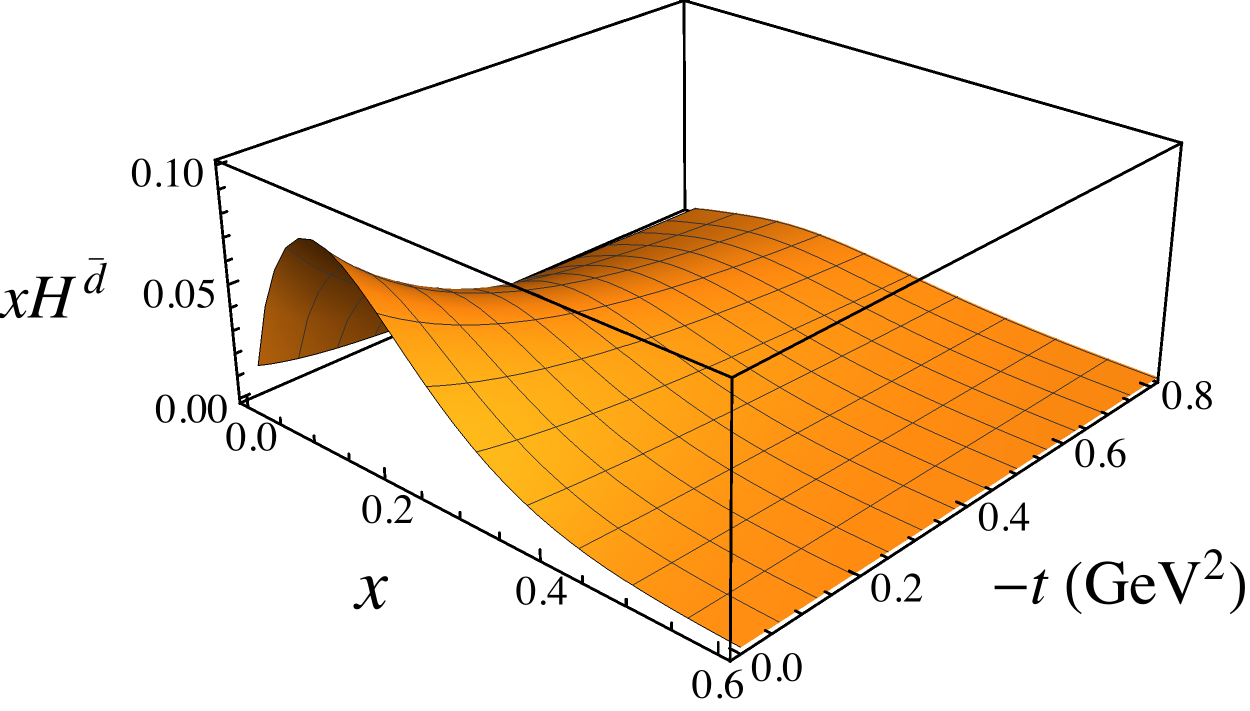}
        }
        \centerline{\small{\bf{(c)}}}
    \end{minipage}
        \begin{minipage}{0.45\linewidth}
        \centering
        \centerline{
        \includegraphics[width=1\textwidth]{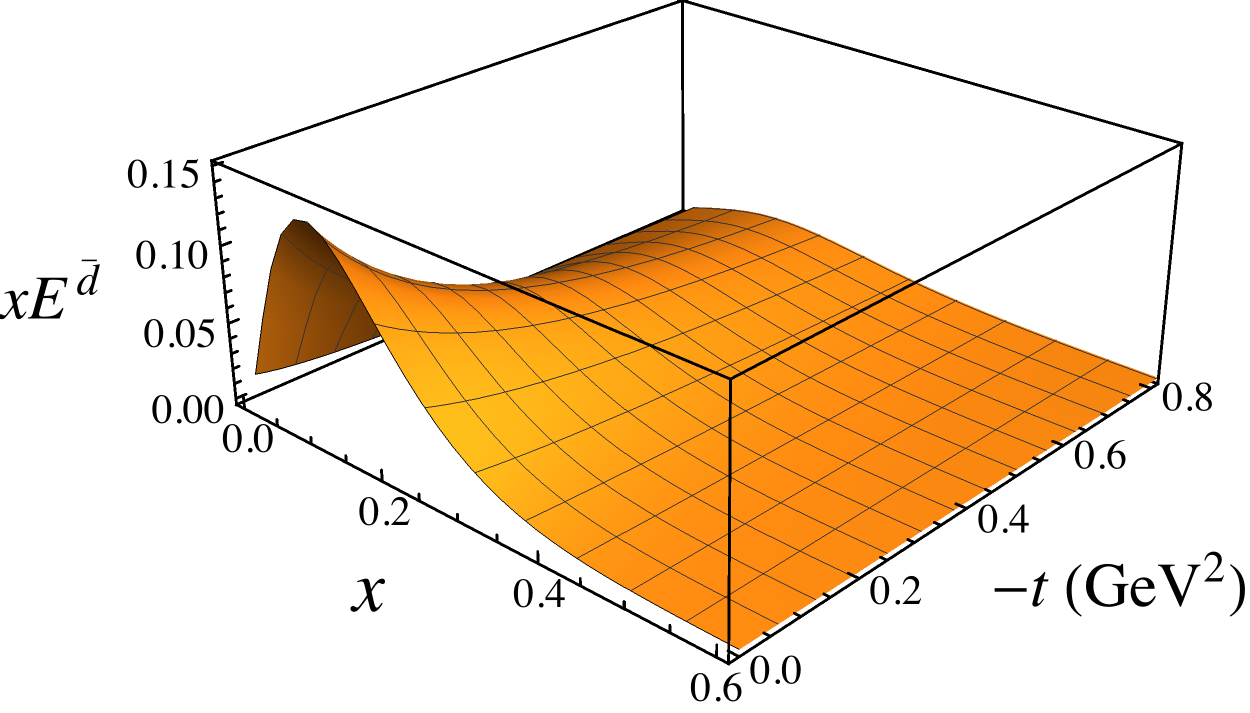}
        }
        \centerline{\small{\bf{(d)}}}
    \end{minipage}
\caption{Electric and magnetic GPDs for light antiquarks: 
{\bf (a)}~$xH^{\bar{u}}$, 
{\bf (b)}~$xE^{\bar{u}}$, 
{\bf (c)}~$xH^{\bar{d}}$, and 
{\bf (d)}~$xE^{\bar{d}}$, versus parton momentum fraction $x$ and four-momentum transfer squared $-t$, for cutoff mass $\Lambda=1$~GeV at a scale $Q=1$~GeV.}
\label{3dubar}
\end{center}
\end{adjustwidth}
\end{figure}

The electric ($H^{\bar q}$) and magnetic ($E^{\bar q}$) GPDs for the light antiquarks in the proton arising from meson loops are shown in Fig.~\ref{3dubar} as a function of the parton momentum fraction $x$ and momentum transfer $-t$ for the $\bar q = \bar u$ and $\bar d$ flavors at the scale $Q=1$~GeV. For $\bar u$ antiquarks, the function $xH^{\bar{u}}$ is positive and peaks at $x \approx 0.1$, roughly independent of the value of $t$. For any fixed value of $x$, $xH^{\bar{u}}$ falls off monotonically with increasing values of $-t$. In contrast, the magnetic $xE^{\bar{u}}$ distribution is negative, with absolute value peaking at slightly smaller $x$ compared with $xH^{\bar{u}}$, and again decreasing in magnitude with increasing $-t$. For $\bar d$ antiquarks, the shape of the $xH^{\bar{d}}$ GPD is similar to that of the $xH^{\bar{u}}$ distribution, although at any given $x$ and $t$ the GPD for the $\bar d$ is larger. This flavor asymmetry stems from the fact that the contribution to $H^{\bar{d}}$ arises from both the octet and decuplet intermediate states, while only the decuplet intermediate states contribute to the $H^{\bar{u}}$ GPD.

The shapes of the magnetic $E^{\bar{q}}$ GPDs reflect the important role played by the orbital angular momentum of the mesons in the intermediate state. For octet baryons, the meson orbital angular momentum tends to be positive, resulting in positive values for $E^{\bar{d}}$. For $\bar u$ quarks, on the other hand, since the intermediate baryons can only be decuplets, the orbital angular momentum of the meson tends to be negative, leading to negative values for $E^{\bar{u}}$. The absolute value of $xE^{\bar{d}}$ is also much larger than $xE^{\bar{u}}$.
Note that the $\delta$-function term in the splitting functions does not contribute to the $H^{\bar{q}}$ and $E^{\bar{q}}$ GPDs, although it does contribute to the lowest moments of these functions.

\begin{figure}[tbph] 
\begin{adjustwidth}{-\extralength}{0cm}
\begin{center}
    \begin{minipage}{0.45\linewidth}
        \centering
        \centerline{
        \includegraphics[width=1\textwidth]{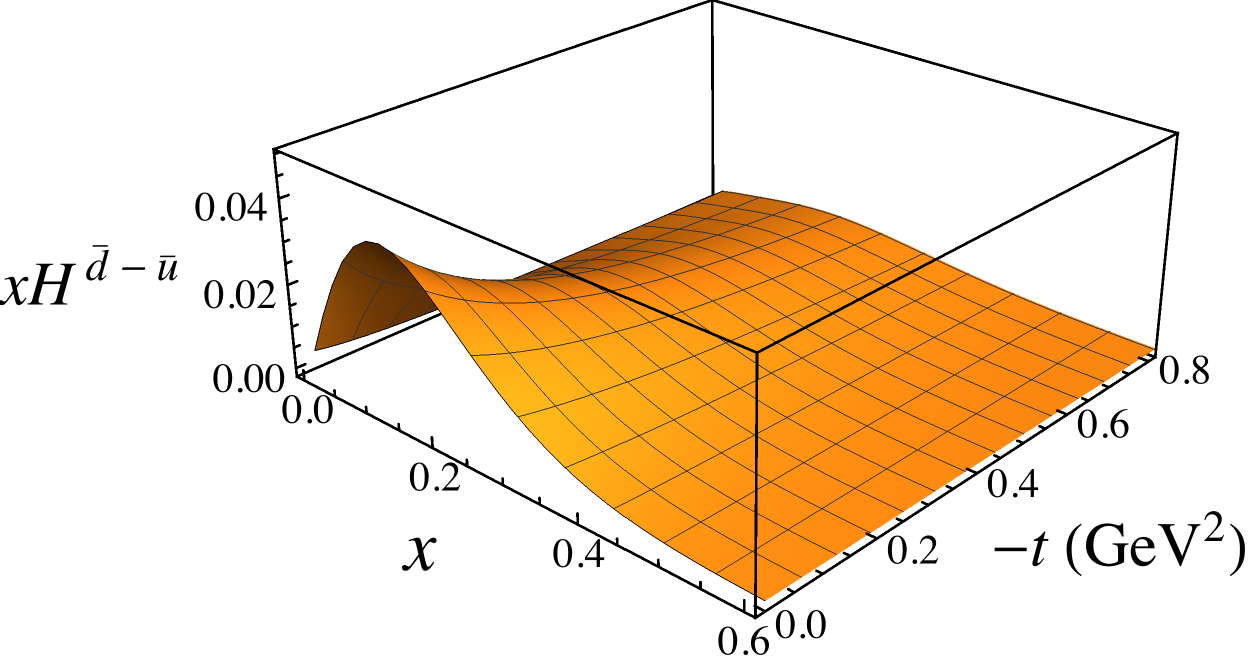}
        }
        \centerline{\small{\bf{(a)}}}
    \end{minipage}
    \begin{minipage}{0.45\linewidth}
        \centering
        \centerline{
        \includegraphics[width=1\textwidth]{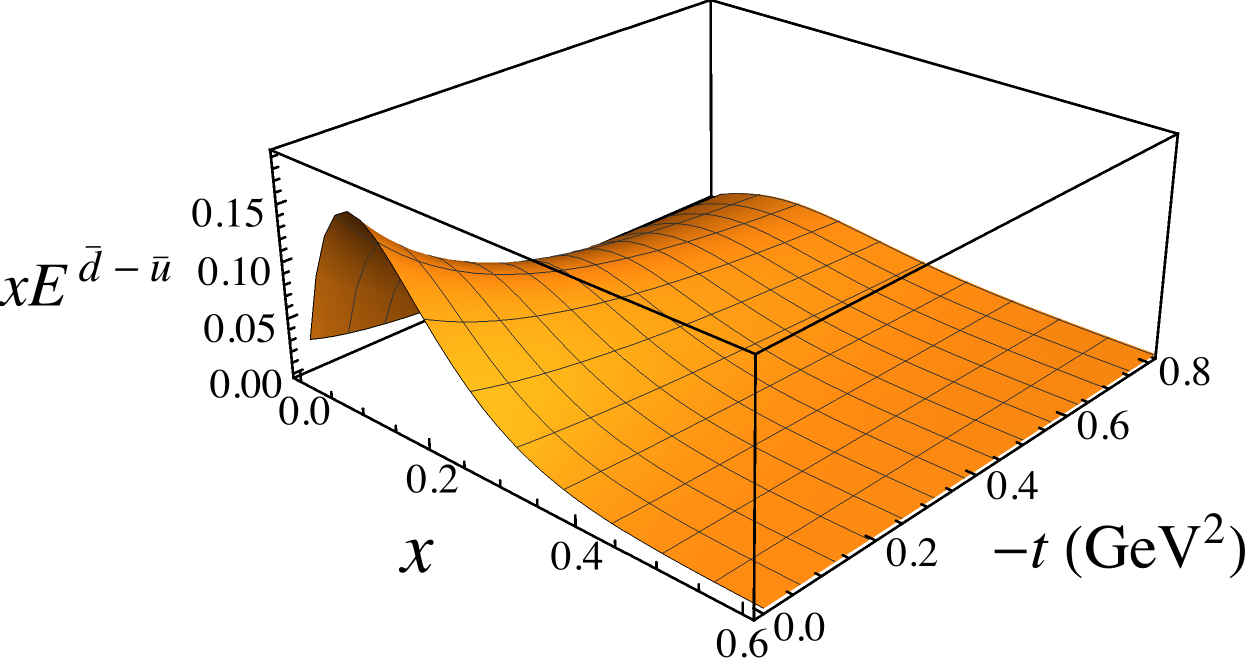}
        }
        \centerline{\small{\bf{(b)}}}
    \end{minipage}
\caption{Light antiquark flavor asymmetry for the {\bf (a)} electric $xH^{\bar{d}-\bar{u}}$ and {\bf (b)}~magnetic $xE^{\bar{d}-\bar{u}}$ GPDs versus parton momentum fraction $x$ and four-momentum transfer squared $-t$, for regulator parameter \mbox{$\Lambda = 1$~GeV.}} 
\label{3du-d} 
\end{center}
\end{adjustwidth}
\end{figure}

Turning now to the light flavor asymmetry of the GPDs, in Fig.~\ref{3du-d} the distributions $xH^{\bar{d}-\bar{u}}$ and $xE^{\bar{d}-\bar{u}}$ are shown versus $x$ and $-t$. Both asymmetries are observed to be positive for all $x$ values, with a peak at $x \approx 0.1$ that decreases with increasing four-momentum transfer squared. At the peak, the magnitude of the magnetic GPD asymmetry $xE^{\bar{d}-\bar{u}}$ is about 4 times larger than the corresponding electric asymmetry $xH^{\bar{d}-\bar{u}}$.

To more clearly illustrate the shape and magnitude of the $\bar{d}-\bar{u}$ asymmetry, in Fig.~\ref{2du-dt0} we plot the electric $xH^{\bar{d}-\bar{u}}$ and magnetic $xE^{\bar{d}-\bar{u}}$ distributions at $t=0$, with the uncertainty bands corresponding to a 10\% uncertainty on the cutoff parameter $\Lambda = 1.0(0.1)$~GeV. The calculated electric asymmetry is compared with a recent parametrization of the $x(\bar d-\bar u)$ PDF from the JAM global QCD analysis of world data~\cite{Cocuzza:2021cbi} at a scale $Q=m_c=1.3$~GeV. The numerical results are in good agreement with the phenomenological parametrization of $x(\bar d-\bar u)$, which is driven mostly by the Drell-Yan proton-proton and proton-deuteron scattering data~\cite{Towell:2001nh, SeaQuest:2021zxb}, and has a maximum of $\approx 0.3-0.4$ at $x \approx 0.05-0.10$. Integrating over~$x$, the values for the two lowest moments of the electric GPD asymmetry are found to be
    $\int_0^1 dx\, H^{\bar{d}-\bar{u}}(x,0) = 0.11(2)$ and 
    $\int_0^1 dx\, xH^{\bar{d}-\bar{u}}(x,0) = 0.009(2)$,
where the uncertainty stems from the range of the cutoff parameter $\Lambda$. The magnetic GPD asymmetry $xE^{\bar{d}-\bar{u}}$ at $t=0$ has a similar shape, but is $\approx 4$ times larger than $xH^{\bar{d}-\bar{u}}$ at the peak. The fact that $xE^{\bar{d}-\bar{u}}$ exceeds $xH^{\bar{d}-\bar{u}}$ is also consistent with the prediction of models based on the large-$N_c$ limit of QCD~\cite{Goeke:2001tz}. After integrating over $x$, one finds
    $\int_0^1 dx\, E^{\bar{d}-\bar{u}}(x,0) = 1.1(2)$ and
    $\int_0^1 dx\, xE^{\bar{d}-\bar{u}}(x,0) = 0.034(6)$.
A large magnitude for the magnetic asymmetry augurs well for future efforts to determine this asymmetry experimentally.

\begin{figure}[htbp] 
\begin{adjustwidth}{-\extralength}{0cm}
\begin{minipage}[b]{.45\linewidth}
\hspace*{-0.3cm}\includegraphics[width=1.08\textwidth, height=5.5cm]{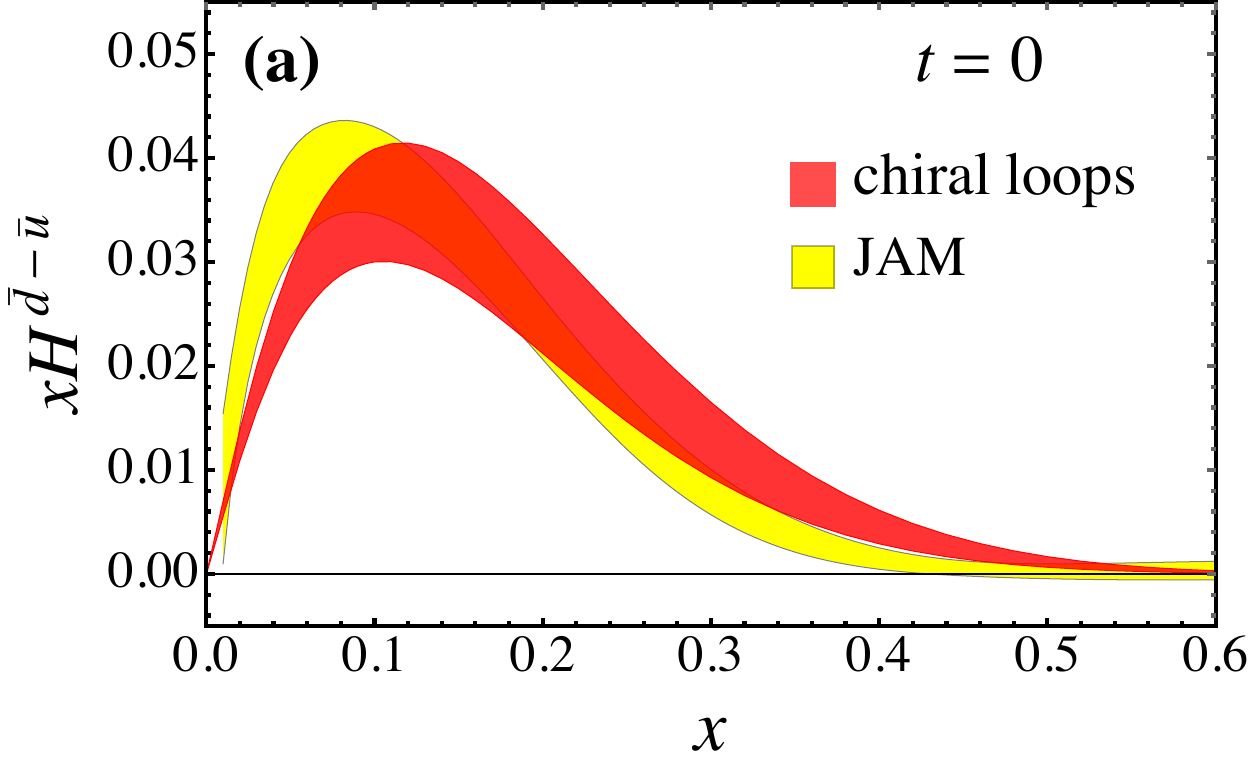}
  \vspace{3pt}
\end{minipage}
\hfill
\begin{minipage}[b]{.45\linewidth}   
\hspace*{-0.8cm} \includegraphics[width=1.09\textwidth, height=5.5cm]{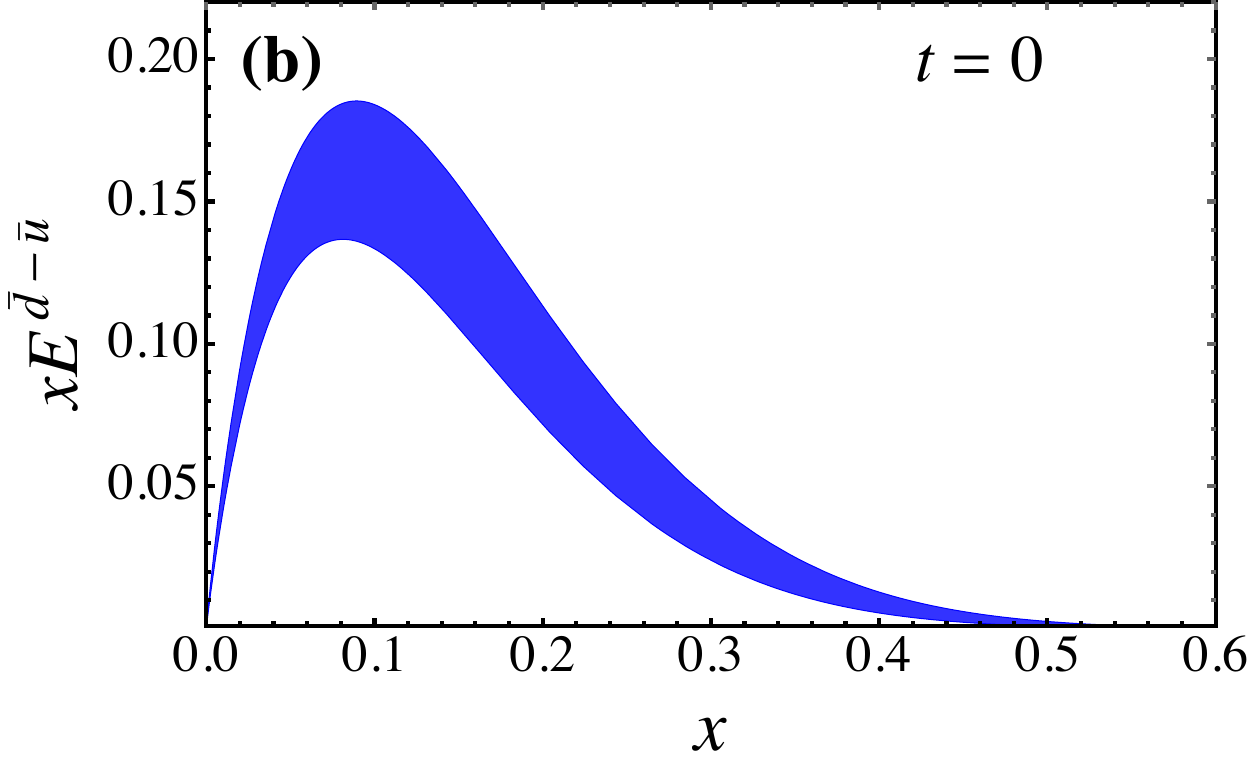} 
   \vspace{3pt}
\end{minipage}  
\\[-0.4cm]
\begin{minipage}[t]{.45\linewidth}
\hspace*{-0.5cm}\includegraphics[width=1.1\textwidth, height=5.5cm]{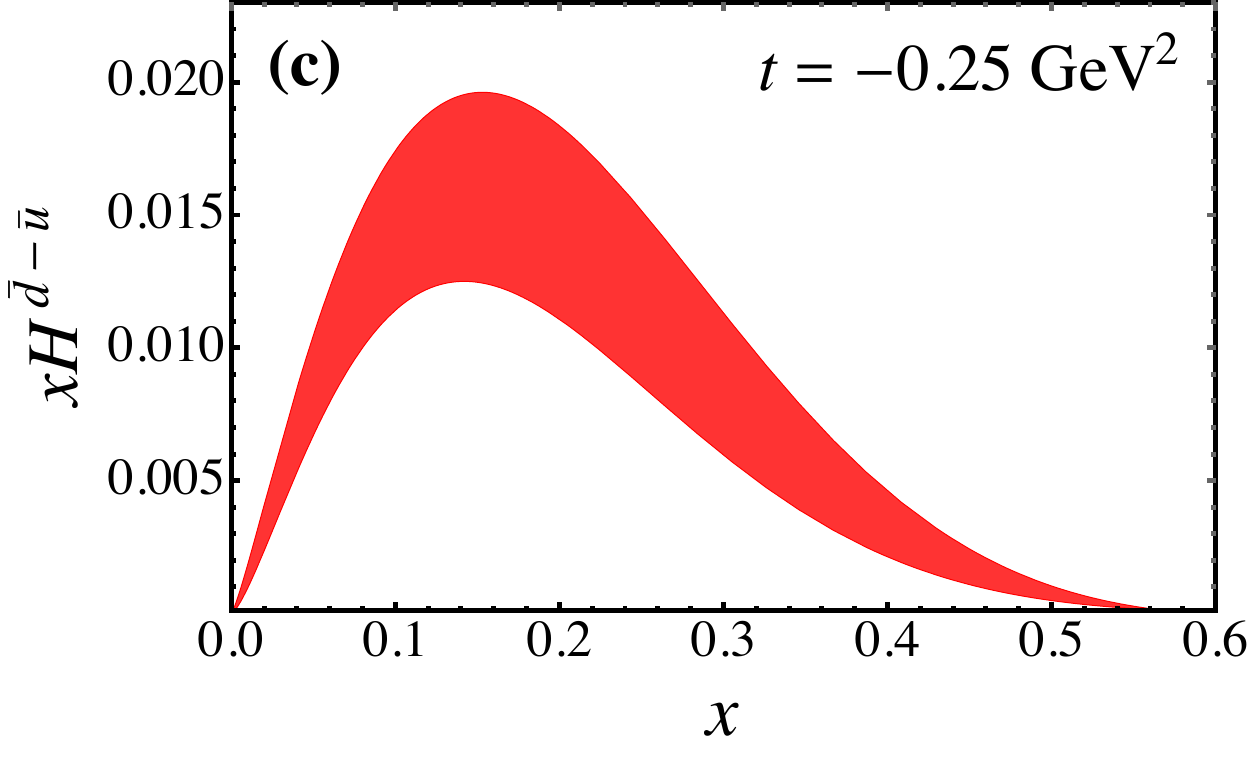}  
  \vspace{-12pt}
\end{minipage}
\hfill
\begin{minipage}[t]{.45\linewidth}   
\hspace*{-0.85cm} \includegraphics[width=1.1\textwidth, height=5.5cm]{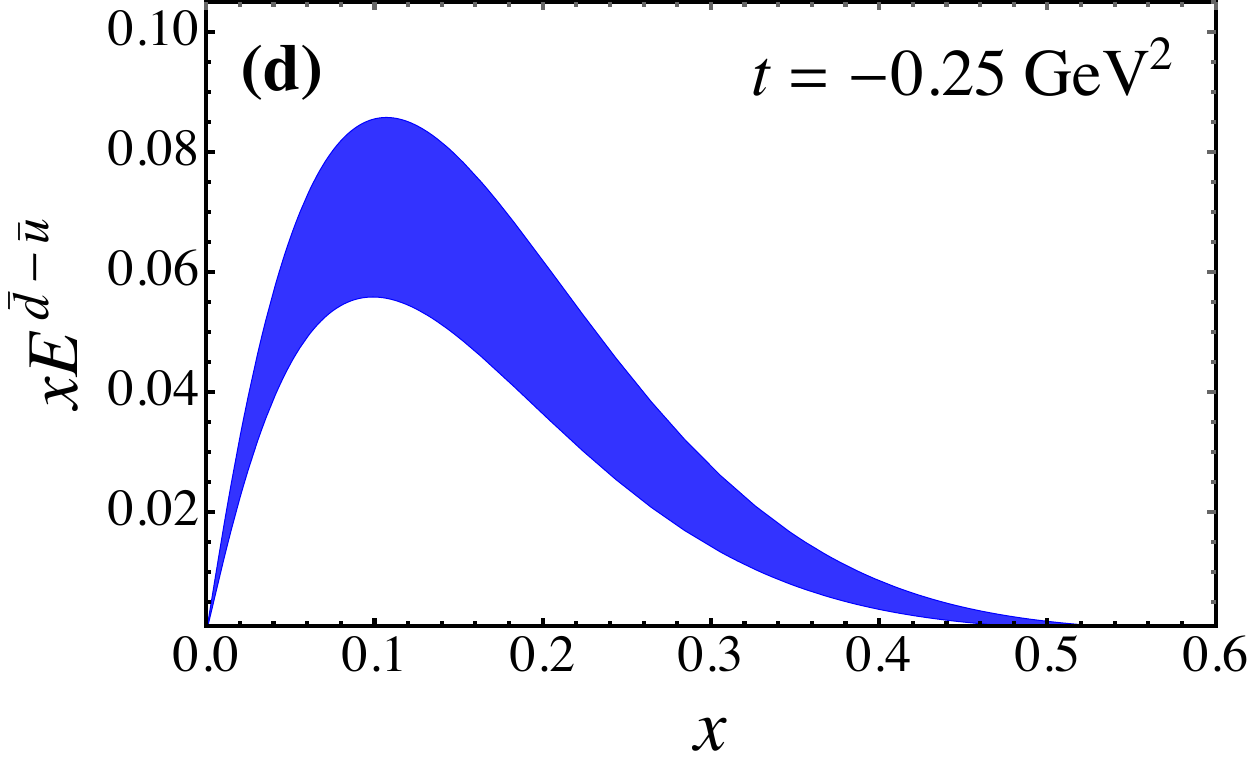} 
   \vspace{-12pt}
\end{minipage} 
\caption{Light antiquark asymmetries for the electric $xH^{\bar{u}-\bar{d}}$ (red bands) and magnetic $xE^{\bar{u}-\bar{d}}$ (blue bands) GPDs versus parton momentum fraction $x$ at four-momentum transfer squared of $t=0$ [{\bf (a)}, {\bf (b)}] and $t=-0.25$~GeV$^2$ [{\bf (c)}, {\bf (d)}], for cutoff parameter $\Lambda = 1.0(1)$~GeV. The asymmetries are shown at the scale $Q=1$~GeV, except for the electric asymmetry at $t=0$, which is compared with the $x(\bar d-\bar u)$ PDF asymmetry from the JAM global QCD analysis~\cite{Cocuzza:2021cbi} (yellow band) at the scale $Q=m_c$.}
\label{2du-dt0}
\end{adjustwidth}
\end{figure}

The $xH^{\bar{d}-\bar{u}}$ and $xE^{\bar{d}-\bar{u}}$ GPD asymmetries at finite $t$ are also shown in Fig.~\ref{2du-dt0}, for $-t = 0.25$~GeV$^2$. As expected from the 3-dimensional plots in Fig.~\ref{3du-d}, the distributions are suppressed at larger $-t$ values, with the magnitudes of the functions about half of those at $t=0$. This is consistent with the GPD inequality $H^q(x,t) \leq H^q(x,0)$~\cite{Pobylitsa:2002gw, Radyushkin:1998es}. The peaks in both functions also shift to slightly larger $x$ values with increasing four-momentum transfer squared.

\begin{figure}[] 
\begin{adjustwidth}{-\extralength}{0cm}
\begin{center}
    \begin{minipage}{0.45\linewidth}
        \centering
        \centerline{
        \includegraphics[width=1\textwidth]{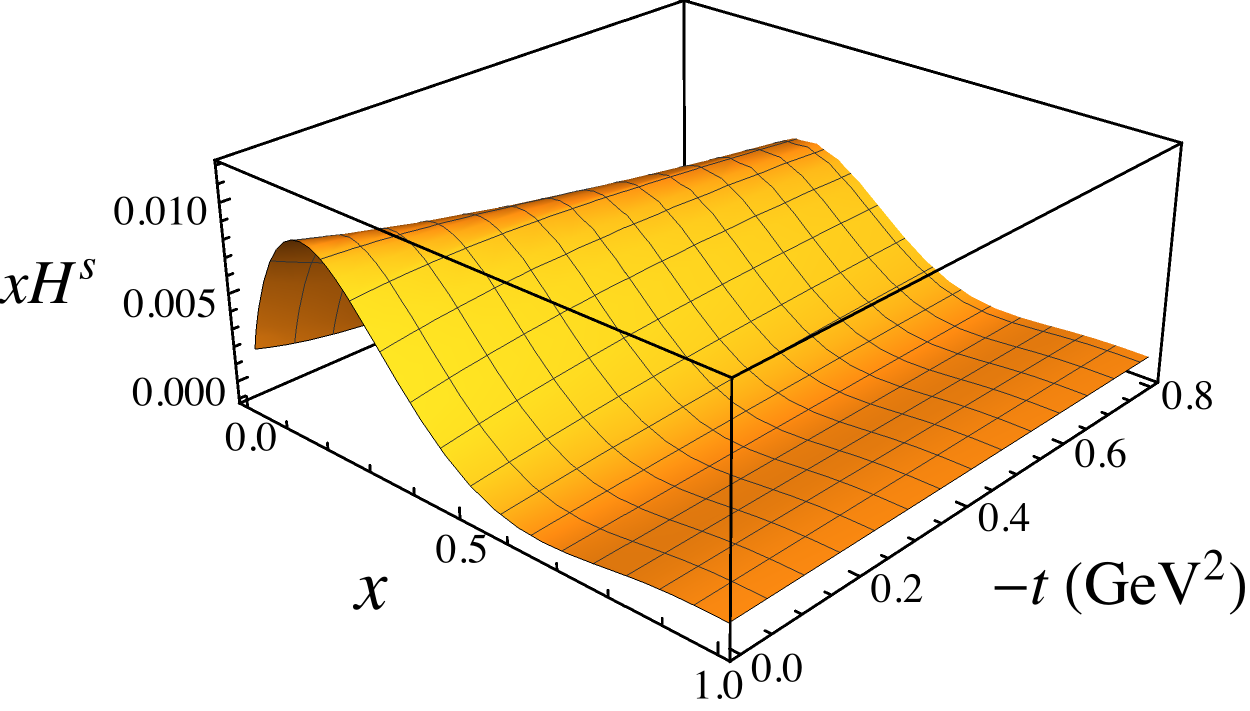}
        }
        \centerline{\small{\bf{(a)}}}
    \end{minipage}
    \begin{minipage}{0.45\linewidth}
        \centering
        \centerline{
        \includegraphics[width=1\textwidth]{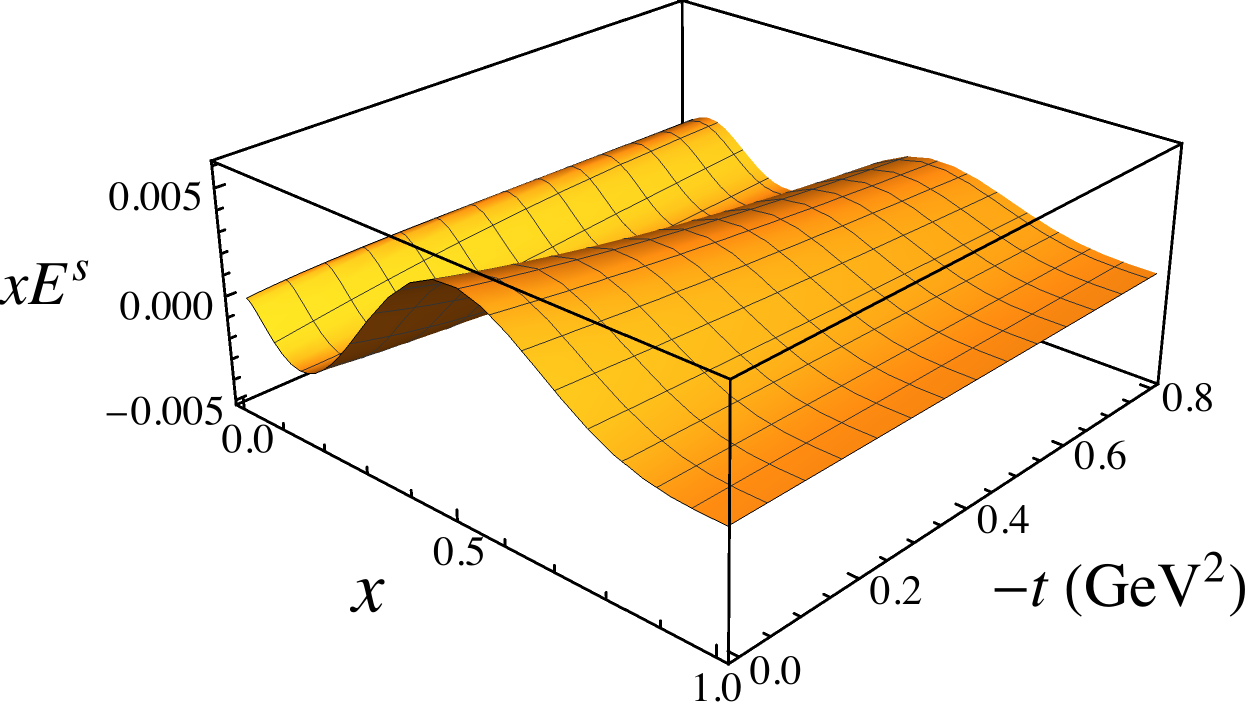}
        }
        \centerline{\small{\bf{(b)}}}
    \end{minipage}    
\\[0.4cm]
        \begin{minipage}{0.45\linewidth}
        \centering
        \centerline{
        \includegraphics[width=1\textwidth]{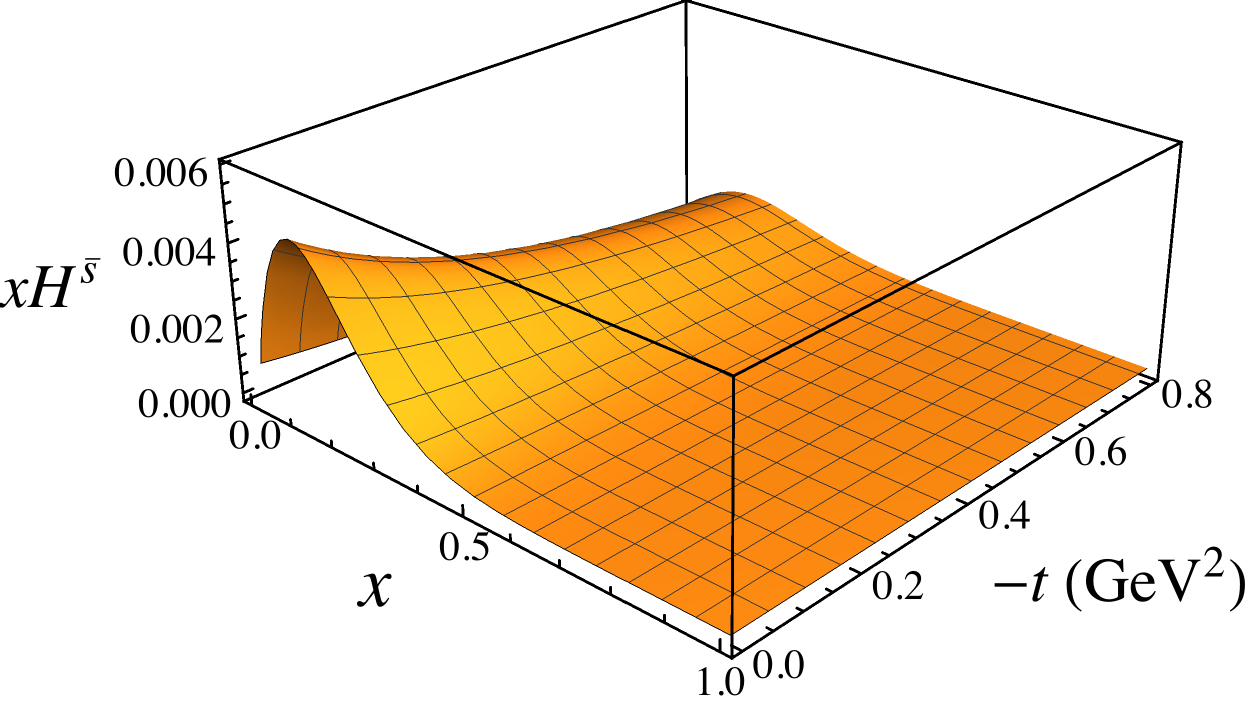}
        }
        \centerline{\small{\bf{(c)}}}
    \end{minipage}
        \begin{minipage}{0.45\linewidth}
        \centering
        \centerline{
        \includegraphics[width=1\textwidth]{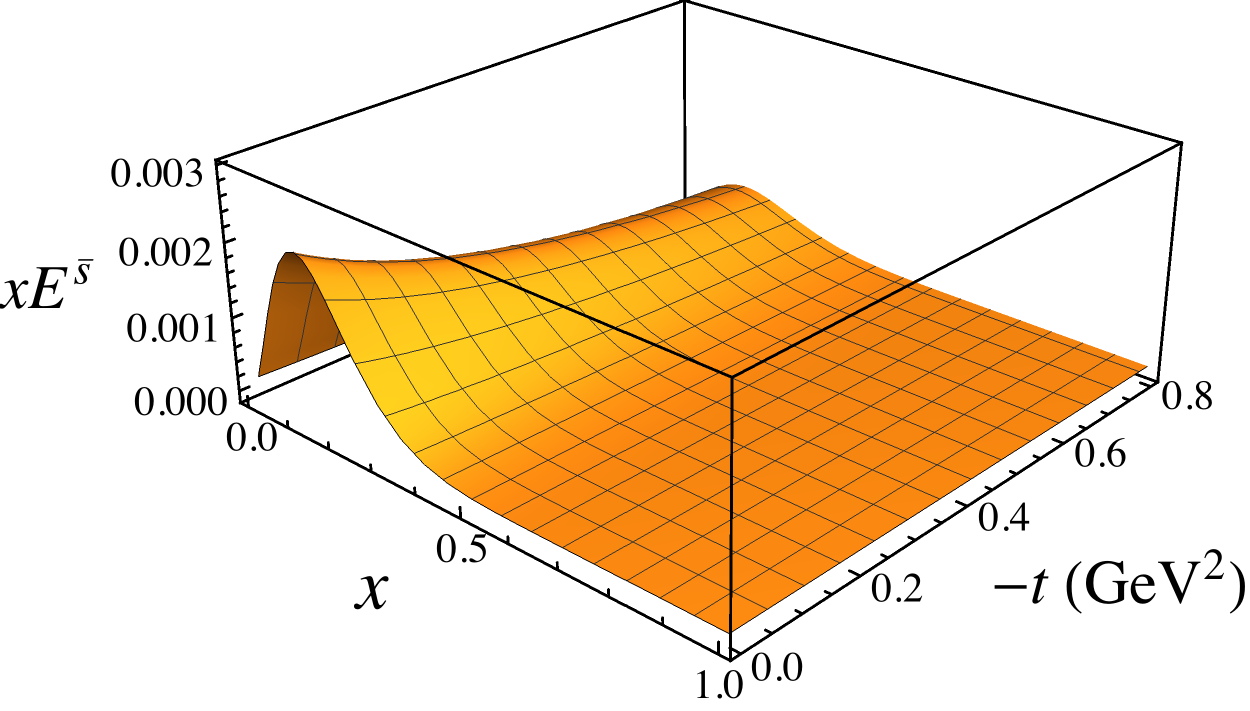}
        }
        \centerline{\small{\bf{(d)}}}
    \end{minipage}
\caption{Electric and magnetic GPDs for the strange and antistrange quarks: {\bf (a)}~$xH^s$, {\bf (b)}~$xE^s$, {\bf (c)}~$xH^{\bar{s}}$, and {\bf (d)}~$xE^{\bar{s}}$ versus the parton momentum fraction $x$ and four-momentum transfer squared $-t$, for $\Lambda=1$~GeV, at the scale $Q=1$~GeV.}
\label{3dssbar}
\end{center}
\end{adjustwidth}
\end{figure}

The kaon loop contributions to the strange quark GPDs are shown in Fig.~\ref{3dssbar}. Compared with the GPDs for the light antiquarks, the strange GPDs are smaller in magnitude, but display some interesting features. As for the light antiquark GPDs, the signs of the electric GPDs $H^s$ and $H^{\bar{s}}$ are both positive. While the shapes of the $s$ and $\bar{s}$ distributions are expected to be almost identical perturbatively~\cite{Catani:2004nc}, the kaon loop contributions to these can be quite different due to their different origins. Assuming the SU(3) symmetric relations for the GPDs in the hadronic intermediate states \cite{He:2022leb,Salamu:2019dok}, the $\bar s$ antiquark GPD arises from diagrams with a direct coupling to the kaon, as in Fig.~\ref{diagrams}(a), while contributions to the $s$ quark GPD come from couplings to the intermediate state hyperons, such as in~Fig.~\ref{diagrams}(b)~\cite{Signal:1987gz, Melnitchouk:1996fj}.

As evident from Fig.~\ref{3dssbar}, at small values of $x$ the strange $H^s$ GPD is larger than the antistrange $H^{\bar{s}}$, while for larger $x$ values, $x \gtrsim 0.5$, the antistrange contribution exceeds the strange. The $x$ integrals of $H^s$ and $H^{\bar{s}}$ at zero momentum transfer, on the other hand, can be shown to be identical with the inclusion of the $\delta$-function term, as is necessary for the requirement of zero net strangeness in the nucleon. Since the $t$ dependence of $H^s$ is different from that of $H^{\bar{s}}$, at finite values of $t$ the lowest moments of the strange and antistrange GPDs need not be the same, which corresponds to nonzero values of the strange electric form factor at $-t > 0$. The behaviors of the magnetic GPDs $E^s$ and $E^{\bar{s}}$ are, on the other hand, rather different. While the sign of $E^{\bar{s}}$ is the same as that of $E^{\bar{d}}$ because of the positive orbital angular momentum of the meson, the strange GPD $E^s$ changes sign with $x$, from negative at small $x$ values to positive at $x \gtrsim 0.3$.

\begin{figure}[tbph] 
\begin{adjustwidth}{-\extralength}{0cm}
\centering 
    \begin{minipage}{0.45\linewidth}
        \centering
        \centerline{
        \includegraphics[width=1\textwidth]{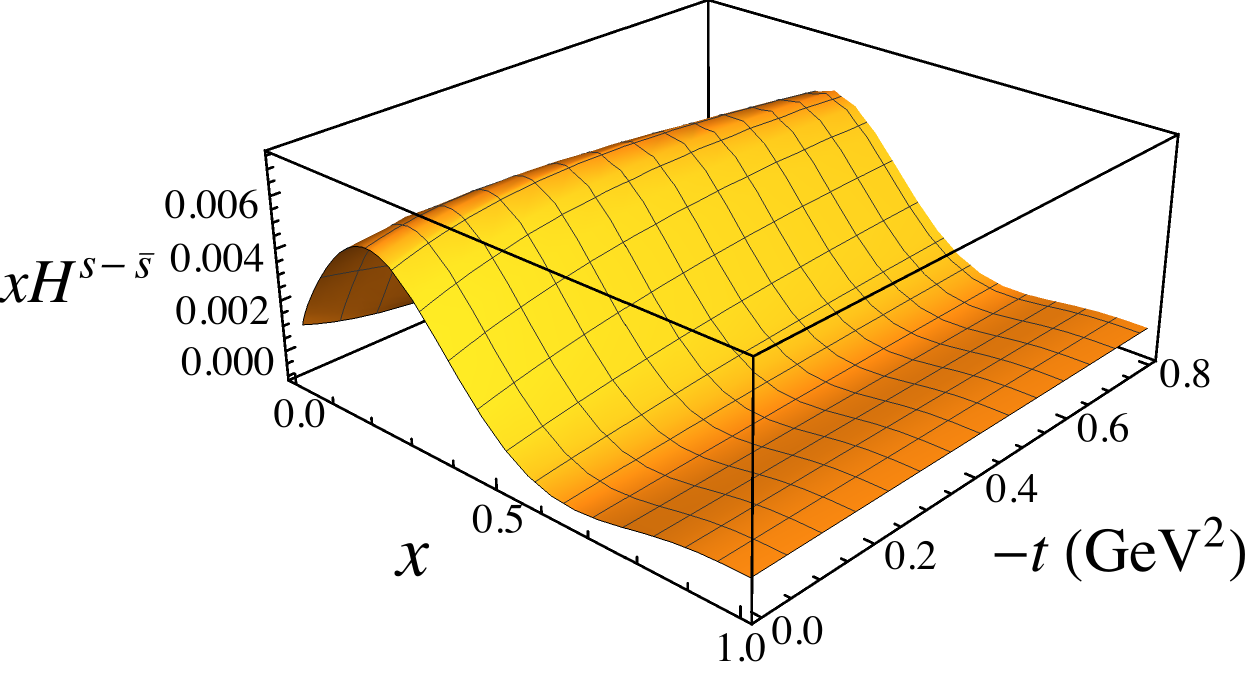}
        }
        \centerline{\small{\bf{(a)}}}
    \end{minipage}
    \begin{minipage}{0.45\linewidth}
        \centering
        \centerline{
        \includegraphics[width=1\textwidth]{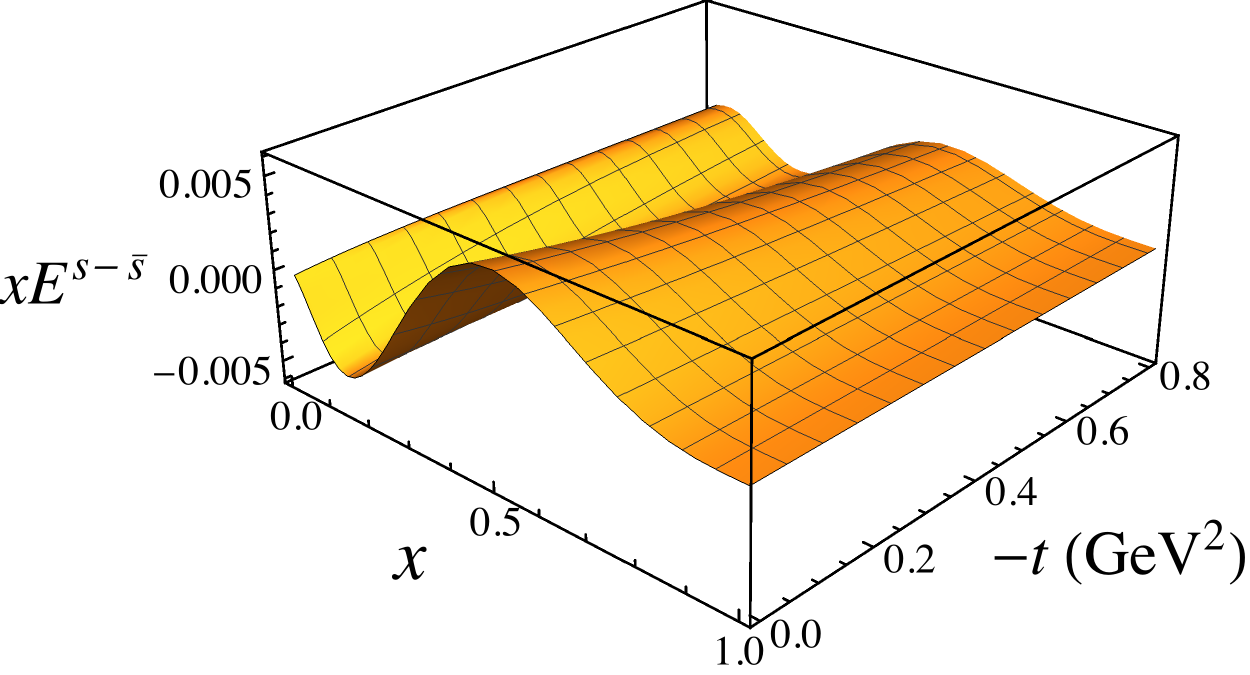}
        }
        \centerline{\small{\bf{(b)}}}
    \end{minipage}
\caption{The strange quark asymmetry for the {\bf (a)} electric $xH^{s-\bar{s}}$ and {\bf (b)} magnetic $xE^{s-\bar{s}}$ GPDs versus momentum fraction $x$ and four-momentum transfer squared $-t$, for $\Lambda = 1$~GeV.}
\label{3ds-sbar}  
\end{adjustwidth}
\end{figure}

In Fig.~\ref{3ds-sbar} we show the strange--antistrange asymmetries $xH^{s-\bar{s}}$ and $xE^{s-\bar{s}}$ versus $x$ and $-t$, for a fixed value of $\Lambda = 1$~GeV. At nonzero values of $x$, the $xH^s$ GPD is generally larger than $xH^{\bar{s}}$, with a maximal asymmetry at $x \approx 0.2-0.3$. Unlike the individual $s$ and $\bar s$ contributions, for a given value of $x$ the asymmetry $xH^{s-\bar{s}}$ does not decrease monotonously with $-t$, and in fact increases at higher $-t$ in some cases. For the magnetic asymmetry $xE^{s-\bar{s}}$, the change of sign with $x$ is driven by the behavior of the strange contribution, $xE^s$. Generally, the $s-\bar{s}$ asymmetry is much smaller than the $\bar{d}-\bar{u}$ asymmetry in the nucleon for both the electric and magnetic GPDs.

In analogy with the $\bar{d}-\bar{u}$ asymmetry in Fig.~\ref{2du-dt0}, in Fig.~\ref{2ds-sbart0} we show the $xH^{s-\bar{s}}$ and $xE^{s-\bar{s}}$ asymmetries at $t=0$ and $-t=0.25$~GeV$^2$ for varying cutoff parameters between $\Lambda=0.9$~GeV and 1.1~GeV. The change in sign of $xH^{s-\bar{s}}$ is evident, with the asymmetry being positive at small $x$, before turning negative at $x \gtrsim 0.5$. The calculated asymmetry is compared with recent PDF parametrizations of $x(s-\bar s)$ from the JAM~\cite{Cocuzza:2021cbi} and NNPDF~\cite{Faura:2020oom} global QCD analyses, which show very large uncertainties relative to the magnitude of the computed result. For the lowest nonzero moment, one finds 
    $\int_0^1 dx\, xH^{s-\bar{s}}(x,0) = 0.0009^{(5)}_{(4)}$
for $\Lambda=1.0(1)$~GeV, which is comparable to other recent estimates of the strange asymmetry~\cite{Bentz:2009yy, Wang:2016ndh, Salamu:2019dok}. For the magnetic asymmetry $xE^{s-\bar{s}}$, the situation is reversed, with the asymmetry trending negative at small $x$ and becoming positive at larger $x$ values, $x \gtrsim 0.3$. For comparison, the analogous integrated magnetic GPD asymmetry is
    $\int_0^1 dx\, xE^{s-\bar{s}}(x,0) = 0.0009^{(12)}_{(8)}$
for the $x$-weighted moment. For the lowest moment of the magnetic asymmetry $E^{s-\bar{s}}$, which corresponds to the strange quark contribution to the proton's magnetic moment, $\mu_s$, one finds
    $\int_0^1 dx\, E^{s-\bar{s}}(x,0) = \mu_s = -0.033^{(11)}_{(13)}$.

\begin{figure}[t] 
\begin{adjustwidth}{-\extralength}{0cm}
\begin{minipage}[b]{.45\linewidth}
\hspace*{-0.55cm}\includegraphics[width=1.125\textwidth, height=5.5cm]{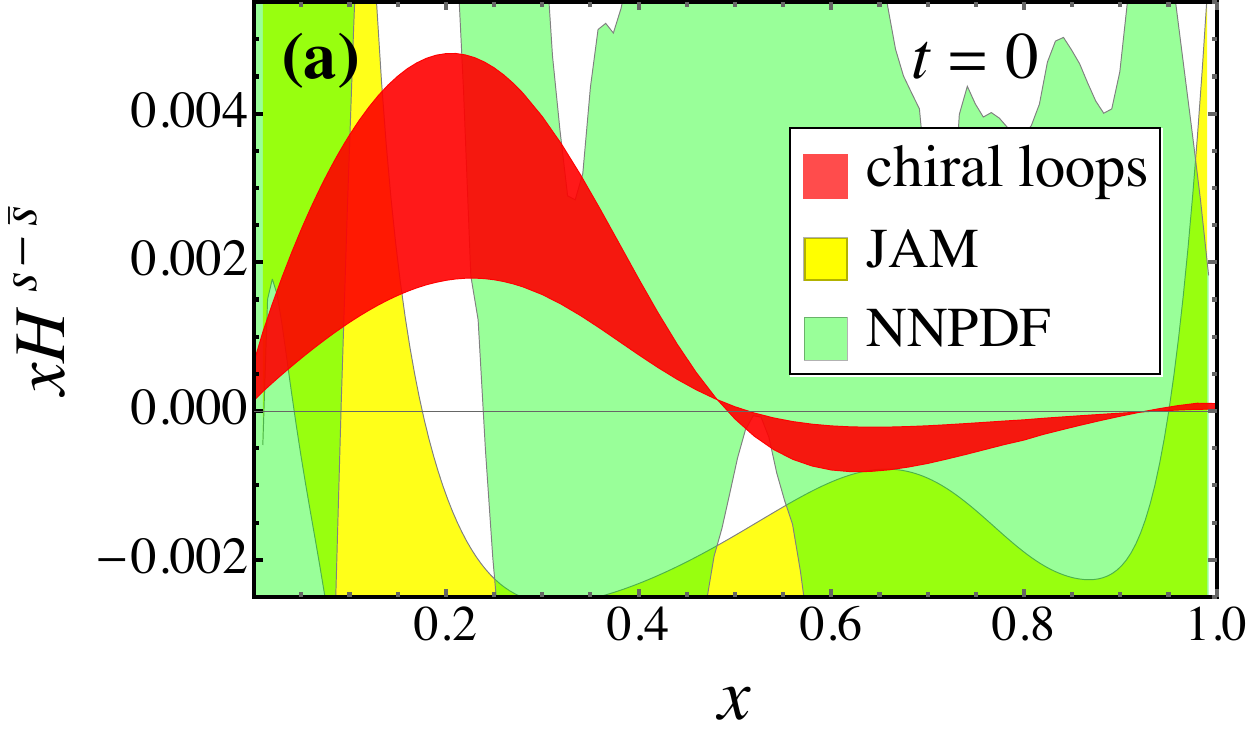} 
 \vspace{0pt}
\end{minipage}
\hfill
\begin{minipage}[b]{.45\linewidth}   
\hspace*{-1.1cm} \includegraphics[width=1.1\textwidth, height=5.5cm]{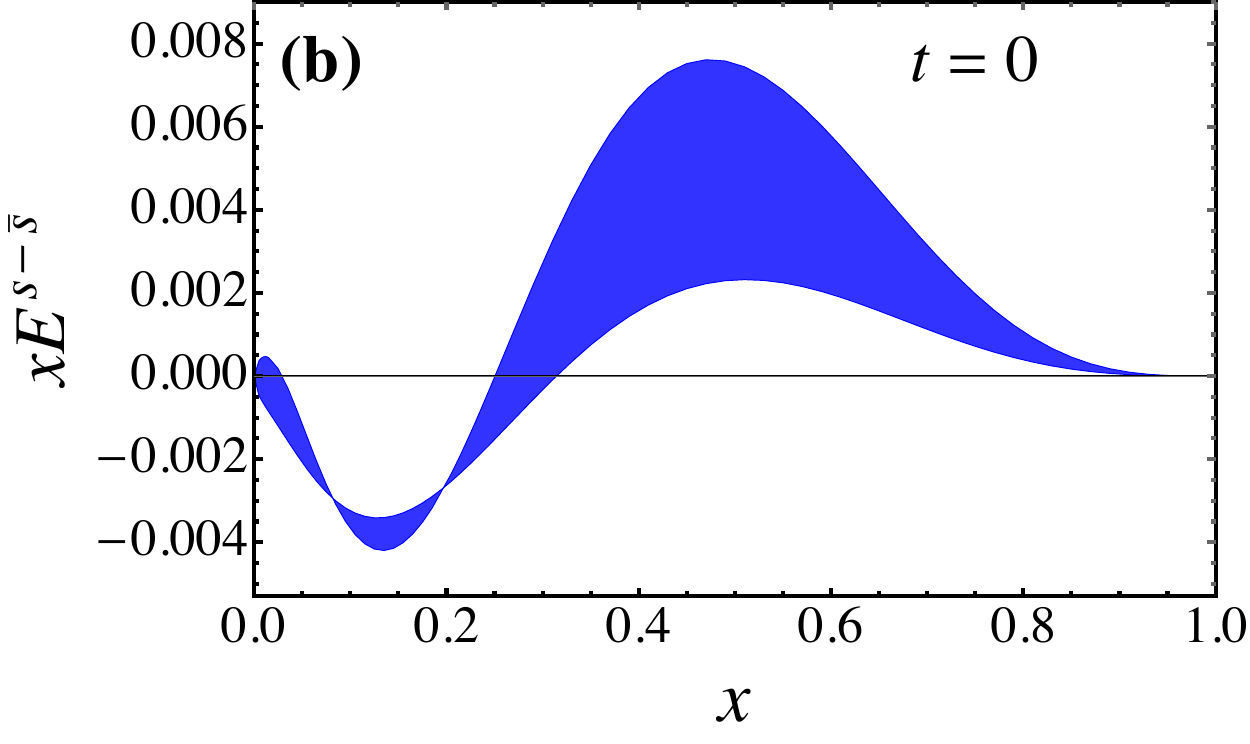} 
   \vspace{14pt}
\end{minipage}  
\\[-0.4cm]
\begin{minipage}[t]{.45\linewidth}
\hspace*{-0.3cm}\includegraphics[width=1.1\textwidth, height=5.5cm]{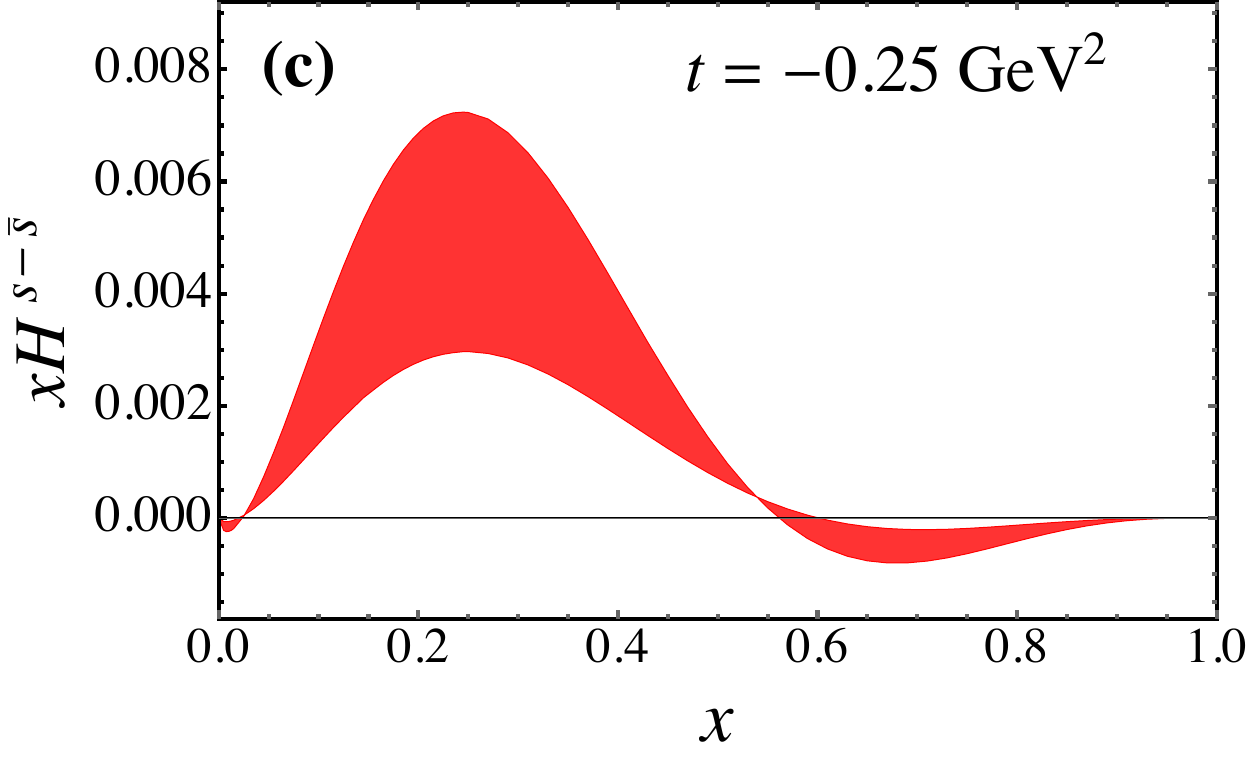}   
 \vspace{-12pt}
\end{minipage}
\hfill
\begin{minipage}[t]{.45\linewidth}   
\hspace*{-1.15cm} \includegraphics[width=1.1\textwidth, height=5.5cm]{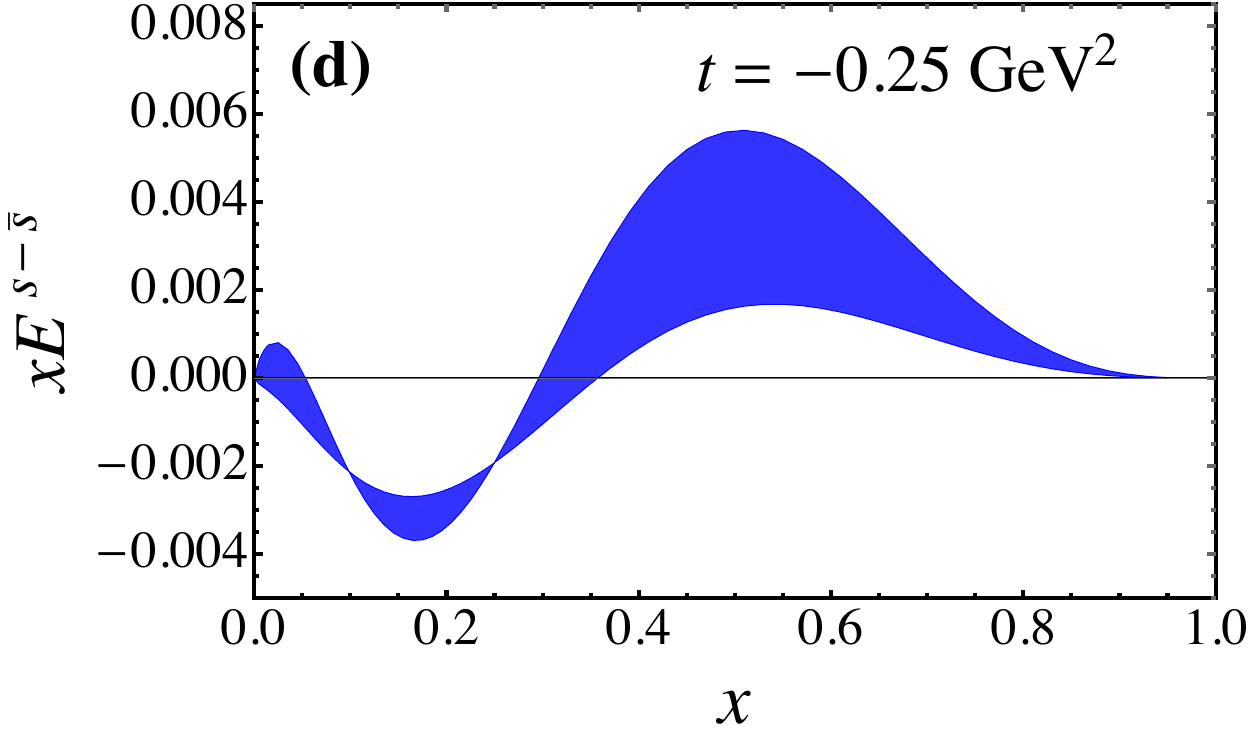}   \vspace{-12pt}
\end{minipage} 
\caption{Strange quark asymmetry for the $xH^{s-\bar{s}}$ (red bands) and $xE^{s-\bar{s}}$ (blue bands) GPDs versus $x$ at squared momentum transfers $t=0$ [{\bf (a)}, {\bf (b)}] and $-t = 0.25$~GeV$^2$ [{\bf (c)}, {\bf (d)}], with the bands corresponding to cutoff mass $\Lambda=1.0(1)$~GeV. The asymmetries are shown at $Q=1$~GeV, except for the strange electric asymmetry at $t=0$, which is compared with PDF parametrizations of $x(s-\bar s)$ from JAM~\cite{Cocuzza:2021cbi} (yellow band) and NNPDF~\cite{Faura:2020oom} (green band) evolved to $Q=m_c$.}
\label{2ds-sbart0}
\end{adjustwidth}
\end{figure}

At nonzero values of $t$, the strange asymmetry is not as strongly suppressed as the nonstrange $\bar d-\bar u$ asymmetry. At $-t=0.25$~GeV$^2$, for instance, as also shown in Fig.~\ref{2ds-sbart0}, the magnetic GPD asymmetry $xE^{s-\bar{s}}(x,t)$ is only slightly smaller in magnitude than that at $t=0$. For the electric GPD asymmetry the peak value of the magnitude of $xH^{s-\bar{s}}(x,t)$ at $-t=0.25$~GeV$^2$ is even larger than that at $t=0$. \\

\subsection{GPDs with nonzero skewness}

In this section we extend the discussion of zero-skewness GPDs in the proton~\cite{He:2022leb} to the nonzero skewness case. As a first step towards a complete calculation to one-loop order, this analysis considers the contributions to the sea quark and antiquark GPDs in the proton from the virtual pseudoscalar meson cloud dressing of the bare baryon. Typically, in calculations of meson loop contributions to sea quark and antiquark asymmetries, assuming that the undressed proton has a flavor symmetric sea~\cite{Salamu01}, the meson coupling diagrams in Fig.~\ref{diagrams}(a), \ref{diagrams}(k), \ref{diagrams}(l) and \ref{diagrams}(m) are taken to be the dominant source of differences between sea quark and antiquark PDFs and GPDs.

For the rainbow diagrams in Figs.~\ref{diagrams}(a) and \ref{diagrams}(m), the relevant splitting functions are nonzero when $y$ is in the region $-\xi \leq y \leq 1$, for positive $\xi$ values. The splitting function in this region can be convoluted with the pion GPD to obtain the quark GPD in the physical nucleon. Figure~\ref{fig:convol_illu} illustrates the decomposition of the convolution formula for the rainbow diagram in Fig.~\ref{diagrams}(a) into different subprocesses. The splitting functions for Figs.~\ref{fig:convol_illu}(a), \ref{fig:convol_illu}(b) and \ref{fig:convol_illu}(d) are all located in the $y>\xi$ region, while the quark GPD in the pion corresponds to the quark DGLAP, ERBL and antiquark DGLAP regions, respectively. After convoluting with the hadronic splitting function, these contribute to the quark GPD of the proton in the quark DGLAP, ERBL, and antiquark DGLAP regions, respectively. For the subprocess in Fig.~\ref{fig:convol_illu}(c) the splitting function is located at $y\in [-\xi,\xi]$, and can be considered as the meson--meson pair annihilation process, analogous to the distribution amplitude-like dynamics at the quark level.

\begin{figure}[t] 
\begin{adjustwidth}{-\extralength}{0cm}
\begin{center}
\includegraphics[scale=1]{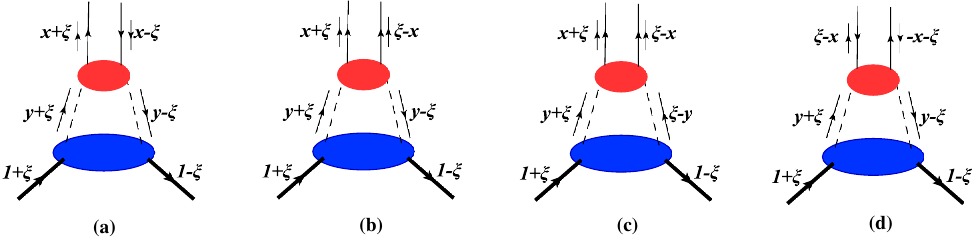}
\caption{Representation of the convolution formula in Eqs.~(\ref{eq:conv_a}), (\ref{eq:conv_b}), (\ref{eq:conv_c}), and (\ref{eq:conv_d}), respectively, with the \{dashed, thick solid, thin solid\} lines representing the \{pseudoscalar meson, proton, quark\}. The processes in diagrams {\bf (a)} and {\bf (d)} represent the DGLAP region for the quark and antiquark, respectively, while the processes in {\bf (b)} and {\bf (c)} contribute to the ERBL region.}
\label{fig:convol_illu}
\end{center}
\end{adjustwidth}
\end{figure}

By combining these various processes, the contribution to the skewness nonzero GPDs from the diagram in Fig.~\ref{diagrams}(a) can be expressed in the convolution form as \\
\begin{small}
\begin{subequations}
\label{eq:convolution_sum}
\begin{empheq}[left={\hspace*{-0.4cm}
H_q^{\rm (rbw)}(x,\xi,t)=\empheqlbrace}]{align}
\label{eq:conv_a}
    &\int_x^1 \frac{dy}{y} f_{\phi B}^{({\rm rbw})}(y,\xi,t)\, H_{q/\phi}\Big(\frac{x}{y},\frac{\xi}{y},t\Big), \hspace*{4.6cm}    
    [\xi < x < y]
\\ \nonumber \\ 
\label{eq:conv_b}
    &\int_\xi^1 \frac{dy}{y} f_{\phi B}^{({\rm rbw})}(y,\xi,t)\, H_{q/\phi}\Big(\frac{x}{y},\frac{\xi}{y},t\Big), \hspace*{4.6cm}    
    [x < \xi < y]
\\ \nonumber \\
\label{eq:conv_c}
	&\int_{-\xi}^{\xi} \frac{dy}{2y}\, f_{\phi B}^{({\rm rbw})}(y,\xi,t)\frac{1}{\pi} \int_{s_0}^\infty\!\! ds
    \frac{{\rm Im}\Phi_{q/\phi}{ \big( \frac12(1\!+\!\frac{x}{\xi}), 
                                       \frac12(1\!+\!\frac{y}{\xi}),
                                      s
                                 \big)}}{s-t+i\epsilon},  \hspace*{1.2cm}    
    [|x|,|y| < \xi]\!\!
\\ \nonumber \\
\label{eq:conv_d}
    &\int_{-x}^1 \frac{dy}{y} f_{\phi B}^{({\rm rbw})}(y,\xi,t)\, H_{q/\phi}\Big(\frac{x}{y},\frac{\xi}{y},t\Big), \hspace*{4.5cm}    
    [\xi < -x < y < 1]
\\ \nonumber 
\end{empheq}
\end{subequations}
\end{small}%
where $H_{q/\phi}$ and $\Phi_{q/\phi}$ represent the valence GPD and generalized distribution amplitude (GDA), respectively, in the intermediate pseudoscalar meson.
The integration variable $s = (2k_\phi +\Delta)^2$ in Eq.~(\ref{eq:conv_c}) represents the four-momentum squared of the produced meson pair with momentum $k_\phi$ and $k_\phi+\Delta$, with threshold value $s_0 = (2 m_\phi)^2$.

Convoluting the splitting functions with the corresponding quark GPDs of the virtual pions, in Fig.~\ref{fig:GPD_3d} the contributions of the chiral loop to the $H$ and $E$ GPDs for $u$ quarks are shown as a function of $x$ and $t$ at a fixed $\xi=0.1$. For the contributions from the couplings to the virtual pion loops, as in Figs.~\ref{diagrams}(a), \ref{diagrams}(k), \ref{diagrams}(l) and \ref{diagrams}(m), the results for the $d$-quark GPDs can be obtained from the $u$-quark distributions using isospin symmetry, $\{H,E\}_d(x,\xi,t)=-\{H,E\}_u(-x,\xi,t)$. The most striking structure is seen in the region of low $x$, $|x| \lesssim 0.2$. For the case of the electric $u$-quark GPD, $x H_u$ is positive in the DGLAP region, and has two valleys in the ERBL region. For the magnetic GPD, $x E_u$ is negative in the $x < 0$ region and positive when $x > 0$. The distributions fall rapidly as $|x| \to 1$, and for increasing values of the momentum transfer squared, $-t$.

\begin{figure}[t] 
\begin{adjustwidth}{-\extralength}{0cm}
\begin{minipage}[b]{.45\linewidth}
\hspace*{-0.3cm}\includegraphics[width=1.1\textwidth, height=5.5cm]{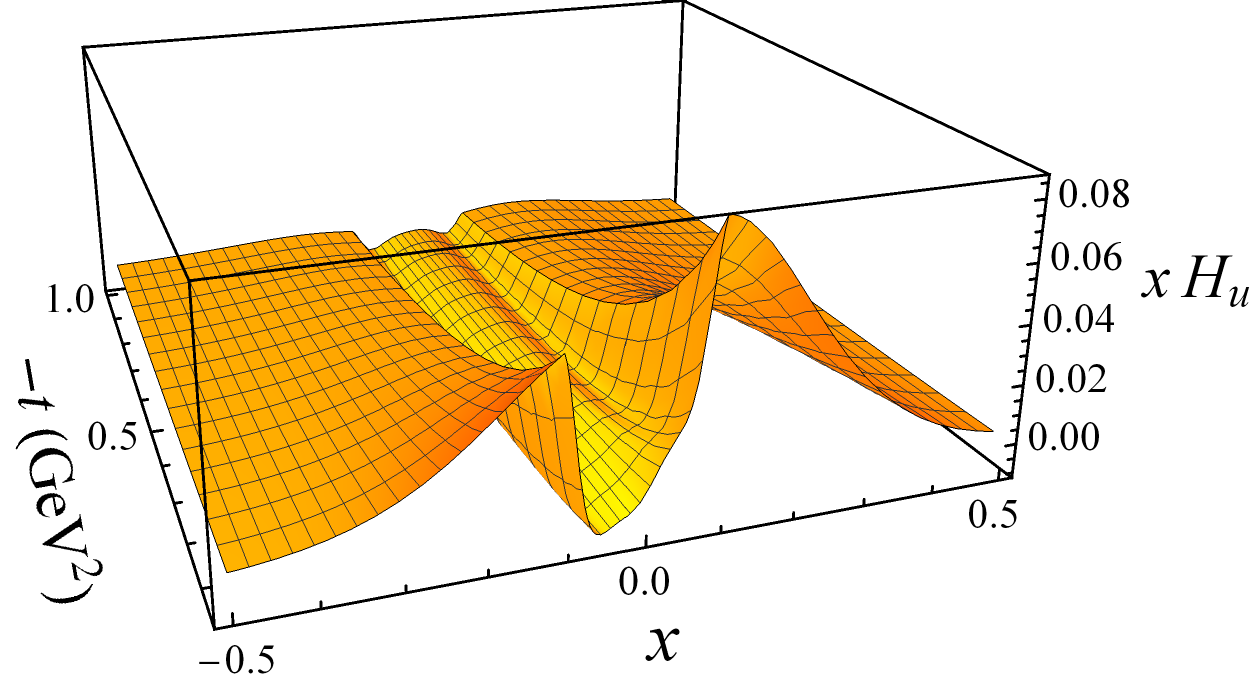}  \vspace{-6pt}
\end{minipage}
\hfill
\begin{minipage}[b]{.45\linewidth}   
\hspace*{-0.95cm} \includegraphics[width=1.1\textwidth, height=5.5cm]{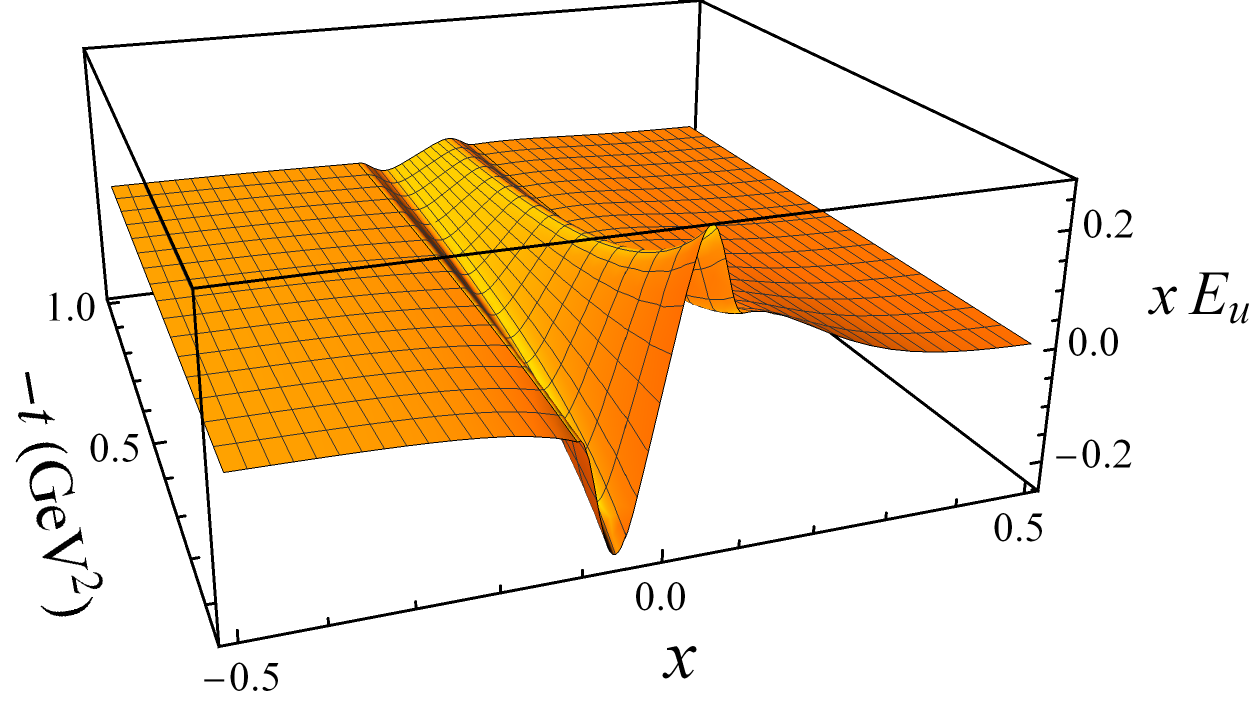} 
\vspace{-6pt}
\end{minipage}  
\\[-0.4cm]
\caption{Total three-dimensional $u$-quark GPDs $xH_u$ and $xE_u$ as functions of $x$ and $t$, for fixed $\xi=0.1$. The corresponding $d$-quark distributions can be obtained from the $u$-quark GPDs using the isospin symmetry relation, $\{H,E\}_d(x,\xi,t) = -\{H,E\}_u(-x,\xi,t)$, which holds for the contributions from virtual pion loops as diagrams in Figs.~\ref{diagrams}(a), \ref{diagrams}(k), \ref{diagrams}(l) and \ref{diagrams}(m).}
\label{fig:GPD_3d}
\end{adjustwidth}
\end{figure}

\begin{figure}[t] 
\begin{adjustwidth}{-\extralength}{0cm}
\begin{minipage}[b]{.45\linewidth}
\hspace*{-0.3cm}\includegraphics[width=1.1\textwidth, height=5.5cm]{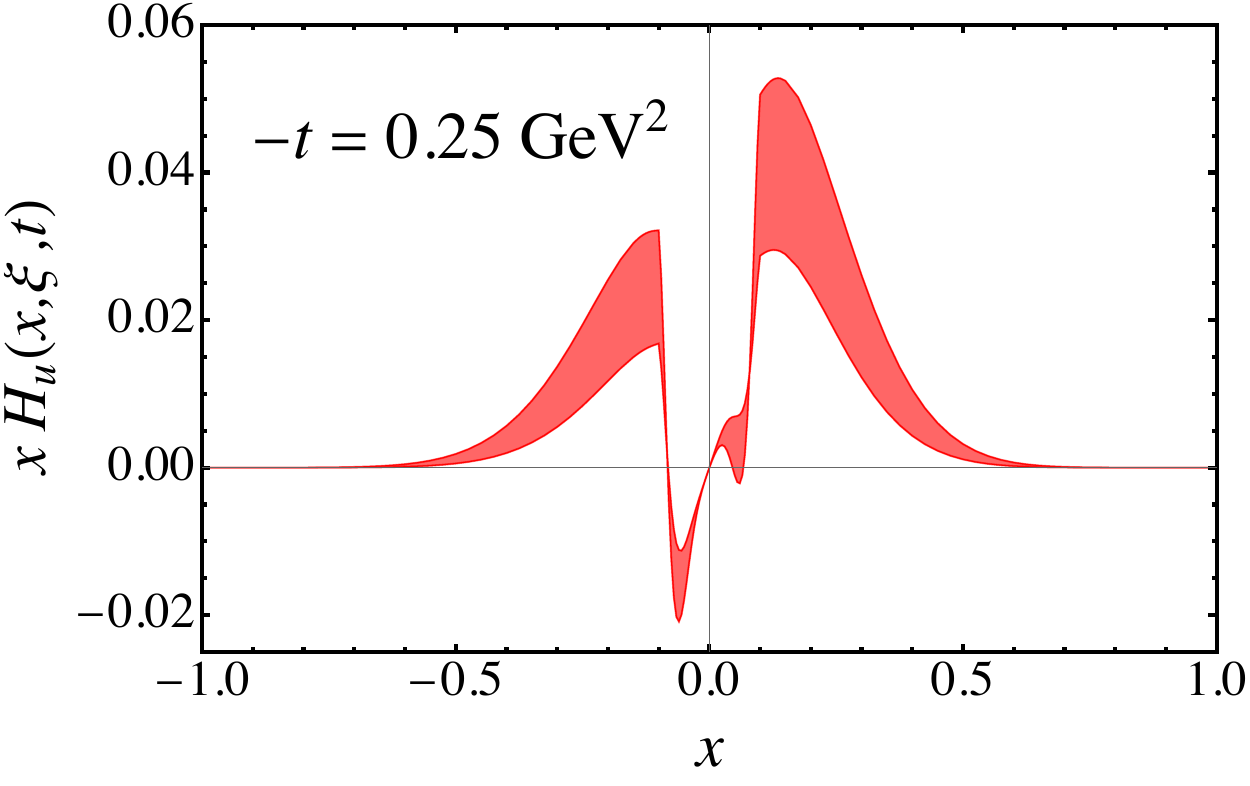}   
 \vspace{12pt}
\end{minipage}
\hfill
\begin{minipage}[b]{.45\linewidth}   
\hspace*{-0.7cm} \includegraphics[width=1.08\textwidth, height=5.5cm]{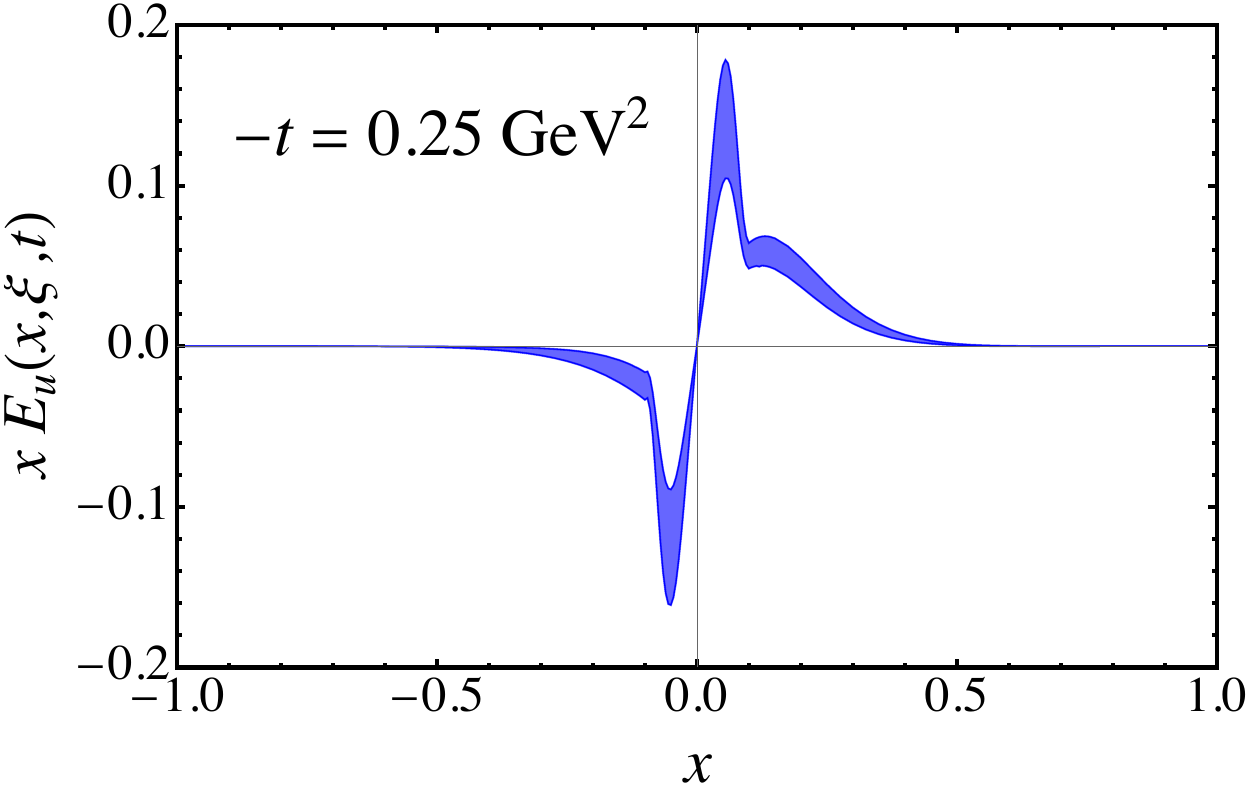} 
   \vspace{12pt}
\end{minipage}  
\\[-0.4cm]
\begin{minipage}[t]{.45\linewidth}
\hspace*{-0.5cm}\includegraphics[width=1.12\textwidth, height=5.5cm]{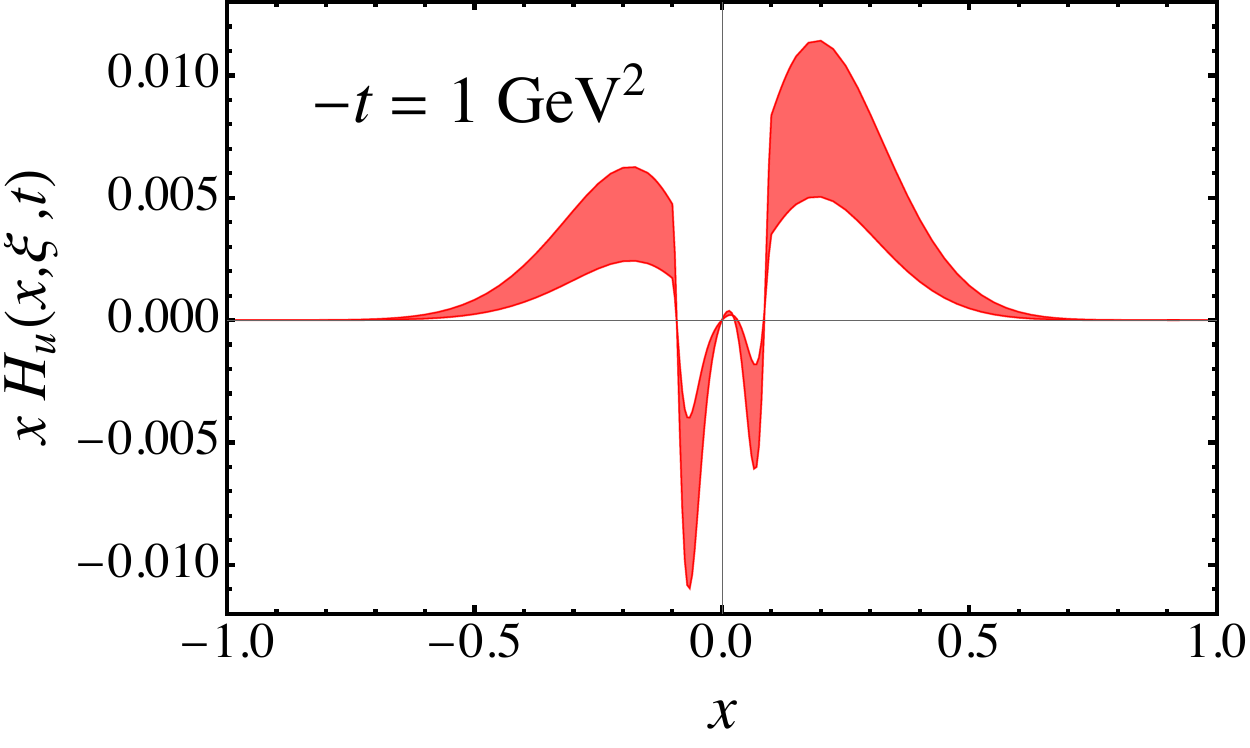} 
 \vspace{-12pt}
\end{minipage}
\hfill
\begin{minipage}[t]{.45\linewidth}   
\hspace*{-0.85cm} \includegraphics[width=1.1\textwidth, height=5.5cm]{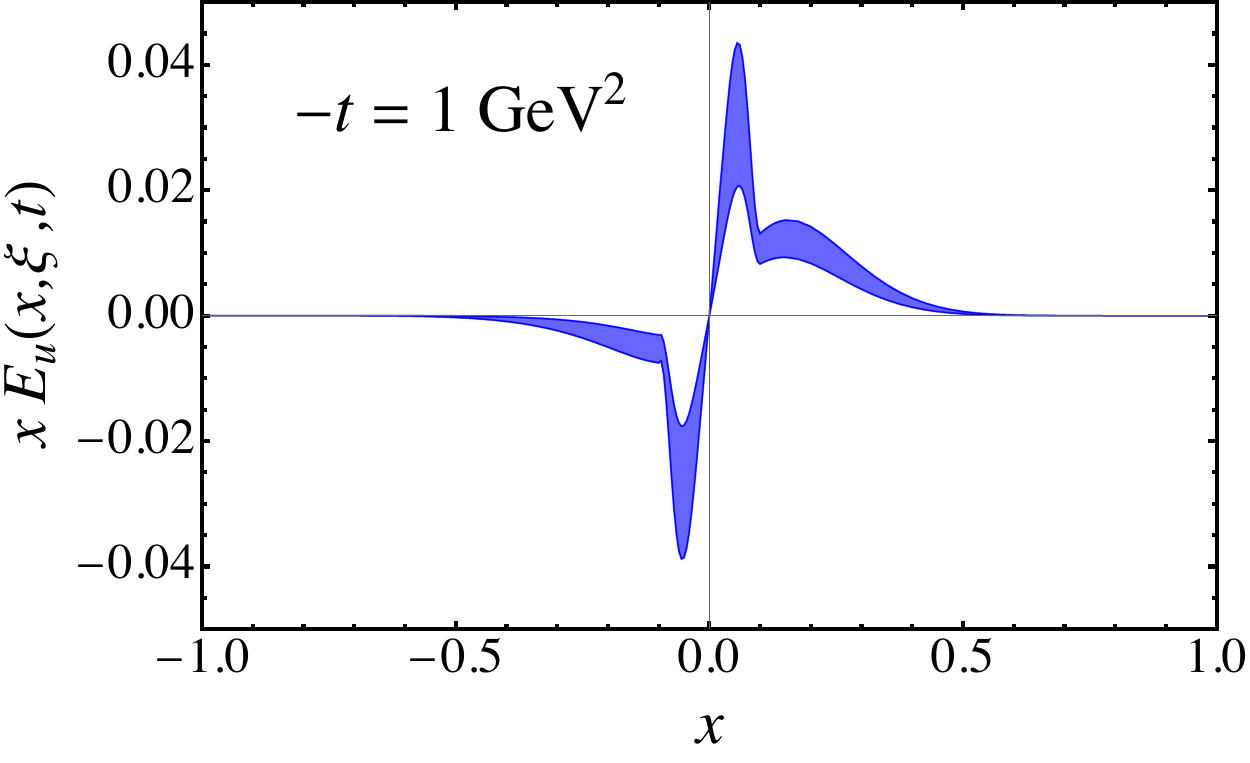}  
   \vspace{-12pt}
\end{minipage} 
\caption{Meson loop contributions to the electric and magnetic GPDs $H_u$ and $E_u$ for $\xi=0.1$ and $-t=1$~GeV$^2$. The corresponding $d$-quark GPDs can be obtained from the relation $\{H,E\}_{d}(x,\xi,t)=-\{H,E\}_{u}(-x,\xi,t)$.}
\label{fig:GPD_2d_1GeV}
\end{adjustwidth}
\end{figure}

To more clearly illustrate the dependence of the GPDs on the parton momentum fraction~$x$, in Fig.~\ref{fig:GPD_2d_1GeV} we show the two-dimensional projections of the electric and magnetic GPDs $x H_u$ and $x E_u$ for $-t=0.25$~GeV$^2$ and $-t=1$~GeV$^2$. Although the splitting function $f_{\pi^+\Delta^0}^{\rm (rbw)}$ is discontinuous at $y=\xi$, the quark GPDs remain continuous at the ridge $x=\pm\xi$. The reason is that the convolution formulas in Eqs.~(\ref{eq:conv_a}), (\ref{eq:conv_b}) and (\ref{eq:conv_d}) are only related to the splitting function in the $y>\xi$ region, and the results obtained using these should be continuous at $\xi$ if the input pion GPD is continuous at $x=\xi$, which guarantees that the amplitudes for DVCS and hard exclusive meson production are finite~\cite{Diehl:2003ny}. However, there are no constraints on the derivatives of GPDs at the point $x=\xi$, and the discontinuity of the derivative can arise from the different integration regions in the $\alpha$--$\beta$ plane for the DGLAP and ERBL regions when using the double distribution parametrization~\cite{Musatov:1999xp, Diehl:2003ny}. As the contribution of the $D$ term in the parametrization only exists in the ERBL region, it also leads to a discontinuity of the first derivative of the GPD at $x=\xi$. On the other hand, the contribution to the quark GPD from Eq.~(\ref{eq:conv_c}) vanishes $x=\xi$ because of the endpoint property of the distribution amplitude, $\Phi_{q/\pi}(1,\kappa,s)=0$, where $\kappa$ is the light-cone fraction $\kappa=p'^+/(p^++p'^+)$.

Comparing the results with different $-t$, the absolute values of the GPDs at $-t = 0.25$~GeV$^2$ are some 4--5 times larger than those at $-t = 1$~GeV$^2$. For the $u$-quark distribution, one finds that the quark GPD $H_u$ in the DGLAP region for $x>0$ has a larger magnitude than at $x<0$ (which by crossing symmetry is equivalent to the $\bar u$ distribution at $x>0$). According to the above isospin relation for the contributions from the pion coupling diagrams, the $u$-quark GPD $xH_u(x>0)$ is identical to the $d$-quark GPD $xH_d(x<0)$, the latter which is equivalent to the $x\bar d$ distribution at $x>0$. The contributions from the loop diagrams Figs.~\ref{diagrams}(a), \ref{diagrams}(k), \ref{diagrams}(l) and \ref{diagrams}(m) naturally give an enhancement of the $\bar d$ distribution compared with the $\bar u$, reminiscent of the empirical result for the $\bar d$ and $\bar u$ PDF asymmetry in the collinear region~\cite{Salamu01}.

Moreover, the result in the $0<x<\xi$ region is positive for $-t = 0.25$~GeV$^2$, but negative for $-t = 1$~GeV$^2$. This can be understood from the fact that the pion GPD or GDA includes the $D_{q/\pi}(t)$ form factor in the ERBL region, which has a different $t$ dependence compared with the pion form factor $F_\pi(t)$. The magnitude of the magnetic GPD $E_u$ is larger than that of the electric GPD $H_u$, and the quark and antiquark distributions in the DGLAP region have opposite signs. This implies that the electric GPD for the $\bar u$ flavor has a different sign to that for $\bar d$, and the absolute value of the magnetic GPD $E$ for $\bar d$ is larger than that for $\bar u$. \\ \\ \\ \\

\section{Nonlocal QED}\label{sec:qed}

The second example of application of nonlocal field theory methods that we discuss in this review is the generalization of QED to the most general nonlocal interaction. We begin with some basics about the nonlocal QED Lagrangian, following by a discussion of solid quantization and gauge invariance, before focusing on the specific problem of the $g-2$ anomaly.

\subsection{Nonlocal QED Lagrangian}
\label{sec-2}

In this section we introduce the general extension of the local QED Lagrangian, where both the free and interacting parts are nonlocal~\cite{Li1, Li2}, and present the Feynman rules for vertices, including the additional interaction generated from gauge links. Recall that the local QED Lagrangian is given by
\begin{equation}
\mathcal{L}^{\text{local}}
= \bar{\psi}(x)\big( i\!\!\not\!\partial - m\big)\psi(x)
- e\,\bar{\psi}(x)\!\not\!\!{A}(x)\psi(x)-\frac14F^{\mu\nu}(x)F_{\mu\nu}(x).
\end{equation}
Based on the same $U(1)$ symmetry, the local QED Lagrangian can be transformed into a nonlocal Lagrangian using the method described in Refs.~\cite{Review, h1, He:2017viu, He:2018eyz, Salamu:2018cny, Salamu:2019dok, Yang:2020rpi, h6, He:2022leb, Gaozhengyang, Li1, Li2, terning, terning2, v3}. The most general nonlocal Lagrangian can be written as 
\begin{align}
\mathcal{L}^{\text{nl}}
&=\int d^4 a\, \bar{\psi}\, \big(x+\tfrac12 a\big) 
I\big(x+\tfrac12 {a},x\big)
\left(i\!\!\not\!{\partial}-m\right) 
\psi\big(x-\tfrac12{a}\big) 
I\big(x,x-\tfrac12{a}\big) F_1(a)
\notag\\
&\quad -\, e\int d^4a\, d^4b\, \bar{\psi}\big(x+\tfrac12{a}\big) I\big(x+\tfrac12{a},x\big)
\!\not\!\!{A}(x+b)\psi\big(x-\tfrac12{a}\big)
I\big(x,x-\tfrac12{a}\big)
F_1(a)\, F_2(a,b)
\notag\\
&\quad-\frac14\int d^4d\, F^{\mu\nu}(x)\, F_{\mu\nu}(x+d)\, F_4(d),
\label{eq-Lagrangian}
\end{align}
where the gauge link,
\begin{equation}\label{eq-link}
I(x,y) \equiv \exp\left(ie\int d^4c \int_{x}^{y} dz^\mu A_\mu(z+c)\, F_3(a,c)\right),
\end{equation}
is introduced to guarantee local gauge invariance. In the nonlocal Lagrangian (\ref{eq-Lagrangian}) the fermion fields $\psi$ and $\bar{\psi}$ are located at spacetime coordinates $x-\tfrac12{a}$ and $x+\tfrac12{a}$, respectively, while the photon field $A_\mu$ is located at $x+b$. The functions $F_1(a)$, $F_2(a,b)$, $F_3(a,c)$ and $F_4(d)$ are the correlation functions, normalized according to
\begin{equation}\label{eq-normalize}
  \int d^4a\, F_1(a)
= \int d^4b\, F_2(a,b)
= \int d^4c\, F_3(a,c)
= \int d^4d\, F_4(d)
= 1,
\end{equation}
where we note the independence of $a$ after the integration over $b$ or $c$ for the functions of $F_2(a,b)$ and $F_3(a,c)$, respectively. In the limits where $F_1(a)=\delta(a)$, $F_2(a,b)=\delta(b)$ and $F_4(d)=\delta(d)$, the nonlocal Lagrangian $\mathcal{L}^{\text{nl}}$ reduces to the local Lagrangian $\mathcal{L}^{\text{local}}$. It is straightforward to show that the most general nonlocal QED Lagrangian \eqref{eq-Lagrangian} is invariant under the gauge transformation
\begin{subequations}
\begin{eqnarray}\label{eq-gaugetrans}
\psi(x)\!\! &\to&\!\! e^{i\alpha(x)}\psi(x),
\\ 
A_\mu(x)\!\! &\to&\!\! A_\mu(x)-\frac{1}{ e}\partial_\mu\alpha'(x),
\end{eqnarray}
\end{subequations}
where 
\begin{equation}\label{eq-alpha}
\alpha(x)
= \int d^4b\, \alpha'(x+b)\, F_2(a,b)
= \int d^4c\, \alpha'(x+c)\, F_3(a,c). 
\end{equation}
Note here $\alpha(x) = \alpha'(x)$ in the local limit.

\begin{figure}[t] 
\begin{adjustwidth}{-\extralength}{0cm}
\centering
\includegraphics{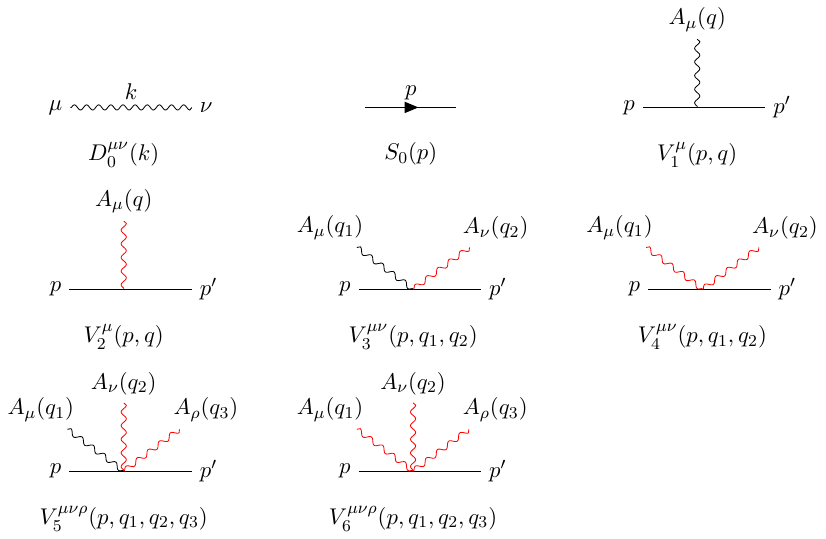}
\caption{Propagators and vertices in nonlocal QED appearing in the calculation of one-loop diagrams. The black and red wavy lines denote photons from minimal substitution and the gauge link, respectively.}\label{fig-v}
\end{adjustwidth}
\end{figure}

From the nonlocal Lagrangian (\ref{eq-Lagrangian}) one can derive the corresponding Feynman rules. The propagators and vertices are illustrated in Fig.~\ref{fig-v}, where in addition to photons generated from the minimal substitution one also has photons arising from the gauge links. Unlike the minimal substitution, which can only generate one photon, two or more photons can be obtained from the gauge link, so that the vertices in nonlocal QED can involve more than one photon. The free fermion and photon propagators in nonlocal QED are therefore modified according to
\begin{subequations}
\label{eq:modprop}
\begin{eqnarray}
S_0(p)\!\!\!&=&\!\!\!\left(\frac{i}{\not\!{p}-m}\right)\, \frac{1}{\widetilde{F}_1(p)}\, ,
\\
D_0^{\mu\nu}(k)\!\!\!&=&\!\!\!\left(\frac{-i g^{\mu\nu}}{k^2}\right)\,\, \frac{1}{\widetilde{F}_4(k)}\, ,
\end{eqnarray}
\end{subequations}
where the functions $\widetilde{F}_1(p)$ and $\widetilde{F}_4(k)$ are Fourier transformations of the correlation functions $F_1(a)$ and $F_4(d)$, respectively. The fermion-photon interaction term in the nonlocal Lagrangian $\mathcal{L}^{\text{nl}}$ generated from the minimal substitution is given by
$\int d^4a\, d^4b\, \bar{\psi}\big(x+\tfrac12{a}\big)\!\not\!\!{A}(x+b)\,\psi\big(x-\tfrac12{a}\big) F_1(a)\, F_2(a,b)$. The corresponding interaction vertex is
\begin{equation}\label{eq:v1}
V_1^\mu(p,q) 
= \gamma^\mu \int\!\!\frac{d^4k}{(2\pi)^4}\,
  \widetilde{F}_1(k)\,\widetilde{F}_2\Big(\frac{p+p'}{2}-k,q\Big) 
= \gamma^\mu\, \widetilde{G}_{2}(p,q),
\end{equation}
where $q$, $p$ and $p'$ are the photon, initial fermion and final fermion momenta, respectively, and $\widetilde{G}_{i}(p,q)$ is defined as
\begin{equation}
\widetilde{G}_{i}(p,q)\equiv\widetilde{G}_{i}(P,q)
= \int\!\!\frac{d^4 k}{(2\pi)^4}\widetilde{F}_1(k)\,\widetilde{F}_i(P-k,q) ~~ (i=2,3),
\end{equation}
with $P \equiv (p+p')/2$.
%
%

In addition to the usual local QED interactions, the nonlocal Lagrangian $\mathcal{L}^{\text{nl}}$ introduces additional interactions involving photons generated from the gauge link \eqref{eq-link}. The method for deducing the Feynman rules for these vertices was discussed in Refs.~\cite{terning, terning2, v3, Review}. The related action for the interaction with one photon from the gauge link can be written as
\begin{equation}
S = -ie \int d^4a\, d^4c\, d^4x\, \bar{\psi}\big(x+\tfrac12{a}\big) G_3(a,c) \big(i\!\!\not\!{\partial}_x-m\big) \psi\big(x-\tfrac12{a}\big)
I\big(x-\tfrac12{a}, x+\tfrac12{a}\big),
\end{equation}
where 
$G_3(a,c) = F_1(a)\, F_3(a,c)$
and 
$I\big(x-\tfrac12{a}, x+\tfrac12{a}\big) = \int_{x-\frac12 a}^{x+\frac12 a} dz^\mu A_\mu(z+c)$.
To obtain the Feynman rule for this vertex requires calculating
$\int d^4a\,d^4c\, G_3(a,c)\, e^{iPa}\, I\big(x-\tfrac12{a}, x+\tfrac12{a}\big)$.
Using the identity 
\begin{eqnarray}
\int d^4a\, d^4c\, G_3(a,c)\, e^{iPa}\, 
I\big(x-\tfrac12{a}, x+\tfrac12{a}\big) 
\!\!&=&\!\! \int d^4a\, d^4c\, d^4k_1\, d^4k_2\, \widetilde{G}_3(k_1,k_2)\, e^{ik_1a} \, e^{ik_2c}\, e^{iPa}\, 
I\big(x-\tfrac12{a}, x+\tfrac12{a}\big)
\nonumber \\
\!\!&=&\!\! \int d^4a\, d^4c\, d^4k_1\, d^4k_2\, \Big(\widetilde{G}_3(-i\partial_a ,k_2) e^{ik_1a}\Big)\, e^{ik_2c}\, e^{iPa}\, 
I\big(x-\tfrac12{a}, x+\tfrac12{a}\big)
\end{eqnarray}
and performing partial integration, one can show that~\cite{terning, v3, Review}
\begin{equation}
\widetilde{G}_{3}(-i\partial_a,k_2)e^{iPa} 
I\big(x-\tfrac12{a}, x+\tfrac12{a}\big) 
= e^{iPa} \widetilde{G}_{3}(-i\mathcal{D}_a,k_2)\,
  I\big(x-\tfrac12{a}, x+\tfrac12{a}\big),
\end{equation}
where $\mathcal{D}_a=\partial_a + iP_a$. Taylor expanding and using the iteration method~\cite{terning, terning2, v3, Review}, one can then write
\begin{equation}
\widetilde{G}_{3}(-i\partial_a,k_2)\, 
I\big(x-\tfrac12{a}, x+\tfrac12{a}\big) 
= i\int d^4q \frac{(P^\mu+q^\mu/2)}{(2P\cdot q+q^2)}
\left( \widetilde{G}_3(p+q,q) - \widetilde{G}_3(p,q) \right)
\left( A(q)e^{iq(x+a/2)} + A(q)e^{iq(x-a/2)} \right).
\end{equation}

The additional electromagnetic vertex with one photon from the gauge link is then given by
\begin{equation}\label{eq:v2}
V_2^\mu(p,q) = (\!\not\!{p}-m)\frac{(q+2P)^\mu}{(q+P)^2-P^2}
\left(\widetilde{G}_{3}(q+p,q)-\widetilde{G}_{3}(p,q)\right).
\end{equation}
Similarly, the electromagnetic vertex with two photons (one from minimal substitution with momentum $q_1$ and the other from the gauge link with momentum $q_2$) can be obtained as~\cite{v3, terning, terning2}
\begin{align}\label{eq:v3}
V_3^{\mu\nu}(p,q_1,q_2) = i\gamma^\mu\frac{(q_2+2P)^\nu}{(q_2+P)^2-P^2} 
\left( \widetilde{G}_{23}(q_2+p,q_1,q_2) - \widetilde{G}_{23}(p,q_1,q_2) \right),
\end{align}
where $\widetilde{G}_{ij}$ is defined as
\begin{equation}\label{eq-Gij}
\widetilde{G}_{ij}(p,q_1,q_2)
= \int\frac{d^4k_1\, d^4k_2}{(2\pi)^8}\, \widetilde{F}_1(k_1)\, 
  \widetilde{F}_i\left(k_2,q_1\right)\, \widetilde{F}_j\big(P-k_1-k_2,q_2\big) ~~ (i,j = 2,3).
\end{equation}
The interaction vertex where the two photons are both from the gauge link is given by
\begin{align}\label{eq:v4}
&V_4^{\mu\nu}(p,q_1,q_2) 
= i(\!\not\!{p}-m) 
\left\{ 2 g^{\mu\nu} 
    \frac{\widetilde{G}_{33}(p+q_1+q_2,q_1,q_2) 
        - \widetilde{G}_{33}(p,q_1,q_2)}{(P+q_1+q_2)^2-P^2}
\right.
\notag\\
& \hspace*{1cm} 
- \frac{\widetilde{G}_{33}(p+q_1+q_2,q_1,q_2) - \widetilde{G}_{33}(p,q_1,q_2)}{(P+q_1+q_2)^2-P^2} 
  \Bigg[ \frac{(2P+q_1)^\mu (2P+2q_1+q_2)^\nu}{(P+q_1+q_2)^2 - (P+q_1)^2}
    +(\mu\leftrightarrow \nu, q_1\leftrightarrow q_2)
  \Bigg]
\notag\\
&\left. \hspace*{1cm}
+ \Bigg[ \frac{\big( \widetilde{G}_{33}(p+q_1,q_1,q_2)
                    - \widetilde{G}_{33}(p,q_1,q_2) 
               \big)(2P+q_1)^\mu(2P+2q_1+q_2)^\nu}{\left((P+q_1)^2-p^2)((P+q_1+q_2)^2-(P+q_1)^2\right)}+(\mu\leftrightarrow \nu, q_1\leftrightarrow q_2)
  \Bigg]
\right\}.
\end{align}
In the above and following equations, when $q_i \leftrightarrow q_j$ only the first argument in the function $\widetilde{G}$ changes. While higher-order interactions with additional photons can be generated from the expansion of the gauge link, for the study of the lepton anomalous magnetic moments at one-loop level interactions up to three photons are needed. The interaction vertex with three photons, with one from minimal substitution and the other two from the gauge link, can be written as
\begin{align}\label{eq:v5}
&V_5^{\mu\nu\rho}(p,q_1,q_2,q_3)
= \gamma^\mu
\left\{
2 g^{\nu\rho} \frac{\widetilde{G}_{233}(p+q_2+q_3,q_1,q_2,q_3)-\widetilde{G}_{233}(p,q_1,q_2,q_3)}{(P+q_2+q_3)^2-P^2}
\right.
\notag\\
& \hspace*{0.5cm}
- \frac{\widetilde{G}_{233}(p+q_2+q_3,q_1,q_2,q_3)-\widetilde{G}_{233}(p,q_1,q_2,q_3)}{(P+q_2+q_3)^2-P^2}
\Bigg[\frac{(2P+q_2)^\nu(2P+2q_2+q_3)^\rho}{(P+q_2+q_3)^2-(P+q_2)^2} 
+ (\nu \leftrightarrow \rho, q_2\leftrightarrow q_3)
\Bigg]
\notag\\
&\left. \hspace*{0.5cm}
+ \Bigg[\frac{\big(\widetilde{G}_{233}(p+q_2,q_1,q_2,q_3)-\widetilde{G}_{233}(p,q_1,q_2,q_3)\big)(2P+q_2)^\nu(2P+2q_2+q_3)^\rho}{((P+q_2)^2-P^2)((P+q_2+q_3)^2-P^2)}
+ (\nu\leftrightarrow \rho, q_2\leftrightarrow q_3)
\Bigg]
\right\},
\end{align}
where the function $\widetilde{G}_{ijk}(p,q_1,q_2,q_3)$ is defined as
\begin{equation}
\widetilde{G}_{ijk}(p,q_1,q_2,q_3)
= \int\frac{d^4k_1\, d^4k_2\, d^4k_3}{(2\pi)^{12}}\,
\widetilde{F}_1(k_1)\, 
\widetilde{F}_i(k_2,q_1)\, 
\widetilde{F}_j(k_3,q_2)\,
\widetilde{F}_k(P-k_1-k_2-k_3,q_3) ~~ (i,j,k = 2,3).
\end{equation}
The vertex for three photons which are all from the gauge link is more complicated. We can separate it into two terms according to 
\begin{equation}\label{eq:v6}
V_6^{\mu\nu\rho}(p,q_1,q_2,q_3) 
= V_{6,a}^{\mu\nu\rho}(p,q_1,q_2,q_3) 
+ V_{6,b}^{\mu\nu\rho}(p,q_1,q_2,q_3),
\end{equation}
where the first term $V_{6,a}^{\mu\nu\rho}$ is given by
\begin{align}
&V_{6,a}^{\mu\nu\rho}(p,q_1,q_2,q_3)
= 2 (\!\not\!{p}-m)g^{\mu\nu}
\left\{ -\frac{[\widetilde{G}_{333}(p+q_1+q_2,q_1,q_2,q_3)-\widetilde{G}_{333}(p,q_1,q_2,q_3)](2P+2q_1+2q_2+q_3)^\rho}{[(P+q_1+q_2)^2-P^2][(P+q_1+q_2+q_3)^2-(P+q_1+q_2)^2]}\right.
\notag\\
&+\frac{\widetilde{G}_{333}(p+q_1+q_2+q_3,q_1,q_2,q_3)-\widetilde{G}_{333}(p,q_1,q_2,q_3)}{(P+q_1+q_2+q_3)^2-P^2}\left[\frac{(2P+2q_1+2q_2+q_3)^\rho}{(P+q_1+q_2+q_3)^2-(P+q_1+q_2)^2}\right.
\notag\\
&+\frac{(2P+q_3)^\rho}{(P+q_1+q_2+q_3)^2-(P+q_3)^2}\bigg]
-\frac{[\widetilde{G}_{333}(p+q_3,q_1,q_2,q_3)-\widetilde{G}_{333}(p,q_1,q_2,q_3)](2P+q_3)^\rho}{[(P+q_3)^2-P^2][(P+q_1+q_2+q_3)^2-(P+q_3)^2]}\Bigg\}
\notag\\
& \hspace*{1cm}
+ (\mu\to\nu,\nu\to\rho,\rho\to\mu,q_1\to q_2,q_2\to q_3,q_3\to q_1)
\notag\\
& \hspace*{1cm}
+ (\mu\to\rho,\nu\to\mu,\rho\to\nu,q_1\to q_3,q_2\to q_1,q_3\to q_2),
\end{align}
and the second term $V_{6,b}^{\mu\nu\rho}$ is
\begin{align}
&V_{6,b}^{\mu\nu\rho}(p,q_1,q_2,q_3)
= (\!\not\!{p}-m)
\Bigg\{\frac{\widetilde{G}_{333}(p+q_1+q_2+q_3,q_1,q_2,q_3)-\widetilde{G}_{333}(p,q_1,q_2,q_3)}{(P+q_1+q_2+q_3)^2-P^2}
\notag\\
&\quad\times \bigg[\frac{(2P+2q_2+2q_3+q_1)^\mu(2P+2q_3+q_2)^\nu(2P+q_3)^\rho}{[(P+q_1+q_2+q_3)^2-(P+q_2+q_3)^2][(P+q_1+q_2+q_3)^2-(P+q_3)^2]}+(\mu\leftrightarrow \nu, q_1\leftrightarrow q_2)\bigg]
\notag\\
&-\left[\frac{\widetilde{G}_{333}(p+q_2+q_3,q_1,q_2,q_3)-\widetilde{G}_{333}(p,q_1,q_2,q_3)}{(P+q_2+q_3)^2-P^2}\right. 
\notag\\
&\quad\times\left.\frac{(2P+2q_2+2q_3+q_1)^\mu(2P+2q_3+q_2)^\nu(2P+q_3)^\rho}{[(P+q_1+q_2+q_3)^2-(P+q_2+q_3)^2][(P+q_2+q_3)^2-(P+q_3)^2]}
+(\mu\leftrightarrow \nu, q_1\leftrightarrow q_2)\right]
\notag\\
&-\frac{\widetilde{G}_{333}(p+q_3,q_1,q_2,q_3)-\widetilde{G}_{333}(p,q_1,q_2,q_3)}{(P+q_3)^2-P^2}\bigg[
\frac{(2P+2q_2+2q_3+q_1)^\mu(2P+2q_3+q_2)^\nu(2P+q_3)^\rho}{[(P+q_1+q_2+q_3)^2-(P+q_2+q_3)^2][(P+q_1+q_2+q_3)^2-(P+q_3)^2]}
\notag\\
&\quad\left.-\frac{(2P+2q_2+2q_3+q_1)^\mu(2P+2q_3+q_2)^\nu(2P+q_3)^\rho}{[(P+q_1+q_2+q_3)^2-(P+q_2+q_3)^2][(P+q_2+q_3)^2-(P+q_3)^2]}+(\mu\leftrightarrow \nu, q_1\leftrightarrow q_2)\right]
\notag\\
& \hspace*{2cm}
+ (\mu\to\nu,\nu\to\rho,\rho\to\mu,q_1\to q_2,q_2\to q_3,q_3\to q_1)
\notag\\
& \hspace*{2cm}
+ (\mu\to\rho,\nu\to\mu,\rho\to\nu,q_1\to q_3,q_2\to q_1,q_3\to q_2)
\Bigg\}.
\end{align}
With these Feynman rules one can proceed to calculate the lepton magnetic form factors from the nonlocal Lagrangian. \\

\subsection{Solid quantization}

The previous section described how the photon and lepton propagators are modified by the nonlocal Lagrangian. Alternatively, the propagators can also be obtained from canonical quantization. In Refs.~\cite{Quantization1, Quantization2} new quantization conditions---referred to as solid quantization---were proposed,
\begin{subequations}
\begin{eqnarray}
\left[\phi(\vec{x},t), \phi(\vec{y},t)\right] 
&=& \left[\pi(\vec{x},t), \pi(\vec{y},t)\right] = 0,
\\
\left[\phi(\vec{x},t), \pi(\vec{y},t)\right] 
&=& i\Phi\left(\vec{x}-\vec{y}\right),
\end{eqnarray}
\label{sq}
\end{subequations}
where the function $\Phi\left(\vec{x}-\vec{y}\right)$ describes the
correlation between fields at spatial points $\vec{x}$ and $\vec{y}$. For the case of point particles, the function $\Phi\left(\vec{x}-\vec{y}\right)$ is replaced by a 3-dimensional $\delta$ function, $\delta^{(3)}\left(\vec{x}-\vec{y}\right)$. 
For the non-point particle case, particles at different positions could be partially superimposed, so that there exists some probability that particles and antiparticles are created at different positions.

Expanding the field $\phi$ as
\begin{equation}
\phi(\vec{x},t) = \int\!\frac{d^3p}{(2\pi)^2\, 2\omega_p}
\Big[ 
  A(\vec{p})\, e^{i\vec{p}\cdot\vec{x}-i\omega_p t} 
+ A^\dag(\vec{p})\, e^{-i\vec{p}\cdot\vec{x}+i\omega_p t}
\Big],
\end{equation}
the creation and annihilation operators satisfy the relations
\begin{subequations}
\begin{eqnarray}
\Big[ A(\vec{p}), A(\vec{q}) \Big] 
\!\!&=&\!\! \Big[ A^\dag(\vec{p}), A^\dag(\vec{q}) \Big] 
 = 0,
\\
\Big[ A(\vec{p}), A^\dag(\vec{q}) \Big] 
\!\!&=&\!\! (2\pi)^3\, 2\omega_p\,\delta^{(3)}(\vec{p} - \vec{q})\, \Psi(\vec{p}).
\end{eqnarray}
\end{subequations}
The fields $\Phi\left(\vec{x}\right)$ and $\Psi\left(\vec{p}\right)$ obey the  relations
\begin{subequations}
\begin{eqnarray}
\Phi\left(\vec{x}\right)
\!\!\!&=&\!\! \frac12 \int\frac{d^3p}{(2\pi)^3}\Psi(\vec{p})
(e^{i\vec{p}\cdot\vec{x}}+e^{-i\vec{p}\cdot\vec{x}}),
\\
\Psi\left(\vec{p}\right)
\!\!\!&=&\!\! \frac12 \int d^3x\, \Phi(\vec{x})
(e^{i\vec{p}\cdot\vec{x}}+e^{-i\vec{p}\cdot\vec{x}}),
\end{eqnarray}
\end{subequations}
and are normalized according to
\begin{subequations}
\begin{eqnarray}
\Phi(0)\!\!\! &=&\!\!\! \int\frac{d^3p}{(2\pi)^3}\, \Psi(\vec{p}),
\\
\Psi(0)\!\!\! &=&\!\!\! \int d^3x\, \Phi(\vec{x})\ =\ 1.
\end{eqnarray}
\end{subequations}
Compared with the usual commutation relation where $\Phi\left(\vec{x}\right) = \delta^{(3)}(\vec{x})$, here we have $\Phi\left(\vec{x}\right)$ normalized 1, while $\Psi(\vec{p})$ is normalized to $\Phi(0)$. With the new quantization, the fields can be written in terms of usual creation and annihilation operators as
\begin{equation}
\phi(\vec{x},t) = \int\!\frac{d^3p}{(2\pi)^2\, 2\omega_p} \sqrt{\Psi(\vec{p})} 
\left[ a(\vec{p}) e^{i\vec{p}\cdot\vec{x}-i\omega_p t}
     + a^\dag(\vec{p}) e^{-i\vec{p}\cdot\vec{x}+i\omega_p t}
\right].
\end{equation}
To obtain the Feynman propagator of the scalar field in the solid quantization, we recall that the propagator is formally defined as
\begin{eqnarray}
\Delta_F(x'-x)\!\! &=&\!\! \langle 0|T\phi(x')\phi(x)|0\rangle
 =  \int\!\frac{d^3k}{(2\pi)^2\, 2\omega_k}
    \left[ \theta(t'-t)\, e^{ik\cdot(x'-x)} + \theta(t-t')\, e^{-ik\cdot(x'-x)}
    \right].
\end{eqnarray}
Using the integral representation of the step function,
\begin{equation}
\theta(t) 
= \text{lim}_{\epsilon\to 0^+} 
  \int\frac{d\tau}{2\pi i} \frac{e^{i\tau t}}{\tau-i\epsilon},
\end{equation}
the Feynman propagator is given by
\begin{equation}
\Delta_F (x'-x) 
= \int \frac{d^4k}{(2\pi)^4} 
  \frac{i\Psi(\vec{k})\, e^{-ik\cdot(x'-x)}}{k^2 - m^2 + i\epsilon}.
\end{equation}

For spin-1/2 fermion fields, the nonzero anticommutation relation is
\begin{equation}
\Big\{\psi_\alpha(\vec{x},t), \bar{\psi}_\beta(\vec{y},t)\Big\}
= \gamma^0_{\alpha\beta}\, \Phi(\vec{x}-\vec{y}),
\end{equation}
and the corresponding field can be written as
\begin{eqnarray}
\psi(\vec{x},t) \!\!\!&=&\!\!\!
\sum_{s=\pm} \int\!\frac{d^3p}{(2\pi)^2\, 2\omega_p} \sqrt{\Psi(\vec{p})}\,
\left[ b_s(\vec{p}) u_s(\vec{p}) e^{i\vec{p}\cdot\vec{x}-i\omega_p t}
     + d_s^\dag(\vec{p}) v_s(\vec{p}) e^{-i\vec{p}\cdot\vec{x}+i\omega_p t}
\right],
\end{eqnarray}
where $b$ and $d^\dag$ are usual annihilation and creation operators, and $u_s(\vec{p})$ and $v_s(\vec{p})$ are the Dirac spinors. The propagator for the spin-1/2 field is then
\begin{equation}
S_F(x'-x) = \int\frac{d^4k}{(2\pi)^4}\,
\frac{i\Psi(\vec{k}) (\!\not\!k+m)\, e^{-ik\cdot(x'-x)}}{k^2-m^2+i\epsilon}.
\end{equation}

For photon fields $A^\mu$, one can expand
\begin{eqnarray}
A^\mu(\vec{x},t) 
\!\!\!&=&\!\!\!
\sum_{\lambda=\pm} \int\!\frac{d^3p}{(2\pi)^2\, 2\omega_p} \sqrt{\Psi(\vec{p})}\,
\left[ 
  a_\lambda(\vec{p}) \epsilon^\mu(\vec{p},\lambda)\, 
  e^{i\vec{p}\cdot\vec{x}-i\omega_p t} 
+ a^\dag_\lambda(\vec{p}) \epsilon^{*\mu}(\vec{p},\lambda)\,
  e^{-i\vec{p}\cdot\vec{x}+i\omega_p t}
\right],
\end{eqnarray}
where $\epsilon^\mu(\vec{p},\lambda)$ is the photon polarization vector. The photon propagator can then be written as
\begin{equation}
D_F^{\mu\nu}(x'-x) 
= \int\frac{d^4k}{(2\pi)^4}
  \frac{-i\Psi(\vec{k})\, g^{\mu\nu}\, e^{-ik\cdot(x'-x)}}{k^2-m^2+i\epsilon}.
\end{equation}

Note that, in principle, the functions $\Psi(\vec{p})$ and $\Phi(\vec{x}-\vec{y})$ depend on the details of the particles, such as the mass and width, and with the new quantization conditions the Feynman rules should be modified accordingly. In particular, the new propagator of the field should be multiplied by a factor $\Psi(\vec{k})$, and the external field multiplied by a factor $\sqrt{\Psi(\vec{k})}$.

In contrast to the nonrelativistic case, for the relativistic version of the solid quantization we consider the field with a distribution in four-dimensional spacetime. For a scalar field $\phi(x)$, we can write
\begin{equation}
\phi(x) = \int \frac{d^4p}{(2\pi)^4} H(p^2)
\left[ \alpha_p\, e^{-ip\cdot x} + \alpha_p^\dag\, e^{ip\cdot x} \right],
\end{equation}
where the function $H(p^2)$ describes the four dimensional distribution for non-point particle in relativistic case.
The operators $\alpha_p$ and $\alpha_p^\dag$ obey the commutation relations
\begin{subequations}
\begin{eqnarray}
\big[\alpha_p, \alpha_q\big]
\!\! &=&\!\! \big[\alpha_p^\dag,\alpha_q^\dag\big] = 0, 
\\
\big[\alpha_p,\alpha_q^\dag\big]
\!\!&=&\!\! (2\pi)^4\, \delta^{(4)}(p-q).
\end{eqnarray}
\end{subequations}
The commutation relations for the scalar field and its conjugate are
\begin{eqnarray}\nonumber
\left[\phi(\vec{x},t),\pi(\vec{y},t)\right]\!\!
&=&\!\! \int\frac{d^4p}{(2\pi)^4} H^2(p^2)\, 
ip_0\, (e^{i\vec{p}\cdot\vec{x}} + e^{-i\vec{p}\cdot\vec{x}})
\\
\nonumber \!\!
&=&\!\! \int\frac{d^3p}{(2\pi)^3} \frac{i\Psi(\vec{p})}{2}(e^{i\vec{p}\cdot\vec{x}}+e^{-i\vec{p}\cdot\vec{x}})
\\
&\equiv& i\Phi(\vec{x}-\vec{y}),
\end{eqnarray}
where
\begin{equation}
\Psi(\vec{p}) = \int\frac{dp_0}{\pi}H^2(p^2)p_0.
\end{equation}
For a point-particle with mass $m$, one has $\Psi(\vec{p})=1$, and $H^2(p^2) = 2\pi\delta(p^2-m^2)$. Note that $H(p^2)$ then is proportional to $\delta^{1/2}(p^2-m^2)$ instead of $\delta(p^2-m^2)$, since the field is expanded in terms of $\alpha_p$ and $\alpha^\dag_p$ instead of $a_p$ and $a_p^\dag$.

For simplicity, we can rewrite the scalar field as
\begin{eqnarray}
\phi(x)
\!\!&=&\!\! \int\!\frac{d^4p}{(2\pi)^4}\, dM^2\, H(M^2)\, \delta(p^2-M^2)\, 
\left[ \alpha_p\, e^{-ip\cdot x} + \alpha_p^\dag\, e^{ip\cdot x} \right]
\nonumber\\
\!\!&=&\!\! \int\!\frac{d^3p}{(2\pi)^4\, 2\omega_M} dM^2\, H(M^2)\,
\left[ \alpha_{\vec{p},\omega_M}\, e^{i\vec{p}\cdot\vec{x}-i\omega_Mt}
    +  \alpha_{\vec{p},\omega_M}^\dag\, e^{-i\vec{p}\cdot\vec{x}+i\omega_Mt}
\right],
\end{eqnarray}
where $M$ is a mass parameter and $\omega_M=\sqrt{\vec{p}^2+M^2}$. The propagator for the relativistic scalar field can be written as
\begin{eqnarray}\ \nonumber
\Delta_F(x'-x) = \int\!\frac{d^3k}{(2\pi)^4\, 2\omega_M\, 2\omega_{M'}}
dM^2\, dM'^2\, H(M^2)\, H(M'^2)\, 
\delta(\omega_{M'}-\omega_M)
\left[ \theta(t'-t)e^{ik\cdot(x'-x)}+\theta(t-t')e^{ik\cdot(x-x')} \right],
\end{eqnarray}
where $\delta(\omega_{M'}-\omega_M) = 2\omega_M\, \delta(M'^2-M^2)$. With the definition of $\theta$ function, the propagator can then be written as
\begin{equation}
\Delta_F(x'-x) = \int\!\frac{d^4k}{(2\pi)^4} \frac{dM^2}{2\pi} 
\frac{iH^2(M^2)}{k^2-M^2+i\epsilon}e^{-ik\cdot(x'-x)}.
\end{equation}
Again, when $H^2(M^2) = 2\pi\delta(M^2-m^2)$, the propagator reduces to that for a point particle with mass $m$. If $H^2(M^2)$ is chosen to be $2\pi [\delta(M^2-m^2)-\delta(M^2-\Lambda^2)]$, one obtains the result for Pauli-Villars regularization. By comparing with the propagators obtained in solid quantization and those in Eqs.~(\ref{eq:modprop}), one obtains a relationship between $H(p^2)$ and $\widetilde{F}_1(p^2)$,
\begin{equation}
\int \frac{dM^2}{2\pi}\frac{H^2(M^2)}{p^2-M^2}\
=\ \frac{1}{\widetilde {F}_1(p^2)(p^2-m^2)}.
\end{equation}
Again, for consistency with the nonlocal Lagrangian, the canonical quantization condition must be modified to that for the solid quantization in Eqs.~(\ref{sq}).

\subsection{Gauge invariance}

Before discussing the magnetic moments, we first demonstrate that the Ward-Green-Takahashi identity and charge conservation can be obtained from the nonlocal Lagrangian \cite{Li2}. The nonlocal Lagrangian is invariant under the local $U(1)$ transformation,
\begin{equation}\label{eq-suppose1}
\int\!d^4x\, d^4a\, 
\bar{\psi}\big(x+\tfrac12{a}\big) \psi\big(x-\tfrac12{a}\big)\, 
F_1(a)\, \alpha(x)
= \int d^4x\, d^4a\, d^4b\, 
\bar{\psi}\big(x+\tfrac12{a}\big) \psi\big(x-\tfrac12{a}\big)\, F_1(a)\, F_2(a,b)\, \alpha'(x+b),
\end{equation}
which leads to Eq.~(\ref{eq-alpha}) above. For fermions with momenta $k_1$ and $k_2$, Eq.~(\ref{eq-suppose1}) implies that
\begin{equation}\label{eq-suppose2}
\widetilde{F}_1(K)\,\widetilde{\alpha}(k_1-k_2)=\int d^4k_3\,\widetilde{F}_1(k_3)\,\widetilde{F}_2(K-k_3,k_2-k_1)\,\widetilde{\alpha}'(k_1-k_2),
\end{equation}
where $K\equiv (k_1+k_2)/2$, and $\widetilde{\alpha}$ is the Fourier transform of the phase $\alpha$ introduced in Eq.~(\ref{eq-gaugetrans}).
In particular, when $k_1=k_2$, one has
\begin{equation}\label{eq-suppose3}
\widetilde{F}_1(k_1)\, \widetilde{\alpha}(0) 
= \int d^4k_3\, 
  \widetilde{F}_1(k_3)\, \widetilde{F}_2(k_1-k_3,0)\, \widetilde{\alpha}'(0).
\end{equation}
Meanwhile, the Fourier transformations of Eqs.~\eqref{eq-normalize} and \eqref{eq-alpha} are given by
\begin{equation}
\int d^4k\,\widetilde{F}_2(k,0)\,e^{-i k\cdot a}=1
\end{equation}
and
\begin{equation}
\widetilde{\alpha}(k)=\int d^4k'\, \widetilde{F}_2(k',-k)\, \widetilde{\alpha}'(k)\,e^{-i k'\cdot a}\,,
\end{equation}
respectively. Consequently, one finds that $\widetilde{\alpha}(0) = \widetilde{\alpha}'(0)$, and Eq.~(\ref{eq-suppose3}) can therefore be rewritten as
\begin{equation}\label{eq-suppose4}
\widetilde{F}_1(p) = \int d^4k\, \widetilde{F}_1(k)\, \widetilde{F}_2(p-k,0)
\equiv G_2(p,q=0).
\end{equation}
Similarly, we also have the corresponding relation
\begin{equation}\label{eq-suppose5}
\widetilde{F}_1(p)=\int d^4k\,\widetilde{F}_1(k)\,\widetilde{F}_3(p-k,0)\equiv G_3(p,q=0).
\end{equation}
With the definition of $G_{ij}$ in Eq.~(\ref{eq-Gij}), and using Eqs.~(\ref{eq-suppose4}) and (\ref{eq-suppose5}), one has
\begin{align}
\widetilde{G}_{ij}(p,q_1,q_2=0)&=\int\frac{d^4 k_1 d^4 k_2}{(2\pi)^8}\,\widetilde{F}_1(k_1)\, \widetilde{F}_i\left(k_2,q_1\right)\, \widetilde{F}_j\left(p-k_1-k_2,0\right)
= \widetilde{G}_{i}(p,q_1)\label{eq-Gij-Gi},
\end{align}
and similarly,
\begin{equation}\label{eq-Gijk-Gij}
\widetilde{G}_{ijk}(p,q_1,q_2,q_3=0) = \widetilde{G}_{ij}(p,q_1,q_2).
\end{equation}
Note that Eqs.~\eqref{eq-Gij-Gi} and \eqref{eq-Gijk-Gij} are valid for $q_i=0$ for any $i$.

\begin{figure}[tbph] 
\begin{adjustwidth}{-\extralength}{0cm}
\centering
\includegraphics[scale=1.0]{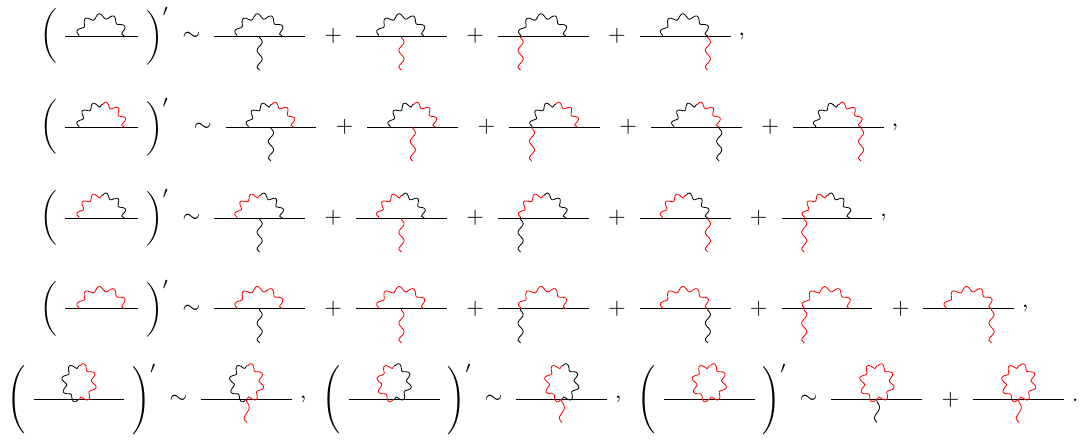}
\caption{Relations between the lepton's self-energy diagrams and lepton-photon vertex diagrams at one-loop level in nonlocal QED. The black and red wavy lines represent photons arising from minimal substitution and the gauge link, respectively.}
\label{fig-ward}
\end{adjustwidth}
\end{figure}

From the above equations, after some algebra one can obtain the following identities,
\begin{subequations}
\label{eq-dall}.
\begin{align}
\frac{ d S_0(p)}{ d p_\mu} 
&= i\lim_{q\to0}S_0(p)
\big[ V_1^\mu(p,q)+V_2^\mu(p,q)\big] S_0(p),
\label{eq-d1}
\\
\frac{\partial V_1^\mu(p,q_1)}{\partial p_\nu} 
&= i\lim_{q_2\to0}V_3^{\mu\nu}(p,q_1,q_2),
\label{eq-d2}
\\
\frac{\partial V_2^\mu(p,q_1)}{\partial p_\nu} 
&= i\lim_{q_2\to0}
\big[ V_3^{\nu\mu}(p,q_2,q_1)+V_4^{\mu\nu}(p,q_1,q_2) \big],
\label{eq-d3}
\\
\frac{\partial V_3^{\mu\nu}(p,q_1,q_2)}{\partial p_\rho} 
&= i\lim_{q_3 \to 0} V_5^{\mu\nu\rho}(p,q_1,q_2,q_3),
\label{eq-d4}
\\
\frac{\partial V_4^{\mu\nu}(p,q_1,q_2)}{\partial p_\rho}
&= i\lim_{q_3\to0}
\big[ V_5^{\rho\mu\nu}(p,q_3,q_1,q_2)+V_6^{\mu\nu\rho}(p,q_1,q_2,q_3) \big]
\label{eq-d5}.
\end{align}
\end{subequations}
The identities in Eqs.~(\ref{eq-dall}) allow the relationship between the self-energy and vertex to be established. At one-loop level, there are 7 self-energy diagrams and 24 vertex diagrams, as illustrated in Ref.~\cite{Li2}. The total self-energy $\Sigma(p)$ and vertex $\Gamma^\mu(p,q)$ can be written,
\begin{subequations}
\begin{align}
\Sigma(p) &= \sum_{i=1}^7 \Sigma_i(p),
\\
\Gamma^\mu(p,q) &= V_1^\mu + V_2^\mu + \sum_{i=1}^{24} \Gamma_i^\mu(p,q),
\end{align}
\end{subequations}
where $\Sigma_i(p)$ and $\Gamma_i^\mu(p,q)$ are given explicitly in Ref.~\cite{Li2}. One can further show that 
\begin{subequations}
\begin{align}
-\frac{ d \Sigma_1(p)}{ d p_\mu}&=\lim_{q\to0}\big[\Gamma^\mu_1(p,q)+\Gamma^\mu_2(p,q)+\Gamma^\mu_3(p,q)+\Gamma^\mu_5(p,q)\big],\\
-\frac{ d \Sigma_2(p)}{ d p_\mu}&=\lim_{q\to0}\big[\Gamma^\mu_6(p,q)+\Gamma^\mu_7(p,q)+\Gamma^\mu_8(p,q)+\Gamma^\mu_9(p,q)+\Gamma^\mu_{11}(p,q)\big],\\
-\frac{ d \Sigma_3(p)}{ d p_\mu}&=\lim_{q\to0}\big[\Gamma^\mu_{4}(p,q)+\Gamma^\mu_{12}(p,q)+\Gamma^\mu_{13}(p,q)+\Gamma^\mu_{14}(p,q)+\Gamma^\mu_{16}(p,q)\big],\\
-\frac{ d \Sigma_4(p)}{ d p_\mu}&=\lim_{q\to0}\big[\Gamma^\mu_{10}(p,q)+\Gamma^\mu_{15}(p,q)+\Gamma^\mu_{17}(p,q)+\Gamma^\mu_{18}(p,q)+\Gamma^\mu_{19}(p,q)+\Gamma^\mu_{20}(p,q)\big],\\
-\frac{ d \Sigma_5(p)}{ d p_\mu}&=\lim_{q\to0}\Gamma^\mu_{22}(p,q),\\
-\frac{ d \Sigma_6(p)}{ d p_\mu}&=\lim_{q\to0}\Gamma^\mu_{21}(p,q),\\
-\frac{ d \Sigma_7(p)}{ d p_\mu}&=\lim_{q\to0}\big[\Gamma^\mu_{23}(p,q)+\Gamma^\mu_{24}(p,q)\big].
\end{align}
\end{subequations}
These relations correspond to the diagrams in Fig.~\ref{fig-ward}. The derivative of the self-energy diagrams gives rise to the vertex diagrams, generating one external photon field to be attached to the self-energy diagram at all possible places. For example, for the first rainbow self-energy diagram in Fig.~\ref{fig-ward}, a photon from minimal substitution or from the gauge link can be attached to the internal lepton line, while only a photon generated from the gauge link can be attached to the vertex. For the fourth rainbow diagram, both minimal substitution and gauge link photons can be attached to the vertex.

The dressed fermion propagator $S(p)$ can be written in terms of the free fermion propagator and the self-energy as 
$S(p) = S_0(p) 
      + S_0(p)\Sigma(p)S_0(p) 
      + S_0(p)\Sigma(p)S_0(p)\Sigma(p)S_0(p) 
      + \cdots$. 
It is then straightforward to obtain 
\begin{equation}\label{eq-q0loop}
\lim_{q\to0} S(p+q)\big[-i\Gamma^\mu(p+q,p)\big]S(p) = -\frac{dS(p)}{d p_\mu},
\end{equation}
which corresponds to the Ward-Green-Takahashi identity,
\begin{equation}\label{eq-Ward}
\lim_{q\to 0} \big[-iq_\mu\Gamma^\mu(p+q,p)\big] 
= \lim_{q\to 0} \big[S^{-1}(p+q)-S^{-1}(p)\big].
\end{equation}
The dressed propagator can be also written as
\begin{equation}
S(p)=\frac{i Z_2}{\not\!{p}-m},
\end{equation}
where $Z_2$ is the wave function renormalization constant, and $Z_2-1 = d\Sigma(p)/d\!\!\!\not{\!\!p}\big|_{\not{p}=m}$. The Ward-Green-Takahashi identity \eqref{eq-Ward} also implies that
\begin{equation}
Z_2\Gamma^\mu(p,p)=\gamma^\mu.
\end{equation}
As usual, the Dirac and Pauli form factors are defined as \cite{pes}
\begin{equation}
Z_2\Gamma^\mu(p+q,p)=\gamma^{\mu}F_1(q^2)+\frac{i\sigma^{\mu\nu}q_\nu}{2m}F_2(q^2),
\end{equation}
with normalization $F_1(0)=1$. This is consistent with the renormalized lepton charge being unity. In the next section, we will use the nonlocal QED to calculate the lepton magnetic moments to explore the $g-2$ anomaly.

\subsection{$g-2$ anamaly}\label{qed:g-2}

The current theoretical prediction for the muon anomalous magnetic moment $a_\mu$ in the SM is $a_\mu^{\text{SM}} = 116\,591\,810(43) \times 10^{-11}$ \cite{Aoyama3}. The recent measurement of $a_\mu$ in the E989 experiment at Fermilab (FNAL) found 
\begin{equation}
\Delta a_\mu^{\text{FNAL}} \equiv a_\mu^{\text{FNAL}} - a_\mu^{\text{SM}} = (230 \pm 69) \times 10^{-11},
\end{equation}
which is a $3.3 \sigma$ discrepancy from the SM prediction \cite{Abi}. Combined with the previous E821 measurement at Brookhaven National Laboratory (BNL) \cite{Bennett}, the result revealed a $4.2 \sigma$ deviation from the SM prediction \cite{Abi}
\begin{equation}
\Delta a_\mu = a_\mu^{\text{FNAL}+\text{BNL}} - a_\mu^{\text{SM}} = (251 \pm 59) \times 10^{-11}.
\end{equation}

For the electron, the theoretical prediction for $a_e$ is $a_e^{\text{SM,B}} = 1\,159\,652\,182.032(720) \times 10^{-12}$ \cite{Aoyama}, where the superscript ``B'' refers to the fine structure constant $\alpha$ being measured at Berkeley with $^{137}{\text{Cs}}$ atoms \cite{Parker}. The most accurate measurement of $a_e$ was made by the Harvard group, and the discrepancy from the SM was $2.4 \sigma$ \cite{Hanneke},
\begin{equation}\label{eq:e1}
\Delta a_e^{\text{B}}=a_e^{\text{exp}}-a_e^{\text{SM,B}}=(-87\pm36)\times10^{-14}.
\end{equation}
However, a new determination \cite{Morel} of the fine structure constant $\alpha$, obtained from the measurement at LKB with $^{87}{\text {Rb}}$, improves the accuracy by a factor of 2.5 compared to the previous best measurement at Berkeley \cite{Parker}. With the new $\alpha$ value, the SM prediction for the electron magnetic moment is $1.6 \sigma$ lower than the experimental data,
\begin{equation}\label{eq:e2}
\Delta a_e^{\text{LKB}} = a_e^{\text{exp}} - a_e^{\text{SM,LKB}}
= (48\pm30)\times10^{-14}.
\end{equation}
It is interesting to note that the two $\Delta a_e$ discrepancies have similar magnitude but opposite sign, for still unidentified reasons \cite{Morel}. The small difference between the $\alpha$ values does not affect $\Delta a_\mu$ since it is much larger than $\Delta a_e$. As the SM predictions match all other experimental information very well, the deviation in one of the most precisely measured quantities in particle physics provides an enduring hint for new physics.

In this section we will examine the $g-2$ anomaly within nonlocal QED. By inserting the electromagnetic current into the lepton states with initial and final momenta $p$ and $p'$, one can obtain the Dirac and Pauli form factors of the lepton. The corresponding Feynman diagrams are given in Ref.~\cite{Li2}. Using the projection method, one finds
\begin{subequations}
\begin{eqnarray}
&&F_1(q^2) = -\frac{A(4m^2-q^2) - 6m^2B}{4(4m^2-q^2)^2},
\\
&&F_2(q^2) = m^2 \frac{A(4m^2-q^2) - B(2m^2+q^2)}{q^2(4m^2-q^2)^2},
\end{eqnarray}
\end{subequations}
where $A$ and $B$ can be calculated from the traces
\begin{subequations}
\begin{eqnarray}
&&
A = \sum_{\text{spin}}\bar{u}(p')\Gamma_\mu u(p)\bar{u}(p)\gamma^\mu u(p')
  = {\rm Tr}\Big[\Gamma_{\mu}(\not\!{p}+m)\gamma^{\mu}(\not\!{p}'+m)\Big],
\label{eq-A}
\\
&&
B = \sum_{\text{spin}}\bar{u}(p')\Gamma_\mu u(p)\bar{u}(p)\frac{(p+p')^\mu}{m}u(p')
  = {\rm Tr}\Big[\Gamma_{\mu}(\not\!{p}+m)(\not\!{p}'+m)\frac{(p+p')^\mu}{m}\Big].
\label{eq-B}
\end{eqnarray}
\end{subequations}
In this review, we focus on the Pauli form factor $F_2$, which is related to the anomalous magnetic moment of the lepton. Compared with the standard QED theory, in the nonlocal case the one-loop vertices are rather more complicated. In addition to the usual QED diagrams, there are an additional 23 diagrams for the nonlocal theory. For the first rainbow diagram, the vertex $\Gamma_1^\mu(p,q)$ is expressed as
\begin{equation}
\Gamma_1^\mu(p,q) = -e^2\int\frac{d^4k}{(2\pi)^4} V_1^\nu\left(p'-k,k\right) S_0(p-k+q)V_1^\mu(p-k,q) S_0(p-k) V_1^\rho\left(p,-k\right) D_{0\nu\rho}(k).
\end{equation}
The corresponding Pauli form factor for the vertex $\Gamma_1^\mu(p,q)$ is given by
\begin{align}
F_{2,1}(q^2) = &\frac{-8ie^2m^2}{q^2(4m^2-q^2)^2} \int\!\frac{d^4k}{(2\pi)^4}\left[\frac{(4m^2+2q^2)((k\cdot p)^2+(k\cdot p')^2)-8(m^2-q^2)(k\cdot p)(k\cdot p')}{((p'-k)^2-m^2)((p-k)^2-m^2)k^2}\right.\notag\\
&+\left.\frac{(q^4-4m^2q^2)(k\cdot p+k\cdot p'+k^2)}{((p'-k)^2-m^2)((p-k)^2-m^2)k^2}\right] \widetilde{G}_{2}(p'-k,k)\, \widetilde{G}_{2}(p-k,q)\, \widetilde{G}_{2}(p,-k).
\end{align}
When the momentum transfer $q^2=0$, one has 
\begin{equation}
F_{2,1}(0) = ie^2 \int\!\frac{d^4k}{(2\pi)^4} \frac{(3k^2 + 2k \cdot p)m^2-3(k \cdot p)^2}{2k^2(k^2 - 2k \cdot p)^2m^2} \widetilde{G}_{2}(p-k,k)\, \widetilde{G}_{2}(p-k,0)\, \widetilde{G}_{2}(p,-k).
\end{equation}
For all other diagrams, the expressions for $\Gamma_i^\mu(p,q)$ are given in the Appendix of Ref.~\cite{Li2}.

In the numerical calculations, the photon is treated as a point particle, with no modification to its propagator, and the correlation function $F_4(a)$ in the free photon Lagrangian is chosen to be a delta function, $\delta(a)$. For the lepton-photon interaction, the function $F_i(a,b)~(i=2,3)$ is assumed to be factorized as $f(a)\, F_i(b)$ to simplify numerical calculation. According to Eq.~(\ref{eq-normalize}), $f(a)$ is an $a$-independent constant, equal to 1, and $\int d^4 b\, F_i(b)=1$. As a result, the Fourier transforms of the correlators can be written as
\begin{equation}
\widetilde{G}_i(p,q)=\widetilde{F}_1(P)\, \widetilde{F}_i(q).
\end{equation}
For the vertex with two and three photons, the correlators can similarly be factorized as
\begin{align}
\widetilde{G}_{ij}(p,q_1,q_2)
&= \widetilde{F}_1(P)\, \widetilde{F}_i(q_1)\, \widetilde{F}_j(q_2),
\\
\widetilde{G}_{ijk}(p,q_1,q_2,q_3)
&= \widetilde{F}_1(P)\, \widetilde{F}_i(q_1)\, \widetilde{F}_j(q_2)\, \widetilde{F}_k(q_3),
\end{align}
where $i,j,k = 2$ or 3. The subscript 2 here represents a photon from the minimal substitution while the subscript 3 is for a photon from the gauge link. The correlators in the interaction vertex are chosen to be
\begin{equation}
\widetilde{F}_2(k) = \widetilde{F}_3(k) = \frac{\Lambda_2^2}{\Lambda_2^2-k^2}.
\end{equation}
These correlators were proposed in earlier work~\cite{Review, Li1, Li2} on nonlocal effective field theory and the ``minimal" version of nonlocal QED. Here the free Lagrangian for the lepton is also nonlocal, which gives rise to a modified lepton propagator. For $\widetilde{F}_1(p)$ in the free lepton propagator, one can choose
\begin{equation}\label{eq-regulator1}
\widetilde{F}_1(p) = \frac{\Lambda_1^2-p^2}{\Lambda_1^2}.
\end{equation}
Since the correlator is in the denominator of Eq.~(\ref{eq:modprop}), the modified propagator makes the loop integration more convergent.

In the above correlators, the cutoff parameters $\Lambda_1$ and $\Lambda_2$ can be determined from $\Delta a_e$ and $\Delta a_\mu$, which will give a nonzero difference between the nonlocal QED and SM $\Delta a_l^{\text{nl}}$ results. Note that nonlocal QED itself does not determine the form of the correlation functions or the cutoff parameters, which instead reflect the properties of the particles and need to be determined empirically. In this case, $\Lambda_{2e}$ and $\Lambda_{2\mu}$ are determined for a given $\Lambda_1$ using the experimental $\Delta a_e$ and $\Delta a_\mu$ for the chosen correlators.

When $\Lambda_1$ and $\Lambda_2$ are both infinite, the nonlocal QED reverts back to the standard local QED. To render the nonlocal effect negligible in other electromagnetic processes, such as Compton scattering, electron-electron scattering or electron-positron annihilation, large values ($\gtrsim 1$~TeV) are required for the cutoffs. At each vertex, the correlator $\widetilde{F}_1(k)$ makes the loop integral more divergent, while $\widetilde{F}_2(k)$ and $\widetilde{F}_3(k)$ make the integral more convergent. For a given $\Lambda_1$, the smaller the $\Lambda_2$, the smaller the magnetic moment. A lower-limit will therefore exist for the magnetic moment for the smallest $\Lambda_2$ value, and if $\Delta a_l$ is below the lower-limit value, it cannot be explained in nonlocal QED for reasonable cutoffs.

In contrast, for a given $\Lambda_2$, the smaller the $\Lambda_1$, the larger the magnetic moment. Since for any $\Lambda_1$ the magnetic moment of the lepton is infinite when $\Lambda_2 \to \infty$, there will be no upper-limit value for the lepton magnetic moment. For large cutoffs, depending on the specific values of $\Lambda_1$ and $\Lambda_2$, one can obtain lepton magnetic moments larger or smaller relative to the SM results. Since $\Lambda_1$ and $\Lambda_2$ should be process independent, if the $\Lambda$ values determined in one process cannot reproduce experimental results in other processes, the nonlocal QED would need to be modified further.

\begin{figure}[tbp] 
\begin{adjustwidth}{-\extralength}{0cm}
    \centering
    \includegraphics{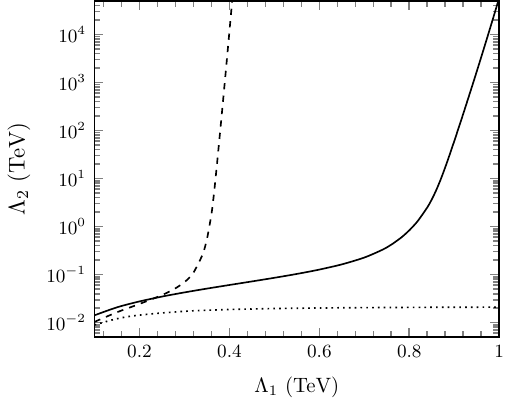}
    \caption{Cutoff parameter $\Lambda_2$ versus $\Lambda_1$ for the muon $\Delta a_\mu$ (solid line), electron $\Delta a_e^{\text{LKB}}$ (dashed line), and electron $\Delta a_e^{\text{B}}$ (dotted line) discrepancies, respectively.}
    \label{fig:Lambda_2-Lambda_1-1}
\end{adjustwidth}
\end{figure}

In Fig.~\ref{fig:Lambda_2-Lambda_1-1} the values of $\Lambda_2$ are plotted versus $\Lambda_1$ for the muon $\Delta a_\mu$, electron $\Delta a_e^{\text{LKB}}$, and electron $\Delta a_e^{\text{B}}$ discrepancies. For a given $\Lambda_1$, one can always find a corresponding $\Lambda_2$ to obtain the experimental discrepancies. For the muon, when $\Lambda_1$ is small ($\lesssim 0.8$~TeV), $\Lambda_2$ increases smoothly with increasing $\Lambda_1$, while $\Lambda_2$ increases rapidly when $\Lambda_1 \gtrsim 0.8$~TeV. For 
    $\Lambda_1 = (0.8, 0.9, 1.0)$~TeV, 
for example, one has the corresponding values 
    $\Lambda_2 = (0.84,\, 60.23,\, 5.82\times 10^4$)~TeV, 
respectively. For the electron case, the results for the two discrepancies $\Delta a_e^{\text{LKB}}$ and $\Delta a_e^{\text{B}}$ are quite different. For the negative $\Delta a_e^{\text{B}}$, the obtained $\Lambda_2^{\text{B}}$ is not sensitive to $\Lambda_1$, and for a broad range of $\Lambda_1$ one finds $\Lambda_2^{\text{B}} \approx 10$--20~GeV, which is unreasonably small. The experimental $e^+ e^- \to \mu^+ \mu^-$ cross section, for example, will not be described for small cutoff parameters $\Lambda_2$. However, for the positive $\Delta a_e^{\text{LKB}}$ result, when $\Lambda_1 \approx 0.35$~TeV, one obtains $\Lambda_2^{\text{LKB}} \approx 0.55$~TeV. The corresponding $\Lambda_2^{\text{LKB}}$ increases rapidly with increasing $\Lambda_1$ when $\Lambda_1 \gtrsim 0.35$~TeV. Since for any $\Lambda_1$ the anomalous magnetic moment of the electron is infinite when $\Lambda_2 \to \infty$, one can always find a value of $\Lambda_2^{\text{LKB}}$ to obtain the correct $\Delta a_e^{\text{LKB}}$ for any large $\Lambda_1$.

The muon $g-2$ discrepancy can therefore be well explained with large $\Lambda$ values, without contradicting other experimental measurements. For the electron, however, $\Delta a_e^{\text{B}}$ cannot be reasonably reproduced in nonlocal QED. As discussed, the correlators $\widetilde{F}_1(k)$ and $\widetilde{F}_{2,3}(k)$ in the vertex have opposite effects. With appropriate choice of cutoffs one can still find negative $\Delta a_e^{\text{nl}}$ with large $\Lambda$ values, although its absolute value is much smaller than $\Delta a_e^{\text{B}}$. For example, the calculated $\Delta a_e^{\text{nl}} = -4.04 \times 10^{-16}$ with $\Lambda_2=1.0$~TeV and infinite $\Lambda_1$. Certainly, for a given $\Lambda_1$, if $\Lambda_2$ is between $\Lambda_2^{\text{B}}$ and $\Lambda_2^{\text{LKB}}$, the calculated $\Delta a_e^{\text{nl}}$ will lie between $\Delta a_e^{\text{B}}$ and $\Delta a_e^{\text{LKB}}$. For $\Lambda_1=1.0$~TeV, if we assume $\Lambda_2$ for the electron is the same as for the muon, the discrepancy $\Delta a_e^{\text{nl}} = 8.19 \times 10^{-14}$.

We should stress that the SM prediction $a_\mu^{\text {SM}}$ has been derived using the leading hadronic vacuum polarization (HVP) contribution to the muon $g-2$, 
    $(a_\mu^{\text {HVP}})_{e^+e^-}^{\text {TI}} = 6\,931(40) \times 10^{-11}$,
based on low-energy cross section data for $e^+e^-\to$ hadrons obtained by the Muon $g-2$ Theory Initiative \cite{Paradisi, Aoyama3}. Recently, the BMW lattice QCD collaboration \cite{Borsanyi} also computed the leading HVP contribution to the muon $g-2$. With careful treatments of critical issues such as scale determination, noise reduction, QED and strong-isospin breaking, and infinite-volume and continuum extrapolations, the obtained HVP contribution was found to be 
    $(a_\mu^{\text {HVP}})_{\text {BMW}} = 7\,075(55)\times 10^{-11}$.
With the larger contribution of $(a_\mu^{\text {HVP}})_{\text {BMW}}$, the discrepancy from the experimental result $\Delta a_\mu$ is reduced to 1.6$\sigma$. Obviously, the findings need to be confirmed by other groups using other discretizations of QCD, which are currently underway. With appropriate choice of the cutoff parameters in the correlation functions, the $1.6\sigma$ discrepancy of the moun $g-2$ can also be obtained in nonlocal QED. For example, for $\Lambda_1 = (1.0, 1.2, 1.4)$~TeV, this can be achieved with $\Lambda_2 = (0.23, 0.68, 71.95)$~TeV, respectively, using the value of the lattice result $(a_\mu^{\text {HVP}})_{\text {BMW}}$ rather than $(a_\mu^{\text {HVP}})_{e^+e^-}^{\text {TI}}$.

\begin{figure}[t] 
\begin{adjustwidth}{-\extralength}{0cm}
\centering
\includegraphics{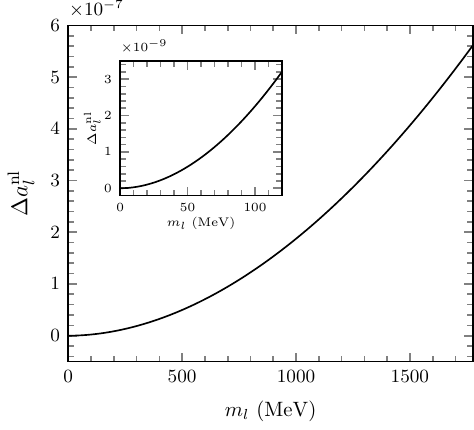}
\caption{Calculated lepton anomalous magnetic moment discrepancy $\Delta a_l^{\text{nl}}$ versus the lepton mass, $m_l$. The cutoff $\Lambda_1$ is fixed to 1~TeV and $\Lambda_2$ is fixed to $\Lambda_{2\mu}$ obtained by the experimental $\Delta a_\mu$ with $\Lambda_1=1$~TeV. The inset illustrates the result at small lepton masses.}
\label{fig:a_l-m}
\end{adjustwidth}
\end{figure}

To illustrate the lepton-mass dependence of the calculated lepton anomalous magnetic moment discrepancy, in Fig.~\ref{fig:a_l-m} we show $\Delta a_l^{\text{nl}}$ versus the lepton mass, $m_l$, for $\Lambda_1=1$~TeV. Since $\Lambda_2$ for the electron is not well determined because of the two different experimental results, for any mass $m_l$, $\Lambda_2$ is chosen to agree with the experimental $\Delta a_\mu$ value for $\Lambda_1=1$~TeV. For $m_l=m_\mu$, the discrepancy $\Delta a_l^{\text{nl}}$ coincides with the experimental discrepancy $\Delta a_\mu$. For $m_l=m_e$, the calculated $\Delta a_l^{\text{nl}}$ is larger than $\Delta a_e^{\text{B}}$, but smaller than $\Delta a_e^{\text{LKB}}$. The discrepancy $\Delta a_l^{\text{nl}}$ increases with increasing lepton mass, and it is $5.62 \times 10^{-7}$ when $m_l$ is at the mass of the $\tau$ lepton. With nonlocal QED, both the muon and electron $g-2$ anomalies can therefore be reasonably well explained. In contrast to other theoretical methods, this calculation does not require the introduction of any new symmetries or new particles. The large positive discrepancy $\Delta a_\tau^{\text{nl}}$ for $\tau$ leptons can be tested by precise experiments in future. \\

\section{Gravitational form factors}
\label{sec:gravity}

While electromagnetic probes have been the primary tool used to study the structure of hadrons, characterizing the gravitational properties of hadrons can also reveal fundamental information about their internal structure. In the final example of a nonlocal field theory, we discuss its application to the calculation of the gravitational form factors, beginning first with a review of some basic elements of nonlocal gravity and the energy-momentum tensor.

\subsection{Nonlocal gravity}

We begin by introducing the nonlocal actions for pion-nucleon interactions in curved spacetime. Theoretically, the action in curved spacetime can be constructed by modifying the covariant derivatives and introducing the metric tensor $g^{\mu\nu}$ and vierbein fields $e^\mu_a$. To illustrate the gauge-field-like nature of gravity, we firstly review the ``gauge'' transformation properties of the nucleon, pion and gravitational fields. Under a general coordinate transformation, these transform according to~\cite{Donoghue:2017pgk, Poplawski:2007iz, Lavrov:2019nuz}
\begin{subequations}\label{eq:gravtransf}
\begin{eqnarray}
x^\mu &\to& x^\mu+\kappa \xi^\mu(x),
\\
\phi(x) &\to& \phi(x), 
\\
g^{\mu\nu}(x) &\to& g^{\mu\nu}(x)+\kappa[\partial^\mu \xi^\nu(x)+\partial^\nu \xi^\mu(x)],
\\
e^\mu_a(x) &\to& e^\mu_a(x)+\kappa\partial_a\xi^\mu(x),
\\
\partial_\mu &\to& \partial_\mu-\kappa\partial_\mu \xi^\alpha(x)\partial_\alpha,
\end{eqnarray}
\end{subequations}
where $\kappa^2=32\pi G$ represents the gravitational coupling constant, $\phi$ denotes the pion field, taking the form  $\phi=(\pi^+,\pi^-,\pi^0)$ and $\xi^\mu(x)$ is a infinitesimal vector parameter. Note that the gravitational metric field $g^{\mu\nu}$ and the vierbein field $e^a_\mu$ include both flat Minkowski and curved spacetime backgrounds. In order to manifestly illustrate the feature of curved spacetime, it is necessary to separate the two backgrounds from each other. This can be done in the weak gravitational background limit, where the metric $g^{\mu \nu}$, vierbein $e^\mu_a$, and $\sqrt{-g}$ fields (where  $g\equiv \det (g_{\mu\nu})$) can be expanded around the flat spacetime as~\cite{Holstein:2006bh, Donoghue:2017pgk, Bjerrum-Bohr:2004qcf}
\begin{subequations}
\begin{eqnarray}
g_{\mu \nu}&\equiv& \eta_{\mu \nu}+\kappa h_{\mu \nu },
\\
g^{\mu \nu}&=& \eta^{\mu \nu}-\kappa h^{\mu \nu }+\mathcal{O}(\kappa^2),
\\
\sqrt{-g}&=& 1+\frac{1}{2}\kappa h+\mathcal{O}(\kappa^2),
\\
e^\mu_a&=& \delta^\mu_a-\frac{\kappa}{2} \eta_{a\lambda }h^{\lambda\mu }+\mathcal{O}(\kappa^2), 
\\
e^a_\mu&=&\delta^a_\mu+\frac{\kappa}{2} \eta^{a\lambda }h_{\lambda\mu }+\mathcal{O}(\kappa^2), 
\end{eqnarray} \label{eq:16}%
\end{subequations}
where $\eta_{\mu \nu}$ is the Minkowski metric, $\delta^\mu_a$ the vierbein in flat spacetime, $h_{\mu \nu}$ the gravitational field, and $h = \eta_{\mu \nu} h^{\mu \nu}$. With Eqs.~(\ref{eq:16}), the scalar curvature, Ricci tensor, Christoffel symbol, and spin connection of Dirac fermion can then be written as~\cite{Donoghue:2017pgk, Holstein:2006bh}
\begin{subequations}
\begin{eqnarray}
R_{\mu \nu}&=&\frac{\kappa}{2}[\partial_\mu \partial^\lambda  h_{\lambda\nu}+\partial_\nu \partial^\lambda  h_{\lambda \mu}-\partial_\mu \partial_\nu h-\partial^2 h_{\mu\nu}]+\mathcal{O}(\kappa^2),
\\
R &=& \kappa[\partial^\mu \partial^\nu h_{\mu\nu }-\partial^2h]+\mathcal{O}(\kappa^2),
\\
\Gamma^\lambda_{\alpha \beta} 
&=& \frac{\kappa}{2}\,\eta^{\lambda\sigma} \left( \partial_\alpha h_{\beta\sigma}
+ \partial_\beta h_{\alpha\sigma} -  \partial_\sigma h_{\alpha\beta} \right)+\mathcal{O}(\kappa^2),
\\
\omega_\mu^{ab} &=&\frac{\kappa}{2}(\partial^bh^a_\mu-\partial^ah^b_\mu)+\mathcal{O}(\kappa^2),
\end{eqnarray}
\label{eq:17}
\end{subequations}
respectively.

Physically, nonlocal interactions involving hadrons can be considered more realistic than local interactions, given the non-pointlike nature of physical hadrons. This feature can be taken into account by defining a nucleon field at a spacetime point $x_\mu$ and displacing the meson or gauge field by a distance $a_\mu$ to spacetime point $x_\mu + a_\mu$ with a correlation function $F(a)$. To guarantee local gauge invariance, the Wilson line operator needs to be introduced. For the electromagnetic case, one has 
\begin{equation}
{\cal G}_\phi^q(x,y)
= \exp\left[ -ie^q_\phi \int_x^y dz^\mu \int\!d^4l\, F(l)\, A_\mu(z+l) \right],
\end{equation}
where $A^\mu(x)$ is the photon field. For the gravitational interaction, the local ``gauge transformation" could be a local coordinate translation, as in Eqs.~(\ref{eq:gravtransf}), with the corresponding ``gauge field" corresponding to the gravitational field. In Refs.~\cite{White:2011yy, Alawadhi:2021uie, Luna:2016idw} it was argued that the gravitational Wilson line is a double copy  of that of the gauge theory. In other words, the gravitational Wilson line operator can be obtained by replacing the gauge field in U(1) gauge link with gravitational field. By analogy with the U(1) gauge link, the gravitational Wilson line operator can be constructed as~\cite{Gravi}
\begin{subequations}
\begin{eqnarray}
W(x,y) &\equiv& \exp
\left[ -\frac{\kappa}{4}
\int^y_x \int d^4l\, F(l)\, h^{\mu\nu}(z+l)\, dz_{\{\mu} \partial_{\nu\}} 
\right], 
\\
W^\mu_\nu(x,y) &\equiv& 
\exp
\left[ \int^y_x \int d^4l\, F(l) 
\left( -\frac{\kappa}{4} \delta^\mu_\nu h^{\alpha\beta}(z+l)\, dz_{\{\alpha} \partial_{\beta\}} -\Gamma^\mu_{\rho\nu} (z+l)\, dz^\rho 
\right)
\right], 
\end{eqnarray}
\end{subequations}
where $F(l)$ is the correlation function for the  nonlocal interaction. The tensor $h^{\mu\nu}$ represents the gravitational field and transforms as
    $h^{\mu\nu} \to h^{\mu\nu} + \kappa \big[ \partial^\mu \xi'^\nu + 
\partial^\nu \xi'^\mu \big]$.
Note that here the infinitesimal parameters $\xi'^\mu$ for the gravitational field are different from $\xi^\mu$ for matter fields, which is similar to the local U(1) transformation. In the nonlocal framework, the pion kinetic term in curved spacetime can be written as~\cite{Alharazin, Gravi}
\begin{equation}
\label{eq:7}
\begin{split}
S^{(2),\rm nl}_{\phi\phi^\dagger}
&=\int d^4x \int d^4l\, F(l)
\sqrt{-g(x+l)}\,
\left[ 
\frac{g^{\mu \nu}(x+l)}{2} \partial_{\{\mu} \phi(x)  \partial_{\nu\}} \phi^\dagger(x)
- m_\phi^2\, \phi(x) \phi^\dagger(x)
\right],
\end{split}
\end{equation}
where $m_\phi$ is the pion mass and we assume the correlation function satisfies $\int\!d^4l\,F(l) = 1$. The mesonic action (\ref{eq:7}) is invariant under the coordinate transformations of Eqs.~(\ref{eq:gravtransf}) if $\xi'^\mu$ and ${\xi}^\mu$ are related by $\int d^4l\, F(l)\, \xi'^\mu(x+l) = \xi^\mu(x)$. The leading-order pion-nucleon action in curved spacetime can be constructed as~\cite{Alharazin, Gravi} 
\begin{eqnarray}
\label{eq:8}
S^{(1),\text {nl}}_{\phi N} 
&=& \int\!d^4x \int\!d^4l\, F(l) \sqrt{-g(x+l)}\,
\Bigg\{
\frac{i}{2} \overline N(x)\, e^\mu_a(x+l)\, \gamma^a \nabla_\mu N(x)
- \frac{i}{2} \nabla_\mu \overline N(x)\, e^\mu_a(x+l)\, \gamma^a N(x)
- M\, \overline N(x) N(x)
\nonumber\\ 
&& \hspace*{4.5cm}
+\ \delta_2
\bigg[ \frac{i}{2} \overline N(x)\, e^\mu_a(x)\gamma^a \nabla_\mu N(x)
- \frac{i}{2} \nabla_\mu \overline N(x)\, e^\mu_a(x) \gamma^a N(x)
\bigg]
- \delta_M\, \overline N(x) N(x)
\Bigg\} 
\nonumber\\
&-& \frac{C_{N\phi}}{f_\phi}\,
\int\!d^4x \int\!d^4a \int\!d^4l\, F(a)\, F(l)\, \sqrt {-g(x+l)}\,
\bar p(x) \gamma^a \gamma^5\, e^\mu_a(x+l)\, B(x)\, 
W^\nu_\mu(x,x+a)\, D_\nu \phi(x+a)
+ {\rm H.c.}
\nonumber\\
&-& i\,\frac{C_{\phi\phi^\dagger}}{2 f^2_\phi} 
\int\!d^4x \int\!d^4a \int\!d^4b \int\!d^4l\,
F(a)\, F(b)\, F(l)\, \sqrt {-g(x+l)}\, 
\bar p(x) \gamma^a\, e^\mu_a(x_l)\, p(x) 
\nonumber\\
&& \times
\Bigg\{  
\Big[ W^\nu_\mu(x,x+a)\, D_\nu\, \phi(x+a) \Big]\,
\Big[ W(x,x+b)\, \phi(x+b) \Big]^\dagger 
-\,
\Big[ W(x,x+b)\, \phi(x_b) \Big]
\Big[ W^\nu_\mu(x,x_a)\, D_\nu\, \phi(x+a) \Big]^\dagger
\Bigg\},
\nonumber\\
&& 
\end{eqnarray}
where $N(x) = [p(x), n(x)]^T$ denotes the intermediate nucleon (proton $p(x)$ or neutron $n(x)$) field, and $M$ is the physical mass of the nucleon. The constants $C_{N\phi}$ and $C_{\phi\phi^\dagger}$ represent the leading-order pion-nucleon coupling constants for different channels, as listed in Table~\ref{tab:11}. Additionally, $\delta_2$ and $\delta_M$ are the mass and wave function renormalization counterterm coefficients, related to the wave function renormalization constant $Z_2$ and bare nucleon mass $M_0$ via $\delta_2 = Z_2 - 1$ and $\delta_M = Z_2 M_0 - M$. The gravitational gauge link operators $W(x,x+a)\, \phi(x+a)$ and $W^\mu_\nu(x,x+a)\, \partial_\mu\phi(x+a)$ can be expanded in powers of the gravitational field $h_{\alpha\beta}(x)$ as
\begin{subequations}
\begin{eqnarray}
\label{eq:exp}
W(x,x+a)\, \phi(x+a)\!\!\!
&=&\!\!\! \phi(x+a) 
\nonumber\\
&-&\!\!\! \frac{\kappa}{4} \int d^4l\, F(l) 
  \int^{x+a}_x h^{\alpha\beta}(z+l)\, dz_{\{\alpha} \partial_{\beta\}}\phi(x+a)
+ {\cal O}(\kappa^2),
\\
\label{eq:Wilson}
W^\mu_\nu(x,x+a)\, \partial_\mu\phi(x+a)\!\!\!
&=&\!\!\! \partial_\nu\phi(x+a)
\nonumber\\
&+& \!\!\! \int d^4l\, F(l)
\int^{x+a}_x
\Big(
- \frac{\kappa}{4} \delta^\lambda_\nu\, h^{\alpha\beta}(z+l)\, dz_{\{\alpha} \partial_{\beta\}}\
-\ \Gamma^\lambda_{\mu\nu}(z+l)\, dz^\mu 
\Big)
\partial_\lambda\phi(x+a)
\nonumber\\
&+& \!\!\!{\cal O}(\kappa^2).
\end{eqnarray}
\end{subequations}
\begin{table}[b] 
\begin{adjustwidth}{-\extralength}{0cm}
\begin{center}
\caption{Effective pion-nucleon coupling constants  (for proton external states) $C_{N\phi}$ and $C_{\phi\phi^\dagger}$ for the leading-order $pN\phi$ and $pp\phi\phi$ interactions, respectively, and $C^{(1)}_{\phi\phi^\dagger}$, $C^{(2)}_{\phi\phi^\dagger}$ and $C^{(3)}_{\phi\phi^\dagger}$ for the next-to-leading order $pp\phi\phi$ interactions.}
\begin{tabular}{cc|c|cc|cc|cc}
\hline   
\hline
\multicolumn{2}{ c|}{ \hspace{0.0cm}$C_{N\phi}$ } \hspace{0.0cm}& 
$C_{\phi\phi^\dagger}$ & \multicolumn{2}{ c|}{ \hspace{0.0cm} $C^{(1)}_{\phi\phi^\dagger}$  } \hspace{-0.1cm}  &  
\multicolumn{2}{ c|}{ \hspace{0.0cm}$C^{(2)}_{\phi\phi^\dagger}$ } \hspace{0.0cm}&
\multicolumn{2}{ c}{ \hspace{0.0cm}$C^{(3)}_{\phi\phi^\dagger}$ } \hspace{0.0cm}  \\
 \hline
$(p\pi^0)$&
$(n\pi^+)$&
$(\pi^+\pi^-)$&
$(\pi^+\pi^-)$& $(\pi^0\pi^0)$&
$(\pi^+\pi^-)$&$(\pi^0\pi^0)$ & $(\pi^+\pi^-)$ & \ $(\pi^0\pi^0)$ \  \\
\hline
 $\frac12 {g_A}$&  $\frac{1}{\sqrt{2}} g_A$ &$\frac{1}{2}$ & $-4$ &$-2$& $-\frac{1}{2}$ & $-\frac{1}{4}$ &  2  &  1   \\
\hline
\hline
\end{tabular}
\label{tab:11}
\end{center}
\end{adjustwidth}
\end{table}
Note that the last two terms of Eq.~(\ref{eq:Wilson}) yield an additional gauge link vertex which is crucial for gauge invariance of the total energy-momentum tensor in the nonlocal case. These terms vanish in the local limit, $F(a) \to \delta^4(a)$, and the nonlocal action (\ref{eq:8}) reduces to the local one. In a similar manner, the nonlocal action for the next-to-leading order  pion-nucleon interaction in curved spacetime is constructed as~\cite{Alharazin, Gravi}
\begin{eqnarray}
\label{eq:10}
S^{(2),\text{nl}}_{\phi N} 
&=& 4 c_1 m^2_\phi 
\int\!d^4x \int\!d^4l\, F(l)\,\sqrt {-g(x+l)}\,
\bar p(x) p(x)
\nonumber\\
&+& c_1 \frac{m^2_\phi C^{(1)}_{\phi\phi^\dagger}}{f^2_\phi}
\int\!d^4x \int\!d^4a \int\!d^4b \int\!d^4l\,
F(a)\, F(b)\, F(l)\, \sqrt{-g(x+l)}\,
\bar p(x) p(x)
\nonumber \\
&& \hspace*{1.5cm} \times
\Big[ W(x,x+a) \phi(x+a) \Big] 
\Big[ W(x,x+b) \phi(x+b) \Big]^\dagger 
\nonumber \\
&+& c_2 \frac{C^{(2)}_{\phi\phi^\dagger}}{M^2 f^2_\phi}
\int\!d^4x \int\!d^4a \int\!d^4b \int\!d^4l\,
F(a)\, F(b)\, F(l)\, \sqrt{-g(x+l)}\, g^{\alpha\mu}(x+l) g^{\beta \nu}(x+l) \nonumber\\
&& \hspace*{1.5cm} \times
\Big(
  \bar p(x) \nabla_\alpha \nabla_\beta p(x)
+ \nabla_\alpha \nabla_\beta \bar p(x) p(x)
\Big) 
\Big[ W^\rho_{\{\mu}(x,x+a)\, D_{\rho}\, \phi(x+a) \Big]
\Big[ W^\lambda_{\nu\}}(x,x+b)\, D_{\lambda}\, \phi(x+b) \Big]^\dagger 
\nonumber\\
&+& c_3 \frac{C^{(3)}_{\phi\phi^\dagger}}{2 f^2_\phi} 
\int\!d^4x \int\!d^4a \int\!d^4b \int\!d^4l\,
F(a)\, F(b)\, F(l)\, \sqrt{-g(x+l)}\,
\bar p(x) g^{\mu \nu}(x+l) p(x)
\nonumber\\
&& \hspace*{1.5cm} \times
\Big[ W^\alpha_{\{\mu}(x,x+a)\, D_{\alpha}\, \phi(x+a) \Big]
\Big[ W^\beta_{\nu\}}(x,x+b)\, D_{\beta}\, \phi(x+b) \Big]^\dagger,
\\
&& 
\nonumber
\end{eqnarray}
where $C^{(1)}_{\phi\phi^\dagger}$, $C^{(2)}_{\phi\phi^\dagger}$ and $C^{(3)}_{\phi\phi^\dagger}$ are the pion-nucleon coupling constants at next-to-leading order for different channels, listed in Table~\ref{tab:11}. Finally, the nonlocal version of nonminimal coupling between the nucleon and the gravitational field can be written as~\cite{Alharazin, Gravi}
\begin{eqnarray}
\label{eq:14}
S^{\rm nl}_{\rm nonmin}
&=& \int\!d^4x \int\!d^4l\,
F(l)\, \sqrt{-g(x+l)}\,
\biggl\{
\frac{c_8}{8}\, R(x+l) \overline N(x) N(x)
\nonumber\\
&& \hspace*{1.5cm}
+\ i\frac{c_9}{M} R^{\mu\nu}(x+l) 
\Big[
  \overline N(x)\, e_\mu^a(x+l)\, \gamma_a \nabla_\nu N(x)
- \nabla_\nu \overline N(x)\, e_\mu^a(x+l)\, \gamma_a N(x) 
\Big]
\biggr\}.
\\
&& 
\nonumber
\end{eqnarray}
Expanding the nonlocal actions around the flat spacetime background using Eqs.~(\ref{eq:16}) and (\ref{eq:17}), and substituting these into Eqs.~(\ref{eq:7}), (\ref{eq:8}), (\ref{eq:10}) and (\ref{eq:14}), one can obtain explicit actions for the interaction between the matter field and the gravitational field $h^{\mu \nu}$. For example, the leading order mesonic action of Eq.~(\ref{eq:7}) can be rewritten as 
\begin{eqnarray}
\label{eq:18}
S^{(2),{\rm nl}}_{\phi\phi^\dagger}
&=& \int\!d^4x \int\!d^4l\, F(l) 
\biggl\{
\partial_\mu \phi(x)\, \partial^\mu\phi^\dagger(x)
- m_\phi^2\, \phi(x) \phi^\dagger(x) 
\nonumber\\
&&
+\ \frac12 \kappa h(x+l)
\Big[
  \partial_\mu \phi(x)\, \partial^\mu \phi^\dagger(x)
  - m_\phi^2\, \phi(x) \phi^\dagger(x)
\Big]
- \frac12 \kappa h^{\mu\nu}(x_l)
  \partial_{\{\mu} \phi(x)\, \partial_{\nu\}} \phi^\dagger(x) 
\biggr\}
\nonumber\\
&&
+\ {\cal O}(\kappa^2),
\end{eqnarray}
where the first two terms represent the pion kinetic action in flat spacetime, and the last three terms represent the lowest-order interaction between the pion and the gravitational field. Similarly, the leading-order pion–nucleon interaction in Eq.~(\ref{eq:8}) can be expanded as
\begin{eqnarray}
\label{eq:19}
S^{(1),\rm nl}_{\phi N}
&=& \int\!d^4x \int\!d^4l\, F(l)
\Bigg\{
- (M + \delta_M)  
\Big( 1+\frac12\kappa h(x+l) \Big) 
\overline N(x) N(x)
\nonumber\\
&& \hspace*{2.5cm}
+ \frac{i}{2}\, (1+\delta_2) \Big( 1+\frac12\kappa h(x+l) \Big)
\bigg[
  \overline N(x) \gamma^\mu \partial_\mu N(x)
- \partial^\mu \overline N(x) \gamma_\mu N(x)
\bigg]
\nonumber\\
&& \hspace*{2.5cm}
- \frac{i}{4} (1+\delta_2) \kappa h^{\mu\nu}(x+l) 
\bigg[
  \overline N(x) \gamma_\mu \partial_\nu N(x)
- \partial_\mu \overline N(x)\, \gamma_\nu N(x)
\bigg]
\Bigg\}
\nonumber\\
&-& \frac{C_{N\phi}}{f_\phi} 
\int\!d^4x \int\!d^4l \int\!d^4a\,
F(a) F(l)\, \bar p(x) \gamma^\mu \gamma^5 N(x)
\Bigg\{
\Big( 1+\frac12\kappa h(x+l) \Big) 
\nonumber\\
&& \hspace*{2.cm}
\times 
\bigg[
\partial_\mu \phi(x+a)
- \frac12 \kappa h^{\nu\lambda}(x+l) \eta_{\lambda\mu}\partial_\nu \phi(x+a) 
\nonumber\\
&& \hspace*{2.5cm}
- \frac{\kappa}{4} \int^{x+a}_x h^{\alpha\beta}(z+l)\, dz_{\{\alpha} \partial_{\beta \}} \partial_\mu\phi(x+a)
\nonumber\\
&& \hspace*{2.5cm}
- \int^{x+a}_x \Gamma^{\alpha}_{\beta\mu}(x+l)\, dz^\beta\partial_\alpha \phi(x+a) 
+ {\rm H.c.}
\bigg]
\Bigg\}
\nonumber\\
&-& i\frac{C_{\phi\phi^\dagger}}{2 f^2_\phi} 
\int\!d^4x \int\!d^4l \int\!d^4a \int\!d^4b\,
F(a)\, F(b)\, F(l) \,\bar p(x) \gamma^\mu p(x)
\nonumber\\
&& \hspace*{1cm}
\times 
\Bigg\{
\Big( 1+\frac12 \kappa h(x+l) \Big)
\bigg[
  \partial_\mu\phi(x+a) \phi^\dagger(x+b)
- \phi(x+b)\, \partial_\mu\phi^\dagger(x+a)
\bigg]
\nonumber\\
&& \hspace*{1.5cm}
- \frac12 \kappa \eta_{\mu\lambda} h^{\nu\lambda}(x+l)
\bigg[
  \partial_\nu\phi(x+a) \phi^\dagger(x+b)
- \phi(x+b)\, \partial_\nu\phi^\dagger(x+a)
\bigg]
\nonumber\\
&& \hspace*{1.5cm}
- \phi^\dagger(x+a) 
\int_x^{x+b}\, dz^\lambda\, \Gamma^\alpha_{\lambda\mu}(z+l)\, \partial_\alpha \phi(z+b)
\nonumber\\
&& \hspace*{1.5cm}
- \frac14\, \partial_\mu\, \phi(x+b) \int^{x+a}_x\, dz_{\beta}\, h^{\alpha\beta}(x+l)\, \partial_\alpha \phi^\dagger(x+a)
\nonumber\\
&& \hspace*{1.5cm}
- \frac14 \phi^\dagger(x+b) \int^{x+a}_x\, dz_{\beta}\, h^{\alpha\beta}(x+l)\, \partial_\mu \partial_\alpha \phi(x+a)
+ {\rm H.c.}
\Bigg\}
\nonumber\\
&+& {\cal O}(\kappa^2),
\\
&& 
\nonumber
\end{eqnarray}
where the $\kappa$-independent terms represent the leading-order strong interaction between pion and nucleon in flat spacetime, while terms proportional to $\kappa$ are interactions between the gravitational and matter fields. The terms with the path integral from $x$ to $x+a$ in Eq.~(\ref{eq:19}) are from the  gravitational Wilson line operator of Eq.(\ref{eq:Wilson}). Such terms will generate additional energy-momentum tensor vertices in the nonlocal case. In the same way, the weak field expansion of the next-to-leading order action of Eq.~(\ref{eq:10}) yields
\begin{eqnarray}
\label{eq: c3link}
S^{(2),\rm nl}_{\phi N}
&=& 4\,c_1 m^2_\phi
\int\!d^4x \int\!d^4l\, F(l)\,
\Big( 1 + \frac12 \kappa h(x+l) \Big)
\bar p(x) p(x)
\nonumber\\
&+& c_1\, m^2_\phi\, \frac{C^{(1)}_{\phi\phi^\dagger}}{f^2_\phi} \int\!d^4x \int\!d^4l
\int\!d^4a \int\!d^4b\, F(a)\, F(b)\, F(l) \,
\bar p(x) p(x) 
\nonumber\\
&& 
\times
\biggl\{  
\Big[ 1 + \frac{\kappa h(x+l)}{2} \Big] 
  \phi(x+a) \phi^\dagger(x+b)
- \frac{\kappa}{4} \phi(x+a)
  \int^{x+b}_x h^{\alpha\beta}(z+l)\,
  dz_{\{\alpha} \partial_{\beta\}} \phi^\dagger(x+b)
+ {\rm H.c.}
\biggl\}
\nonumber\\
&+& c_2 \frac{C^{(2)}_{\phi\phi^\dagger}}{M^2 f^2_\phi}
\int\!d^4x \int\!d^4l\, \int\!d^4a \int\!d^4b\, F(a)\, F(b)\, 
F(l)\, 
\biggl\{
\Big[ 
\bar p(x) \partial^{\alpha } \partial^{\beta} p(x) 
+ \partial^{\beta} \partial^{\alpha} \bar p(x) p(x)
\Big]
\nonumber\\
&& \times
\biggl[ 
\Big( 1 + \kappa \frac{h(x+l)}{2} \Big)
\partial_{\{\alpha}\phi(x+a)\partial_{\beta\}}\phi^\dagger(x+{b}) 
\nonumber\\
&& \hspace*{0.5cm} 
-\ \kappa\, h^{\beta \nu}(x+l)\partial_{\{\nu}\phi(x+a)\partial^{\alpha\}}
\phi^\dagger(x+b)  
-\ \kappa\, h^{\alpha \mu}(x_l) \partial^{\{\beta} \phi(x+a) \partial_{\mu\}} \phi^\dagger(x+b) 
\nonumber\\
&& \hspace*{0.5cm} 
-\ \frac{\kappa}{4} 
  \int^{x+a}_x h^{\mu \nu}(z+l) dz_{\{\mu }\partial_{\nu}\} \partial^{\{\alpha}\phi(x+a)\partial^{\beta\}}
\phi^\dagger(x+b)
\nonumber\\
&& \hspace*{0.5cm} 
-\ \kappa \int^{x+a}_x \Gamma^{\lambda}_{\sigma\{\alpha}(z+l) dz^\sigma \partial_{\lambda} \phi(x+a) \partial_{\beta\}} \phi^\dagger(x+b)
+ {\rm H.c.} 
\biggr]
\nonumber\\
&& +
\biggl[
- \frac{\kappa}{2} 
\left( 
  \partial_\alpha h_{\beta\sigma} 
+ \partial_\beta h_{\alpha\sigma} 
- \partial_\sigma h_{\alpha\beta} 
\right)(x+l)
\bar p(x) \partial^{\sigma} p(x)
\nonumber\\
&& \hspace*{0.5cm}
- \frac{\kappa}{2}
\left( 
  \partial_\alpha h_{\beta\sigma}
+ \partial_\beta h_{\alpha\sigma} 
- \partial_\sigma h_{\alpha\beta} 
\right)(x+l)
\partial^{\sigma} \bar p(x) p(x) 
\nonumber\\
&& \hspace*{0.5cm} 
+\ \kappa
\Big(
\frac{i}{2} \bar p(x) \partial^b h^a_{\{\alpha}(x+l) \sigma_{ab} \partial_{\beta\}} p(x)
- \frac{i}{2}\partial_{\{\beta} \bar p(x) \partial^b h^a_{\alpha\}}(x_l)   \sigma_{ab} p(x)
\Big)
\biggr]
\partial^{\{\alpha} \phi(x+a) \partial^{\beta\}} \phi^\dagger(x+b)
\biggr\}
\nonumber\\
&+& c_3\, \frac{C^{(3)}_{\phi\phi^\dagger}}{2f^2_{\phi}}
\int\!d^4x \int\!d^4l\, \int\!d^4a \int\!d^4b\, F(a)\, F(b)\, F(l)\, \bar p(x) p(x)
\nonumber\\
&& \times
\biggl\{
2 \Big( 1 + \frac{\kappa}{2} h(x+l) \Big)
\partial^\mu\phi(x+a) \partial_\mu\phi^\dagger(x+b) 
- \kappa \partial_{\{\nu}\phi(x+a)\partial_{\mu\}} \phi^\dagger(x+b) h^{\mu\nu}(x+l) 
\nonumber\\
&& \hspace*{0.5cm} 
-\ \frac{\kappa}{2} \partial^\mu\phi(x+a) 
\int^{x+b}_x h^{\alpha \beta}(z+l) dz_{\{\alpha} \partial_{\beta\}} \partial_\mu\phi^\dagger(x+b) 
\nonumber\\
&& \hspace*{0.5cm} 
-\ 2\partial^\mu\phi(x+a) 
\int^{x+b}_x \Gamma^{\alpha}_{\beta\mu}(z+l)  dz^{\beta} \partial_{\alpha} \phi^\dagger(x+b) + {\rm H.c.}
\biggr\}
\nonumber\\
&+& {\cal O}(\kappa^2).
\\
&&
\nonumber
\end{eqnarray}
Similarly, the weak gravitational field expansion of the nonlocal nonminimal action of Eq.~(\ref{eq:14}) yields
\begin{eqnarray}
S^{\rm nl}_{\rm nonmin}
&=& \int\!d^4x \int\!d^4l\, F(l)\,
\Bigg\{ 
\kappa\frac{c_8}{8}
\Big[
  \partial^\mu \partial^\nu h_{\mu\nu }(x+l)
- \partial^2 h(x+l)
\Big] \overline N(x) N(x)
\nonumber\\
&&
+\ \kappa \frac{i c_9}{2M}  
\Big[ 
  \partial_\mu \partial^\lambda h_{\lambda\nu}(x+l)
+ \partial_\nu \partial^\lambda h_{\lambda \mu}(x+l)
- \partial_\mu \partial_\nu h(x+l)
- \partial^2 h_{\mu\nu}(x+l)
\Big] 
\nonumber\\
&& \hspace*{0.9cm} \times
\Big[
  \overline N(x) \gamma^\mu \partial^\nu N(x) 
- \partial^\mu \overline N(x) \gamma^\nu N(x) 
\Big]
\Bigg\}
\nonumber\\
&+& {\cal O}(\kappa^2).
\label{eq:27}
\end{eqnarray}

\subsection{Energy-momentum tensor}

Having established these actions in curved spacetime, the next goal is to derive the nucleon energy–momentum tensor within the nonlocal framework. The electromagnetic interaction between a gauge field and a current can be written as $\mathcal{L}_{\rm em} = -eJ^\mu A_\mu$. Similarly, the gravitational interaction between a matter field and the graviton can be expressed as 
$\mathcal{S}_{\rm G} 
= -\frac12 \kappa \int d^4x\, \sqrt{-g}\, T_{\mu\nu}(x)\, h^{\mu\nu}(x)$ \cite{Holstein:2006bh, Bjerrum-Bohr:2004qcf}. The symmetric energy–momentum tensor for the matter field can thus be obtained from the nonlocal actions in curved spacetime by minimizing the action with respect to the gravitational field~$h^{\mu\nu}$,
\begin{equation}
T_{\mu\nu}(x) = -\frac{2}{\kappa} \frac{\delta S_{\rm G} }{\delta h^{\mu\nu}(x)}.
\label{eq:tensor}
\end{equation}
Using this expression for the energy–momentum tensor, the pionic energy–momentum tensor can be obtained as
\begin{eqnarray}
T^{(2),\rm nl}_{\mu\nu,\phi\phi}(x) 
&=& 
\int d^4 F(l)
\Big\{ \partial_{\{\mu} \phi(x-l) \partial_{\nu \}} \phi^\dagger(x-l)
- \eta_{\mu\nu} 
  \big[ 
  \partial^\alpha \phi(x-l) \partial_\alpha \phi(x-l) 
  - m^2_\pi \phi(x-l) \phi^\dagger(x-l) 
  \big]
\Big\}.
\label{eq: eq22}
\end{eqnarray}
Similarly, from Eqs.~(\ref{eq:19}), (\ref{eq: c3link}) and (\ref{eq:27}), one can obtain the energy–momentum tensor associated with the leading- and next‑to‑leading‑order pion–nucleon interactions, as well as the nucleon–gravity nonminmal couplings. The explicit expressions can be found in Ref.~\cite{Gravi}.

\subsection{Gravitational form factors}

In both experimental and theoretical analyses, the amplitude for a gauge field scattered by incident particles can be parameterized in terms of a generalized vertex between the gauge field and the composite particle to be detected. The generalized vertex operator can be expressed in terms of Lorentz‑covariant scalar functions and their corresponding Lorentz structures. These Lorentz‑invariant scalar functions, or form factors, encode internal information about the dynamical properties of particles, such as their charge, mass, pressure, and shear‑force distributions.

If the graviton is treated as a gauge field, the nucleon–nucleon–graviton vertex operator can likewise be parameterized in terms of Lorentz‑invariant gravitational form factors. Similar to the case of electromagnetic form factors, the matrix elements of the energy–momentum tensor, or equivalently the nucleon–nucleon–graviton vertex for the nucleon, are parameterized as
 \cite{Polyakov:2002yz, Polyakov:2018zvc, Diehl:2003ny}
\begin{eqnarray}
\langle p'| T_{\mu\nu}| p \rangle 
&=& \bar u(p')\, \Gamma_{\mu\nu}(p,q)\, u(p)
\notag\\
&=& \bar u(p') 
\Big[ A(t) \frac{\gamma_{\{\mu} P_{\nu\}}}{2} 
  + B(t) \frac{P_{\{\mu} i\sigma_{{\nu\}}\alpha} q^\alpha }{4M}
  + D(t) \frac{q_\mu q_\nu-\eta_{\mu\nu} q^2}{4M}
  + M \bar c(t) \eta_{\mu \nu} 
\Big] u(p),
\label{eq:e1}
\end{eqnarray}
where $P = \frac12 (p + p')$, $q = p' - p$, and $t = q^2 = -Q^2$, with $\eta_{\mu \nu}$ the flat‑spacetime metric. The functions $A(t)$, $B(t)$, $D(t)$, and $\bar{c}(t)$ are the GFFs of the nucleon. Conservation of the energy–momentum tensor, $q_\mu T^{\mu \nu} = 0$, requires that the total $\bar{c}(t)$ from loop contributions vanish, $\sum_i \bar{c}_{i}(t) = 0$. At zero momentum transfer squared, $t=0$, the GFFs satisfy the normalization conditions,
\begin{equation}
A(0)=1,
\qquad
B(0)=0,
\qquad
J(0)=\frac{1}{2}[A(0)+B(0)] = \frac12.
\end{equation}
The GFFs can be projected from the energy momentum tensor using projection operators \cite{Knecht:2001qf, Czarnecki:1996rx, Brodsky:1966mv},
\begin{subequations}
\begin{eqnarray}
P^{\mu\nu}_A\!\!
&=&\!\! (\!\not\! p+M)\left(-\frac{\gamma^{\{\mu} P^{\nu\}}}{\left(4 M^2-t\right)^2}+\frac{20 M  P^{\mu } P^{\nu }}{\left(4 M^2-t\right)^3}+\frac{M q^{\mu } q^{\nu }}{t \left(4 M^2-t\right)^2}-\frac{M \eta ^{\mu \nu }}{\left(4 M^2-t\right)^2}\right)(\!\not\! p'+M),
\\
P^{\mu\nu}_B\!\!
&=&\!\! (\!\not\!p+M)\left(\frac{4 M^2 \gamma^{\{\mu} P^{\nu\}} }{t \left(4 M^2-t\right)^2}-\frac{4M \left(8 M^2+3 t\right)  P^{\mu } P^{\nu }}{t \left(4 M^2-t\right)^3}-\frac{M q^{\mu } q^{\nu }}{t \left(4 M^2-t\right)^2}+\frac{M \eta^{\mu \nu }}{\left(4 M^2-t\right)^2}\right)(\!\not\! p'+M),
\\
P^{\mu\nu}_D\!\!
&=&\!\! (\!\not\!p+M)\left(\frac{4M P^{\mu }  P^{\nu }}{t \left(4 M^2-t\right)^2}+\frac{3 M q^{\mu } q^{\nu }}{t^2 \left(4 M^2-t\right)}-\frac{M\, \eta^{\mu \nu }}{t \left(4 M^2-t\right)}\right)(\!\not\! p'+M),
\\
P^{\mu\nu}_{\bar c}\!\!
&=&\!\! (\!\not\!p+M) \frac{q^{\mu } q^{\nu } }{2 M t \left(4 M^2-t\right)} (\!\not\!p'+M),
\label{eq:project}
\end{eqnarray}
\end{subequations}
such that
\begin{subequations}
\begin{eqnarray}
A(t) &=& {\rm Tr}[P^{\mu \nu }_A\Gamma_{\mu \nu }],
\\
B(t) &=& {\rm Tr}[P^{\mu \nu }_B\Gamma_{\mu \nu }],
\\
D(t) &=& {\rm Tr}[P^{\mu \nu }_D\Gamma_{\mu \nu }],
\\
\bar c(t) &=& {\rm Tr}[P^{\mu \nu }_{\bar c}\Gamma_{\mu \nu }].
\end{eqnarray}
\end{subequations}

\begin{figure}[t] 
\begin{adjustwidth}{-\extralength}{0cm}
\begin{minipage}[b]{.45\linewidth}
\hspace*{-0.3cm}\includegraphics[width=1.1\textwidth, height=5.5cm]{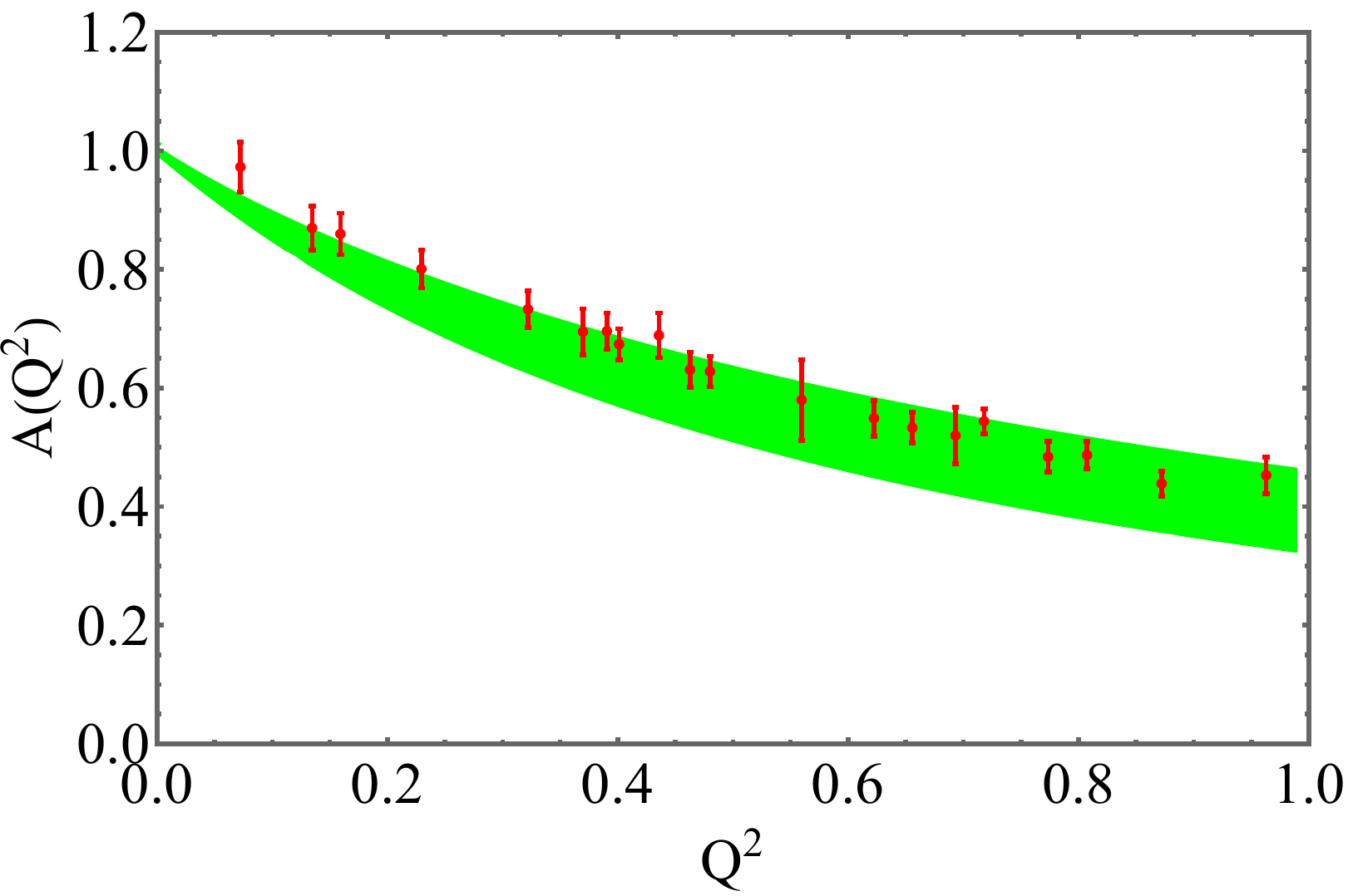} 
 \vspace{6pt}
\end{minipage}
\hfill
\begin{minipage}[b]{.45\linewidth}   
\hspace*{-0.95cm} \includegraphics[width=1.1\textwidth, height=5.5cm]{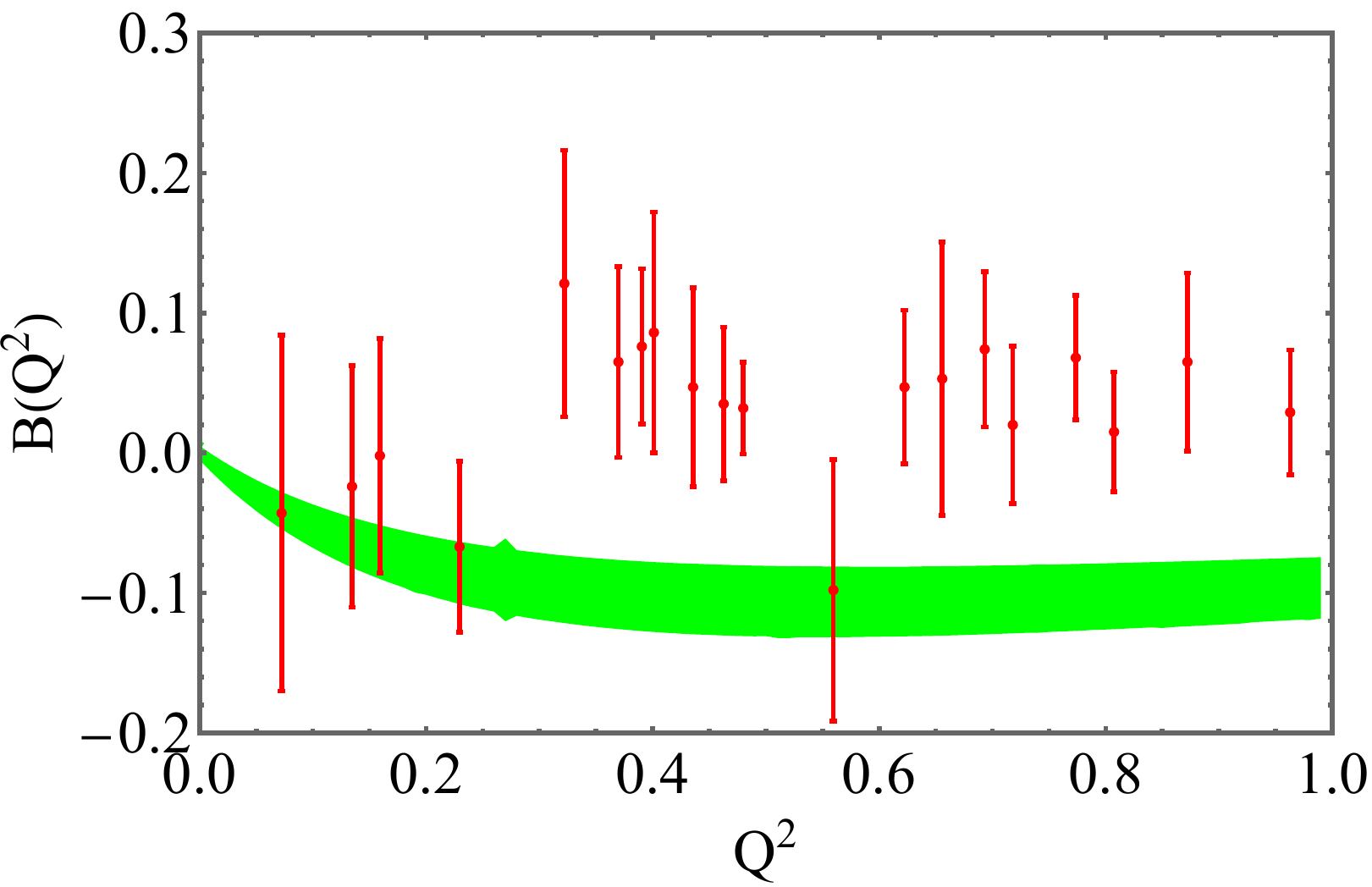} 
   \vspace{6pt}
\end{minipage}  
\\[-0.4cm]
\begin{minipage}[t]{.45\linewidth}
\hspace*{-0.35cm}\includegraphics[width=1.12\textwidth, height=5.5cm]{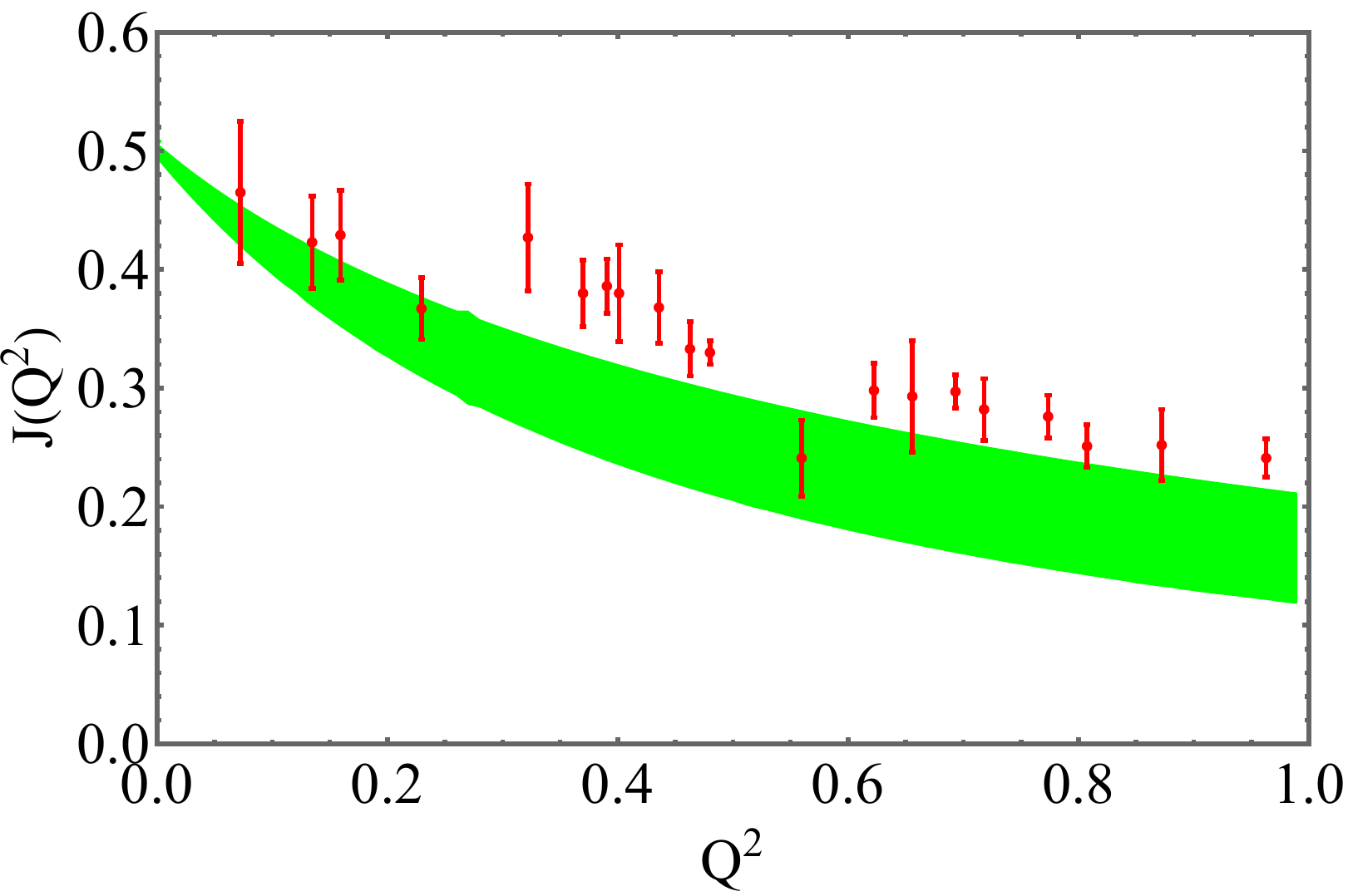}   
 \vspace{-6pt}
\end{minipage}
\hfill
\begin{minipage}[t]{.44\linewidth}   
\hspace*{-0.88cm} \includegraphics[width=1.1\textwidth, height=5.35cm]{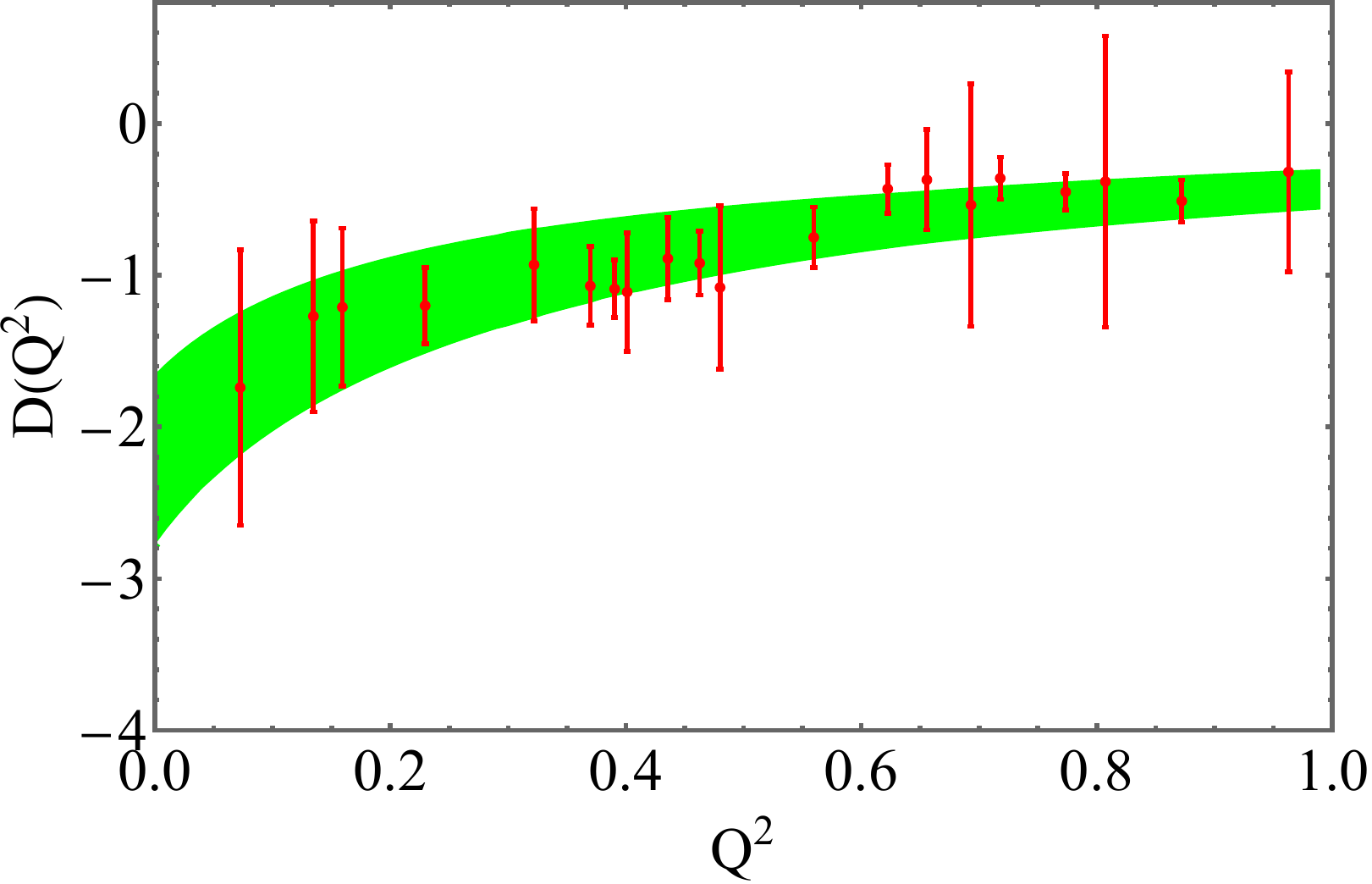}  
   \vspace{-6pt}
\end{minipage} 
\caption{Gravitational form factors $A$, $B$, $D$ and the angular momentum $J$ as a function of $Q^2 = -t$ [in units of GeV$^2$] from the nonlocal calculation (green bands) compared with lattice data \cite{Hackett:2023rif} (red points). The band represents the uncertainty on the cutoff parameter, $\Lambda$.}
\label{fig:fig8}
\end{adjustwidth}
\end{figure}

The GFFs can be calculated numerically using the nonlocal renormalization method. In the nonlocal case, in addition to the chiral low energy constants and nonminimal coupling constants, there is a further cutoff parameter, $\Lambda=(1.0 \pm 0.1)$~GeV~\cite{Salamu:2019dok}, that is determined phenomenologically. The low energy constants 
$c_1=-0.22~\rm GeV^{-1}$,
$c_2 =-0.43~\rm GeV^{-1}$, and
$c_3=-0.11~\rm GeV^{-1}$ are obtained by fitting the lattice data for the nucleon mass, and two additional coupling constants $c_8$ and $c_9$ are determined by the lattice data on the GFFs. With these parameters, in Fig.~\ref{fig:fig8} we show the comparison between the nonlocal model results and recent lattice data \cite{Hackett:2023rif}. The nonlocal model results for the $A$ and $D$ GFFs are in good agreement with the lattice data over a wide region $0 \leq Q^2 \leq 1$~GeV$^2$. This indicates that with the non-point particle assumption, the model can explain the data over a wider region.

In contrast with the lattice data and the local result, the nonlocal $B$ form factor is always negative and decreases with the momentum transfer squared~$Q^2$, which may provide a new opportunity for testing the nonlocal effect of the pion and graviton. Moreover, the $Q^2$ distribution of the angular momentum of the proton lies slightly below the lattice data due to the negativity of the $B$ form factor. By performing least $\chi^2$ analysis, we found that the best values for the nonminimal coupling constants are
$c_8=(-1.20 \pm 0.47)~\rm GeV^{-1}$ and
$c_9=(0.33 \pm 0.05)~\rm GeV^{-1}$, 
giving an overall $\chi^2=0.64$ and $\chi^2=0.42$, respectively. This gives rise to a best fit value for the GFF $D$ as $D(0)=-2.21 \pm 0.56$. \\

\section{Summary}\label{sec:summary}

Nonlocal Lagrangians have been a useful tool for more naturally accommodating the extended nature of hadrons than is possible via local Lagrangians. An important advantage of a nonlocal Lagrangian formulation is the presence of correlation functions describing the nonlocal behavior, whose Fourier transforms are momentum-dependent regulators that render loop integrals finite and provide a gauge-invariant way for removing divergences. In this review we have presented several applications of nonlocal effective field theory that have been of recent phenomenological interest, including nonlocal chiral effective theory of baryons and mesons, nonlocal QED, and an extension of nonlocal effective theory to curved spacetime.

We have described the use of nonlocal chiral effective theory to study hadron properties up to relatively large momentum transfers, beyond the traditional power-counting regime of chiral perturbation theory. We focused in particular on the calculation of nucleon GPDs, including baryon octet and decuplet intermediate states in the computation of the chiral splitting functions at one-loop level, for both zero and nonzero skewness. To ensure local gauge invariance, gauge links are introduced, which generate additional diagrams that guarantee charge conservation. The derived convolution forms allow one to study the three dimensional structure of the nucleon through GPDs as a function of momentum fraction $x$, skewness $\xi$, and momentum transfer $t$, as well as sea quark flavor asymmetries of electric and magnetic GPDs at zero and nonzero momentum transfers.

The nonlocal chiral effective theory was also applied to the calculation of the nucleon gravitational form factors. As with the electromagnetic form factors which are related to the electromagnetic vector current, the gravitational form factors are related to the current associated with the energy-momentum tensor. Extending the formulation to curved spacetime, the symmetric energy-momentum tensor was obtained from the derivative of the nonlocal effective action with respect to the gravitational field $h_{\mu\nu}$. Again, there are additional diagrams generated from the path integral of $h_{\mu\nu}$, which assures the correct normalization for the GFFs.

The extension of the fundamental QED interaction to the nonlocal case involved constructing the most general nonlocal QED Lagrangian, leading to a modification of lepton and photon propagators. The modified propagators can also be obtained from the canonical quantization, associated with new quantization conditions (solid quantization), where the $\delta$ function in the commutation relation is replaced by a correlation function. The introduction of the gauge link increases the number of diagrams in nonlocal QED compared with the local QED ({\it e.g.}, 7 self-energy and 24 vertex Feynman diagrams at one-loop level in the nonlocal case, compared with one self-energy and one vertex diagram in local QED). The modified Ward-Green-Takahashi identity, crucial for maintaining charge conservation, was derived at one-loop level in the nonlocal formulation. As an application, the nonlocal QED was used to explore the lepton $g-2$ anomalies, with both the electron and muon $g-2$ anomalies able to be accounted for without the introduction of new particles or interactions.

In the future, these investigations can be developed in several directions. The nonlocal chiral effective theory can be extended to the study of both chiral-even (helicity-conserving) and chiral-odd (helicity-flipping) GPDs, as well as to $T$-even and $T$-odd transverse momentum dependent distributions, and higher-twist functions. The calculations can also be generalized to study other external hadronic states, including octet and decuplet baryons, as well as pseudoscalar mesons. For the lepton anomalous magnetic moment anomalies, it is remarkable that the nonlocal QED formulation can provide a reasonable description of the discrepancies, without the need for new particles or interactions. It will be interesting to test the large positive discrepancy $\Delta a_\tau^{\text{nl}}$ for $\tau$ leptons, which may further constrain the nonlocal theory and validate the effectiveness of nonlocal QED.

\vspace{8pt} 

\authorcontributions{Conceptualization, C.J., W.M. and P.W.; methodology, F.H., C.J., W.M., Y.S. and P.W.; software, Z.Y.G., F.H. and Y.S.; writing---original draft preparation, P.W.; writing---review and editing, Z.Y.G., F.H., C.J., W.M. and Y.S.; supervision, C.J., W.M. and P.W.; funding acquisition, C.J., W.M. and P.W. All authors have read and agreed to the published version of the manuscript.}

\funding{This research was supported by NSFC under Grant No.~12475088, National Science Foundation under award PHY-2412963, the DOE Contract No. DE-AC05-06OR23177, under which Jefferson Science Associates, LLC operates Jefferson Lab, DOE Contract No. DE-FG02-03ER41260, and in part within the framework of the Quark-Gluon Tomography (QGT) Topical Collaboration, under Contract No. DE-SC0023646.}

\dataavailability{The data presented in this study are included in the article.} 

\conflictsofinterest{The authors declare no conflicts of interest.} 

\end{adjustwidth}

\begin{adjustwidth}{-\extralength}{0cm}

\clearpage
\reftitle{References}

\PublishersNote{}
\end{adjustwidth}
\end{document}